\documentclass[twoside, 12pt]{book}

\def\mytitle{Personalization and Evaluation of Conversational Information Access}
\def\myauthor{Hideaki Joko}
\def\mydate{\formatdate{22}{03}{2023}}

\usepackage[utf8]{inputenc}
\usepackage[T2A,T1]{fontenc}
\usepackage{lmodern}
\usepackage{microtype}
\usepackage{tipa}
\usepackage{stmaryrd}
\usepackage{textcomp}
\usepackage{titlecaps}
\usepackage{amsmath}
\usepackage{natbib}
\usepackage{bibentry}
\nobibliography*
\usepackage{fnpct}
\usepackage{multicol}

\DeclareMathOperator*{\argmax}{arg\,max}

\usepackage{amssymb}
\usepackage{relsize}
\usepackage{siunitx}
\usepackage{xcolor} 
\usepackage{tikz} 
\usepackage{pdflscape}
\usepackage{changepage}

\definecolor{ruhuisstijlrood}{rgb}{0.745,0.192,0.102}

\usepackage{geometry}
\usepackage{graphicx}
\usepackage{pdflscape}
\usepackage{float}
\usepackage{fancyhdr}
\usepackage{caption}
\usepackage{subcaption}
\usepackage{rotating}
\usepackage{subcaption}
\usepackage{bm}

\usepackage{booktabs}
\usepackage{multirow}
\usepackage{tabularx}
\usepackage{soul}
\usepackage{longtable}
\usepackage{threeparttable}
\usepackage{xltabular}

\definecolor{lightgreen}{HTML}{86f986}
\definecolor{lightred}{HTML}{f98686}
\newcommand{\hlgreen}[1]{\sethlcolor{lightgreen}\hl{#1}}
\newcommand{\hlred}[1]{\sethlcolor{lightred}\hl{#1}}

\usepackage{listings}

\usepackage{xr}
\usepackage{subfiles}
\usepackage{epigraph}
\usepackage{nowidow}
\usepackage{needspace}
\usepackage{rotating}

\usepackage{pgfplots}
\pgfplotsset{width=10cm,compat=1.9}

\usepgfplotslibrary{external}
\usetikzlibrary {shapes.geometric}

\newcommand\summaryname{Abstract}
\usepackage{booktabs}
\usepackage[hidelinks]{hyperref}
\usepackage{bookmark}
\usepackage[noabbrev,capitalize]{cleveref}
\usepackage{enumitem}
\usepackage{tabularx}
\usepackage{multirow}
\usepackage[usestackEOL]{stackengine}
\usepackage{cellspace}
\usepackage{graphicx}
\graphicspath{ {./laps/images/} }
\usepackage{pifont}
\usepackage{balance}
\usepackage{enumitem}
\usepackage{makecell}

\newcommand{\fa}[1]{\textcolor{black}{#1}}

\newcommand{\miniskip}{\vspace*{-.5\baselineskip}}
\newcommand{\shrink}{\vspace*{-.9\baselineskip}}

\newcommand{\seconv}{S2E$_{\text{conv}}$}
\newcommand{\seconvminus}{S2E$^{-}_{\text{conv}}$}

\usepackage{colortbl}
\newcommand{\grayline}{\arrayrulecolor{lightgray}\hline\arrayrulecolor{black}}
\usepackage{mdframed}

\usepackage{algorithm}
\usepackage{algorithmicx}
\usepackage{algpseudocode}

\author{\myauthor}
\title{\mytitle}
\date{\mydate}

\begin{document}

\frontmatter

\hypersetup{pageanchor=false}
\begin{titlepage}
	\begin{center}
		\vspace*{3.5cm}
		
		\LARGE{\bfseries\mytitle}
		
		\vspace*{15pt}
		
		\normalsize
		\myauthor
	\end{center}
\end{titlepage}

\clearpage
\thispagestyle{empty}
\null
\clearpage
\begin{titlepage}
	\begin{center}
		\vspace*{3.5cm}
		
		\LARGE{\bfseries\mytitle}
		
		\vspace*{15pt}

		\vspace*{5pt}
		
		\normalsize
		
		\vspace{2.0cm}
		
		\textbf{Proefschrift}
		
		\vspace{0.5cm}
		
		ter verkrijging van de graad van doctor\\
		aan de Radboud Universiteit Nijmegen\\
		op gezag van de rector magnificus
		prof.~dr.~J.M.\ Sanders,\\
		volgens besluit van het college voor promoties\\
		in het openbaar te verdedigen op
		
		\vspace{0.5cm}
		
		woensdag 8 juli 2026\\
		\vspace{0.2cm}
		om 12.30 uur precies
		
		\vspace{0.5cm}
		
		door
		
		\vspace{0.5cm}
		
		\textbf{\myauthor}
	\end{center}
\end{titlepage}

\newpage%

\thispagestyle{empty}

\begingroup
\small

\noindent\textbf{Promotor:}\\
Prof.\ dr.\ A.P.\ (Arjen)\ de\ Vries\\
\newline
\noindent\textbf{Copromotor:}\\
Dr.\ F.\ (Faegheh)\ Hasibi\\
\newline
\noindent\textbf{Manuscriptcommissie:}\\
Prof.\ dr.\ M.A.\ (Martha)\ Larson\\
Prof.\ dr.\ T.\ (Tetsuya)\ Sakai \hfill (Waseda University, Japan)\\
Prof.\ dr.\ S.\ (Suzan)\ Verberne \hfill (Universiteit Leiden)\\
Dr.\ M.\ (Mohammad)\ Alian Nejadi \hfill (Universiteit van Amsterdam)\\
Dr.\ A.\ (Avishek)\ Anand \hfill (Technische Universiteit Delft) \\

\endgroup

\hypersetup{pageanchor=true}

\setcounter{tocdepth}{2}
\tableofcontents

\mainmatter
\chapter{Introduction}
\label{ch:intro}

The dream of having an intelligent conversation in natural language with a computer has fascinated humanity for decades.
Early intelligent systems, such as IBM Watson~\citep{Ferrucci:2010:BWO}, advanced the field by demonstrating the capability of answering complex natural language questions.
However, its expansion into real-world applications like healthcare remained limited due to the system's highly specialized nature, which hindered cross-domain generalization, and insufficient natural language understanding, which made it difficult to handle complex documents and user interactions in realistic scenarios.

Recent advancements in computational capability and neural network architectures, particularly Transformer-based models~\citep{Vaswani:2017:AAY}, have drastically reshaped the landscape of human-computer interaction.
Large Language Models (LLMs), where Transformer meets scaling, have achieved impressive performance in generating fluent natural language, leading to excitement that human-level artificial intelligence may be within reach.
Despite their impressive fluency, however, Transformer-based conversational systems have significant limitations.
They often generate plausible but incorrect information~\citep{Ji:2023:SHN}, creating the need to ground responses in actual stored and retrieved information~\citep{Soudani:2024:FTR}.

This shift towards conversational interaction, coupled with the necessity for grounding, has profoundly reshaped information retrieval systems, as users increasingly favor direct natural language answers over traditional lists of hyperlinks or information cards in search engines.
This paradigm shift has reinforced the importance of conversational information access (CIA) systems which focus on grounding system responses in retrieved knowledge to provide reliable information~\citep{Zamani:2023:CIS, Gao:2019:NAC,Anand:2019:CS}.

Despite significant progress in conversational information access, several challenges remain, including personalization and evaluation.
These systems typically lack understanding of individual users' personal contexts, limiting their ability to effectively engage users and satisfy diverse information needs~\citep{Zamani:2023:CIS,Salemi:2024:LMP,Yazan:2025:IRP}.
Additionally, evaluating conversational systems accurately and efficiently remains difficult due to the inherent complexity and subjective nature of conversational interactions~\citep{Mehri:2020:USR,Liu:2023:GNE}.
Addressing these challenges is crucial for ensuring conversational interactions are both personally relevant~\citep{Salemi:2024:LMP,Gerritse:2020:BCS} and reliably evaluated~\citep{Zamani:2023:CIS,Gao:2019:NAC,dietz:2025:llmtropes}, thereby enhancing user satisfaction and trust in conversational systems.

This thesis addresses these challenges by focusing on personalization in CIA systems and evaluating these systems effectively.
Specifically, the thesis explores conversational information access systems by researching methods for extracting users' personal context, developing personalized CIA systems, and evaluating these systems.

\section{Research Objectives and Questions}
\label{ch:intro:research}

The main research question of this thesis is:

\vspace{1em}
\noindent
\hspace{2.8em}%
\begin{minipage}[t]{0.83\linewidth}
    \textbf{RQ Main:} \emph{How can we develop personalized CIA systems and evaluate them effectively?}
\end{minipage}
\vspace{1em}

\begin{figure}[t]
    \centering
    \includegraphics[width=1.00\linewidth]{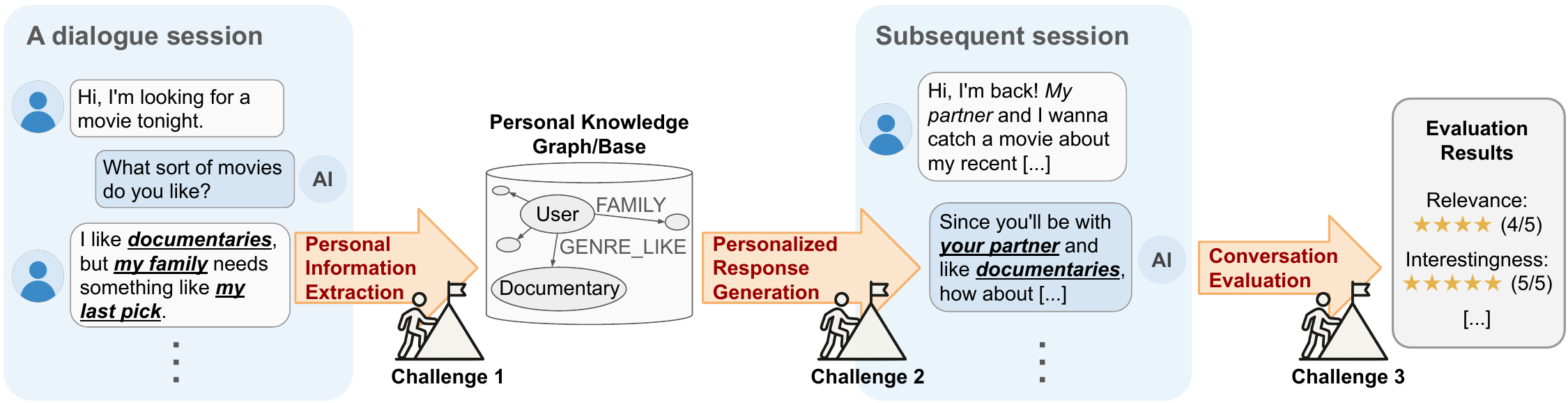}
    \caption{An illustration of a personalized CIA system, highlighting the three core challenges addressed in this thesis: (1) \textbf{Personal Information Extraction}, which focuses on extracting users' personal contexts by accurately identifying personal entities from conversational texts; (2) \textbf{Personalized Response Generation}, which involves developing a CIA model that utilizes the extracted context to generate personalized responses; and (3) \textbf{Conversation Evaluation}, which concerns evaluating CIA systems by considering whole conversation context, diverse interaction trajectories, and provide fine-grained performance insights.}
    \label{fig:intro:research-overview}
\end{figure}

\noindent
To answer this main research question, this thesis investigates personalization in CIA systems by addressing challenges related to (1) extracting personal context, (2) generating personalized responses, and (3) evaluating these systems effectively, as illustrated in Figure~\ref{fig:intro:research-overview}.

\subsection{Personal Information Extraction}

To personalize CIA systems, understanding the user's preferences and personal context is crucial.
This section provides background on extracting and structuring personal information from conversational interactions, as well as the associated research questions.

\subsubsection{Background}

Extracting and storing personal information from conversations is essential for enabling personalization in conversational systems. For instance, a Personal Knowledge Graph (PKG) is used to structure and represent personal entities and their relationships in a graph-based format centered around the user~\citep{Balog:2019:PKG}.
Unlike general-purpose knowledge graphs, PKGs capture entities of personal significance, such as the user's possessions, acquaintances, or preferences; thereby enabling the system to understand and utilize the user's context in conversations, leading to better personalized responses.

A key challenge in creating personal knowledge repositories, such as PKGs, is accurately extracting personal entities from conversational texts and linking them to corresponding entries in existing knowledge bases.
This process is known as \textbf{Entity Linking (EL)}.
Traditional EL typically involves three main steps~\citep{Balog:2018:EOS, vanHulst:2020:REL}: (1) mention detection, identifying text spans that refer to entities, (2) candidate selection, retrieving candidate entities from a knowledge base for each mention, and (3) entity disambiguation, selecting the correct entity from the candidate set.

Traditional EL methods are primarily designed for general documents or short texts and tend to focus on named entities~\citep{Carmel:2014:ERD, Hoffart:2011:RDN}. However, different types of entities contribute to machine understanding in a conversational setting.
For example, in a conversation where a user says ``I love playing jazz on my guitar, a Gibson Les Paul,'' we can identify three types of entities:
(i)~\emph{named entities}, i.e., proper nouns such as ``Gibson Les Paul'';
(ii)~\emph{concepts}, i.e., general categories such as ``jazz''; and
(iii)~\emph{personal entities}, i.e., possessive noun phrases such as ``my guitar,'' which refer to entities of personal significance.
Note that, while personal entity mentions can take many forms, as a first study of its kind in conversational EL, this thesis focuses on personal entities represented as possessive noun phrases.
Although all entity types could contribute to machine understanding of conversations, what EL in conversations entails and how to develop an effective method remain open questions.

\subsubsection{Research Questions}

The first research question this thesis investigates is:

\vspace{1em}
\noindent
\hspace{2.8em}%
\begin{minipage}[t]{0.83\linewidth}
    \textbf{RQ1:} \textit{What does Entity Linking in conversations entail?}
\end{minipage}
\vspace{1em}

\noindent
We address this question by investigating the characteristics of EL in conversations and defining the task.
Specifically, we collect EL annotations on conversations, dubbed ConEL, and investigate the performance of traditional EL systems on those annotations to understand the challenges of EL in conversations.
We found that traditional EL tools are suboptimal for EL in conversational settings, especially struggling to handle concepts and personal entities, which are crucial for personalized conversations.

This calls for the development of a new EL method that can effectively handle conversational settings; therefore, the second research question addressed by this thesis is:

\vspace{1em}
\noindent
\hspace{2.8em}%
\begin{minipage}[t]{0.83\linewidth}
    \textbf{RQ2:} \textit{How can we develop an Entity Linking method to accurately identify important entities for personalized conversational systems?}
\end{minipage}
\vspace{1em}

\noindent
We answer this question by introducing CREL, a conversational entity linking method developed to identify all types of important entities in conversations, including personal entities, concepts, and named entities.
CREL leverages coreference resolution to effectively identify personal references (e.g., ``my dog''), which is crucial for constructing PKGs.
To train and evaluate CREL, we also construct the ConEL-2 dataset, extending ConEL with significantly more dialogues with personal entity annotations.
Through extensive experiments, we found that CREL significantly outperforms traditional EL methods, particularly in handling personal entities.

\subsection{Personalized Response Generation}
\label{ch:intro:research:laps}

Effectively utilizing extracted personal context to generate personalized responses is the next critical step. One of the first steps in achieving this goal is the availability of large-scale personalized conversational datasets 
that reflect user interactions and preferences in the real world.
This section provides background on personalized CIA systems, discusses the need for datasets to train and evaluate personalized response generation, and outlines an associated research question.

\subsubsection{Background}

Personalized response generation in CIA concerns generating responses that satisfy the user's information needs through interactive conversations, considering the conversational context and the user's long-term preferences~\citep{Zamani:2023:CIS}.
While traditionally relying on straightforward rule-based approaches, recent CIA systems employ Transformer-based language models~\citep{Vaswani:2017:AAY}, such as BERT~\citep{Devlin:2019:BER}, T5~\citep{Raffel:2020:ELT}, and GPTs~\citep{Brown:2020:LMF,Achiam:2023:GTR}, which significantly increased the performance by the improvement of natural language understanding and generation capabilities.
However, despite these advancements, effectively capturing and utilizing individual user preferences for response generation remains a challenging task.
Unlike conventional (ad hoc) CIA, which focuses on addressing immediate, one-time conversations, personalized CIA requires systems to effectively elicit, store, and leverage user-specific preferences within and across multiple conversational sessions.

To achieve effective personalization, systems must first understand user preferences, typically through preference elicitation or clarifying questions, either via explicit system-initiated questions or implicit user feedback during conversations.
Once elicited, these preferences need to be remembered and utilized to generate personalized responses in current or subsequent conversation sessions (preference utilization), thereby enhancing relevance and improving the overall user satisfaction.

Training models for personalized response generation requires large-scale datasets that reflect diverse real-world user interactions and preferences~\citep{Salemi:2024:LMP,Salemi:2025:LNF}.
However, existing conversational datasets often fail to capture them due to the complexity of effective preference elicitation and scalability challenges with expert-constructed dialogues.
Even high-quality conversational datasets typically focus on single, relatively short conversations~\citep{Dalton:2019:CAT,Dinan:2018:WWK,Choi:2018:QUA,Eric:2020:MWO}, thus lacking the necessary scale and session-level continuity for modeling personalized interactions over multiple sessions~\citep{Zamani:2023:CIS}.
These gaps highlight the need for a method that can efficiently collect large-scale, multi-session, and multi-domain personalized conversational datasets that reflect real-world user interactions and preferences.

\subsubsection{Research Question}

The third research question this thesis addresses is:

\vspace{1em}
\noindent
\hspace{2.8em}%
\begin{minipage}[t]{0.83\linewidth}
    \textbf{RQ3:} \emph{How can users' personal preferences be utilized to generate personalized responses?}
\end{minipage}
\vspace{1em}

To answer this question, we first propose a model to construct large-scale personalized conversational datasets, dubbed LAPS.
It leverages LLMs to guide human workers in efficiently composing realistic, multi-session, and multi-domain personalized dialogues.
We found that our approach significantly accelerates the dialogue construction process while preserving the diversity and quality of human-written conversations, effectively capturing diverse real user preferences.

Building upon the LAPS dataset, we then investigate how to effectively utilize users' personal preferences to generate personalized responses.
We find that using semi-structured textual representations of user preferences (termed ``preference memory'') enables more accurate utilization of users' preferences for recommendations, improves response explanations, and mitigates LLMs' recall issues compared to using raw dialogue history.
This demonstrates the effectiveness of structured personal context in enhancing personalized response generation.

\subsection{Conversation Evaluation}

Evaluation is crucial for developing CIA systems, as one cannot improve that which one cannot measure~\citep{Voorhees:2005:OTR, Thomson:1889:PLA}.
Although human annotations are the gold standard for evaluating CIA systems, they are expensive and time-consuming; thus, automatic evaluation methods have been proposed to scale up the evaluation process.
In this section, we provide background on automatic evaluation methods for CIA systems, discuss the challenges of evaluating personalized CIA systems, and outline the associated research question.

\subsubsection{Background}

Automatic conversation evaluation is crucial for efficient, scalable, and cost-effective diagnosis of problems and biases during the development of CIA systems.
There are two main types of automatic evaluation methods: \textit{reference-based} and \textit{reference-free}.
Reference-based methods use gold references to evaluate system responses, which include Recall@K, BLEU~\citep{Papineni:2002:BMA}, ROUGE~\citep{Lin:2004:RPA}, and BERTScore~\citep{Zhang:2020:BET}.
While these methods are effective for machine translation and traditional IR tasks, they have limitations in conversation evaluation, as they overlook diverse response possibilities and various evaluation aspects.
These limitations are supported by various studies showing a weak correlation between reference-based methods and human evaluations~\citep{Mehri:2020:USR,Liu:2016:HNE}.

Reference-free methods have been proposed to address these limitations~\citep{Mehri:2020:USR,Zhong:2022:TUM,Liu:2023:GNE}.
These methods evaluate system responses without relying on gold references, consider multiple aspects of system quality, and allow for the assessment of various response possibilities.
For instance, \citet{Mehri:2020:USR} proposed USR that leverages unsupervised methods to evaluate dialogue quality across various aspects, and \citet{Zhong:2022:TUM} proposed UniEval that frames conversation evaluation as a boolean question answering (QA) task.
More recent works use LLMs for evaluation~\citep{Liu:2023:GNE, Upadhyay:2025:LSR, Cook:2024:TBG, Thomas:2024:LLM}.
However, these methods primarily focus on turn-level evaluation with a fixed dialogue history, limiting their ability to assess the whole conversation and capture the diverse user-system conversation trajectories, which are crucial for evaluating system performance in real-world scenarios~\citep{Siro:2022:UUS,Siro:2023:UPU}.
Therefore, there is a need for automatic, reference-free evaluation methods that can assess the entire conversation and capture diverse conversation trajectories, providing fine-grained insights into system performance.

\subsubsection{Research Question}

The fourth research question this thesis addresses is:

\vspace{1em}
\noindent
\hspace{2.8em}%
\begin{minipage}[t]{0.83\linewidth}
    \textbf{RQ4:} \emph{How to automatically evaluate personalized CIA systems in a way that accounts for natural conversation flow and provides fine-grained insights into system performance?}
\end{minipage}
\vspace{1em}

\noindent
We answer this question by introducing FACE, a Fine-grained, Aspect-based Conversation Evaluation method.
FACE is a reference-free evaluation method that handles diverse conversation trajectories and evaluates the entire conversation.
It first decomposes system responses into contextualized conversation particles.
An LLM then evaluates these particles using evaluation instructions optimized via an automatic prompt optimization technique.
The resulting sub-scores are then aggregated into a single score per evaluation aspect, offering interpretable insights into system performance.
Empirical evaluations show that FACE aligns closely with human judgments and provides fine-grained insights into the system's performance across multiple aspects (e.g., relevance and user understanding), thereby providing actionable feedback for system improvement.

\section{Contributions}

In this section, we summarize the main contributions of this thesis.
The contributions are shown as theoretical and methodological contributions, as well as resource contributions.

\vspace{1em}
\medskip \noindent \textbf{Theoretical and Methodological Contributions}

\medskip \noindent \textbf{C1.} We define the problem of EL in conversations, subdividing entities into three categories (named entities, concepts, and personal entities), and analyzing and identifying the unique challenges that make traditional EL methods suboptimal in conversational settings.

\medskip \noindent \textbf{C2.} We provide effective designs for collecting large-scale conversational EL annotations, used in the creation of the ConEL and ConEL-2 datasets.

\medskip \noindent \textbf{C3.} We propose a conversational entity linking method, CREL, which identifies personal entities and incorporates coreference resolution.

\medskip \noindent \textbf{C4.} We propose a scalable method for constructing large-scale personalized conversational datasets, dubbed LAPS, which uses LLMs to guide human workers, efficiently constructing multi-session, multi-domain dialogues that capture real user preferences while maintaining high quality.

\medskip \noindent \textbf{C5.} We investigate effective utilization of users' personal preferences for personalized response generation, showing that semi-structured preference memory improves accuracy, explanations, and mitigates LLMs' recall issues in response generation.

\medskip \noindent \textbf{C6.} We propose an automatic, reference-free, and fine-grained evaluation method for personalized CIA systems. The method, referred to as FACE, assesses entire conversations across diverse interaction paths using optimized instructions and particle-based scoring, providing aspect-based insights that align closely with human judgments.

\vspace{1em}
\medskip \noindent \textbf{Resource Contributions}

\medskip \noindent \textbf{C7.} We release the ConEL datasets, the first-of-its-kind conversational entity linking datasets (ConEL and ConEL-2) with annotations for personal, named, and conceptual entities, enabling empirical analysis of EL in conversations and benchmarking EL systems in conversational contexts.

\medskip \noindent \textbf{C8.} Alongside the ConEL datasets, we release an online resource listing around 130 conversational datasets with a detailed comparison of their characteristics.

\medskip \noindent \textbf{C9.} We release CREL, an open-source conversational entity linking tool that identifies personal entities, named entities, and concepts, suitable for conversational settings.

\medskip \noindent \textbf{C10.} We release the LAPS dataset, a large-scale, multi-session, human-written personalized dialogue dataset for research on preference extraction and personalized CIA development.
The dataset contains 1,406 conversations, with 11,215 preferences extracted from the conversations, ranging across two domains: \textit{recipe} and \textit{movie} recommendations.

\medskip \noindent \textbf{C11.} We release the CRSArena-Eval dataset, a comprehensive evaluation of conversation evaluation (i.e., meta-evaluation) dataset with high-quality multi-aspect human judgments across multiple systems for benchmarking automatic evaluation methods. The dataset contains 467 conversations, with 20,962 human judgments collected across nine systems.

\section{Thesis Overview}

\begin{figure}[t]
    \centering
    \includegraphics[width=1.0\textwidth]{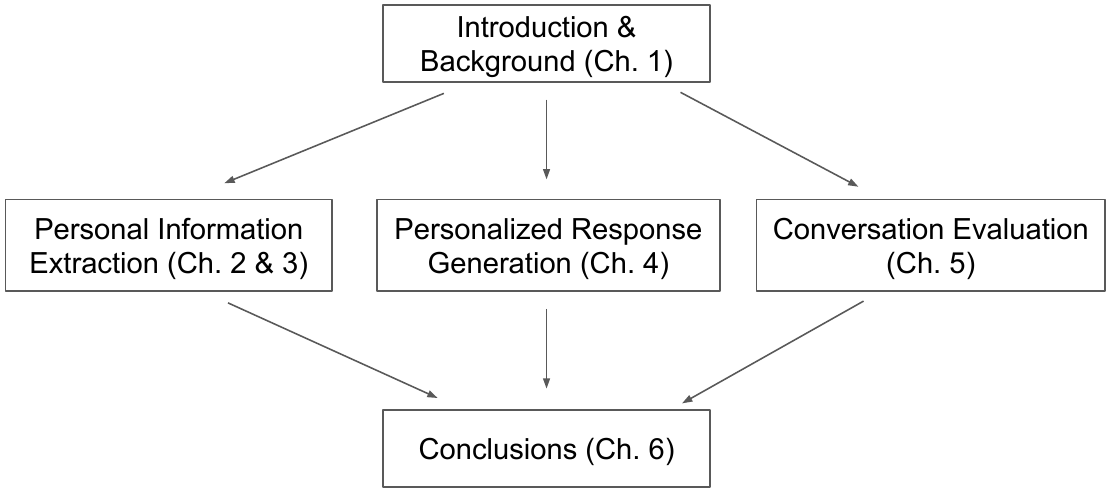}
    \caption{Thesis structure and connections between chapters.}
    \label{fig:thesis_structure}
\end{figure}

This section provides an overview of the thesis structure, describing each chapter and its contributions to the RQs outlined in Section~\ref{ch:intro:research}.
Figure~\ref{fig:thesis_structure} illustrates how each chapter maps to the key research areas throughout the thesis.

\medskip \noindent \textbf{Chapter~\ref{ch:conel}} describes EL in conversations, collecting EL annotations on conversations, dubbed ConEL, and  analyzing the characteristics of EL in conversations.

\medskip \noindent \textbf{Chapter~\ref{ch:crel}} presents a novel EL method (CREL) for personal information extraction in conversations.

\medskip \noindent \textbf{Chapter~\ref{ch:laps}} discusses the construction of large-scale, multi-session human-written personalized conversational datasets, dubbed LAPS, and examines the generation of personalized responses using this data.

\medskip \noindent \textbf{Chapter~\ref{ch:face}} introduces FACE, a fine-grained, aspect-based conversation evaluation method that evaluates personalized CIA systems based on the entire conversation and captures diverse conversation trajectories.

\medskip \noindent \textbf{Scope of the Thesis.}
This thesis focuses specifically on text-based CIA systems. While conversational systems may be multimodal, concentrating on textual conversations lets us focus on core challenges in personalization, dataset construction, and evaluation methods, which can be a basis for future multimodal extensions.

\section{Origins}

The origin of each chapter is described below.

\begin{itemize}
    \item \citep{Joko:2021:CEL} \bibentry{Joko:2021:CEL}. \\ \emph{[Related to RQ1 and C1, C2, C7, C8; included in Chapter~\ref{ch:conel}]}.
    \item \citep{Joko:2022:PEC} \bibentry{Joko:2022:PEC}. \\ \emph{[Related to RQ2 and C3, C9; included in Chapter~\ref{ch:crel}]}.
    \item \citep{Joko:2024:DPL} \bibentry{Joko:2024:DPL}. \\ \emph{[Related to RQ3 and C4, C5, C10; included in Chapter~\ref{ch:laps}]}.
    \item \citep{Bernard:2025:CRS} \bibentry{Bernard:2025:CRS} \\ \emph{[Related to RQ4 and C11; integrated in Chapter~\ref{ch:face}]}.
    \item \citep{Joko:2026:FACE} \bibentry{Joko:2026:FACE}. \\ \emph{[Related to RQ4 and C6, C11; included in Chapter~\ref{ch:face}]}.
\end{itemize}

\medskip \noindent \textbf{Other Publications.}
The papers listed below were also published during the course of the PhD research.
While these papers indirectly contribute to the thesis, they are not direct source material for the thesis.
{
\begin{itemize}
\item \citep{Joko:2022:RUT} \bibentry{Joko:2022:RUT}.
\item \citep{Joko:2026:WIA} \bibentry{Joko:2026:WIA}.
\end{itemize}
}

\chapter{Conversational Entity Linking}
\label{ch:conel}

As discussed in Chapter~\ref{ch:intro}, storing and utilizing a user's personal context through personal knowledge repositories, such as personalized knowledge graphs (PKGs), is crucial for personalizing conversational systems. A key challenge in creating these repositories is accurately extracting entities from dialogue via Entity Linking (EL).
However, traditional EL methods are primarily designed for general documents; their effectiveness in conversational contexts, where entities are often personal or conceptual, remains underexplored.
To close this gap, this chapter addresses the research question:

\vspace{1em}
\noindent
\hspace{2.8em}%
\begin{minipage}[t]{0.83\linewidth}
    \textbf{RQ1:} \textit{What does Entity Linking in conversations entail?}
\end{minipage}
\vspace{1em}

\noindent
To answer this question, this chapter introduces the ConEL dataset, a novel conversational entity linking dataset containing crowdsourced annotations of personal entities, named entities, and concepts, to investigate the unique characteristics of EL in conversations.
The chapter then analyzes these annotations to identify the main properties of conversational EL and evaluate the performance of traditional EL systems on our dataset.
The findings show that existing methods are suboptimal for this task, particularly struggling to identify the personal entities and concepts that are important for personalized interactions.
These results demonstrate the distinct challenges of conversational EL and highlight the importance of further research in this area.

\section{Introduction}
\label{ch:conel:sec:introduction}

Conversational systems are becoming increasingly important with the proliferation of personal assistants, such as Siri, Alexa, Cortana, and the Google Assistant. 
In this realm, understanding user utterances plays a crucial role in holding meaningful conversations with users -- this process is handled by the natural language understanding (NLU) component in traditional task-oriented dialogue systems~\citep{Gao:2019:NAC}. 
A popular text understanding method, which has proven to be effective in various downstream tasks~\citep{Dalton:2014:EQF, Hasibi:2016:EEL, Xiong:2017:WED, Shang:2021:ERO, Hasibi:2017:DFS, Dargahi:2018:QUE}, is \emph{entity linking}: the task of recognizing mentions of entities in text and identifying their corresponding entries in a knowledge graph~\citep{Balog:2018:EOS}. 
In this chapter, we aim to investigate the role of entity linking in conversational systems.

Even though large-scale neural language models (like BERT~\citep{Devlin:2019:BER} and GPT-3~\citep{Brown:2020:LMF}) have repeatedly been shown to achieve high performance in various machine understanding tasks, 
these do not provide replacements for explicit auxiliary information from knowledge graphs.  Rather, the two should be seen as complementary efforts.  Indeed, augmenting neural language models with information from knowledge graphs has been shown to be beneficial in a number of downstream tasks~\citep{Poerner:2020:EBERT, Peters:2019:KnowBERT, Zhang:2019:ERNIE}.  
Specifically, in the context of task-based conversational systems, NLU often relies on entities for fine-grained domain classification and intent determination.
This makes EL for conversational systems even more important.

Despite its importance, research on EL for conversational systems has so far been limited. Traditional EL techniques that are used for documents~\citep{Balog:2018:EOS}, queries~\citep{Hasibi:2015:ELQ, Li:2020:ELQ}, or tweets~\citep{Meij:2012:ASM} are suboptimal for conversational systems for a number of reasons. 
First, unlike documents, conversations are not ``static,'' but are a result of human-machine interaction.  Here, the user can correct the system's interpretation of their request and the system can ask clarification questions to further its understanding of the user's intent. 
Second, conversations are informal and it is common in a conversation to make references to entities by their pronouns; e.g., ``my city,'' ``my guitar,'' and ``its population.''  A conversational system is expected to understand and handle such mentions of personal entities~\citep{Balog:2019:PKG}. 
Third, while entity linking in documents or queries tends to focus on proper noun entities~\citep{Carmel:2014:ERD, Hoffart:2011:RDN}, in a conversational setting all types of entities, including general concepts, can contribute to machine understanding of users' utterances.
In this chapter, we aim to investigate these differences and develop resources to foster research in this area.

\noindent To investigate the main research question of this chapter, we set out to analyze entity linking annotations for existing conversational datasets. We perform a thorough analysis of existing conversational datasets and select four of these for annotation.  These cover the three main categories of conversational problems~\citep{Gao:2019:NAC}: question answering (QA), task-oriented systems, and social chat (cf. Sect.~\ref{ch:conel:sec:dataset_section}).  We aim to annotate ``natural'' conversations, and therefore bias our selection of datasets towards those that are obtained using a Wizard-of-Oz setup.

Annotating dialogues, however, is an inherently complex task, where it is a challenge to keep the cognitive load sufficiently low for crowd workers.
This leads us to a secondary research question:

\vspace{1em}
\noindent
\hspace{2.8em}%
\begin{minipage}[t]{0.83\linewidth}
    \textbf{RQ1a:}  \textit{What are effective designs for collecting large-scale annotations for conversational data?}
\end{minipage}
\vspace{1em}

\noindent Although a large body of research exists on effective designs for collecting large-scale entity annotations~\citep{Adjali:2020:BME, Bhargava:2019:LMW, Finin:2010:ANE, Mayfield:2011:BCL}, to the best of our knowledge, there is no work on conversational data.  We run a number of pilot experiments using Amazon Mechanical Turk (MTurk) to select the best design and instruction. 
Based on these experiments, we develop the \emph{Conversational Entity Linking (ConEL)} dataset, consisting of 100 annotated dialogues (708 user utterances) sampled from the QuAC~\citep{Choi:2018:QUA}, MultiWOZ~\citep{Zang:2020:MWO}, WoW~\citep{Dinan:2018:WWK}, and TREC CAsT 2020~\citep{Dalton:2020:CAT} datasets. 
To enable further study in conversational EL, we also annotate a separate sample of 25 WoW dialogues (containing references to personal entities) and all 25 manually rewritten dialogues of TREC CAsT 2020.

Our findings, obtained by analyzing the annotated dialogues, are as follows. 
\begin{itemize}[leftmargin=2em]
	\item Mentions of personal entities are mainly present in social chat conversations.
	\item While named entities are deemed to be important for text understanding, specific concepts are also found useful for understanding the intents of conversational user utterances.
	\item Traditional EL approaches fall short in providing high precision annotations for both concepts and named entities. This calls for a methodological departure for conversational entity linking, where concepts, named entities, and personal entities are taken into consideration.
\end{itemize}
 
\noindent
In summary, this work makes the following contributions:
\begin{itemize}[leftmargin=2em]
  \item To the best of our knowledge, ours is the first study on EL in conversational systems. We subdivide entities into three categories (named entities, concepts, and personal entities), and analyze the importance of each for conversational data. We further investigate different aspects of EL for three categories of conversational tasks: QA, task-oriented, and social chat.
  \item We investigate effective designs for collecting large-scale EL annotations for conversational data. 
  \item We make the annotated conversational datasets publicly available.\footnote{\url{https://github.com/informagi/conversational-entity-linking}} This data comes with detailed account of the procedure that was followed for collecting the annotations, which can be used for further extension of the collection. 
  \item As an additional (online) resource, we provide a comprehensive list of around 130 conversational datasets released by different research communities with a detailed comparison of their characteristics.  
\end{itemize}

\noindent
The resources provided in this chapter allow for further investigation of entity linking in conversational settings, can be used for evaluation or training of conversational EL systems, and complement existing conversational datasets.

\section{Related Work}

The related work pertinent to this chapter concerns entity linking in documents, queries, and conversational systems, as well as personal entity identification.

\subsection{Entity Linking}
\label{ch:conel:sec:entity_linking}

\medskip \noindent \textbf{Entity linking in documents.}
Entity linking plays an important role in understanding what a document is about~\citep{Balog:2018:EOS}.
TagMe~\citep{Ferragina:2010:TAG} is one of the most popular EL tools, redesigned and improved by \citet{Piccinno:2014:FTM} and renamed to WAT.
\citet{vanHulst:2020:REL} presented REL, which is an open source EL tool based on state-of-the-art NLP research.
Other state-of-the-art EL methods include DeepType~\citep{Raiman:2018:DTM}, Blink~\citep{Wu:2020:BLINK}, and GENRE~\citep{Cao:2021:GENRE}.
Although these approaches are effective for documents, it is known that EL algorithms with high performance on general documents are less effective when applied to short informal texts like queries~\citep{Cornolti:2018:SMA}.

\medskip \noindent \textbf{Entity linking in queries.}
Entity linking in queries poses new challenges due to the short and noisy text of queries, their limited context, and high efficiency requirement~\citep{Hasibi:2017:ELQ, Carmel:2014:ERD, Cornolti:2018:SMA}.
\citet{Cornolti:2018:SMA} tackled some of these challenges by ``piggybacking'' on a web search engine. Relying on external search engines, while being effective, hinders efficiency and sustainability of EL systems. 
\citet{Hasibi:2017:ELQ} studied this challenge with a special focus on striking a balance between effectiveness and efficiency.
These studies consider EL for a single query, while in conversational systems multiple consecutive user turns need to be annotated.

\medskip \noindent \textbf{Entity linking in conversations.}
Research on conversational entity linking has been mainly focused on employing traditional entity linking and named entity recognition methods in conversational and QA systems~\citep{Kumar:2020:MIS, Chen:2016:CIM, Bowden:2018:SNE, Chen:2017:RWA, Vakulenko:2018:MSC, Li:2020:ELQ}. 
Entity linking is also used in multi-party conversations to connect mentions across different parts of dialogues and mapping to their corresponding character~\citep{Chen:2016:CIM}. This is a subtask of entity linking, referred to as character identification.
A close study to our work is~\citep{Bowden:2018:SNE}, where an entity linking tool, focused mainly on named entities,  is developed for open-domain chitchat.
In contrast to these works, we aim to understand EL for conversational systems, annotating conversations with concepts, named entities, and personal entities.

\subsection{Personal Entities}
\label{ch:conel:sec:personal_entities}
Dealing with the mention of personal entities (e.g., ``my guitar'') is important for  personalization of conversational systems.
Consider for example the user utterance ``Do you know how to fix my guitar?''
To answer this question, the system has to know more about the user's guitar type; e.g., ``Gibson Les Paul.'' This information may be available in the conversation history, previous conversations, or other sources (e.g., user's public information in social media). This information can be represented as Resource Description Framework (RDF) triples in the form of subject-predicate-object expressions $\langle e, p, e'\rangle$, e.g., \emph{$\langle$User, guitar, Gibson Les Paul$\rangle$}.
\citet{Li:2014:PKG} proposed a method to detect personal entities ($e$) and their corresponding predefined predicates ($p$) in conversations. Their approach consists of three steps: (1) identifying user utterances that are related to personal entities, (2) predicting entity mentions by classifying those utterances, and (3) finding the  personal entities.
\citet{Tigunova:2019:LBL} address the problem of identifying personal entities from implicit textual clues. They proposed a zero-shot learning method to overcome the lack of sufficient labelled training data~\citep{Tigunova:2020:CHA}.
All these studies focus on identifying predefined classes of predicates.
Extracting RDF triples without predefined relation classes has been studied in the context of open information extraction~\citep{Yates:2007:TRO, Fader:2011:IRO, Angeli:2015:LLS, Cui:2018:NOI}, but not in relation to personal entities.
In this study, we annotate conversations with personal entity mentions and their corresponding entities.

\section{Dataset Selection}
\label{ch:conel:sec:dataset_section}

There exists a large number of conversational datasets released by the natural language processing, machine learning, dialogue systems, and information retrieval communities. 
We made an extensive list of around 130 datasets,\footnote{This list is publicly available at:  \url{https://github.com/informagi/conversational-entity-linking}} extracted from ParlAI~\citep{Miller:2017:PAR} and other dataset comparison lists~\citep{Penha:2019:INT, Choi:2018:QUA, Hauff:2020:datasets}.
These datasets target three conversational problems~\citep{Gao:2019:NAC}:

\begin{itemize}[leftmargin=2em]
	\item \emph{Question answering}, where users ask natural language queries and the system provides answers based on a text collection or a large-scale knowledge repository.
	\item \emph{Task-oriented systems}, which assist users in completing a task, such as making a hotel reservation or booking movie tickets.
	\item \emph{Social chat}, where systems are meant to be AI companions to the users and hold human-like conversations with them.
\end{itemize}

\noindent
To obtain a comprehensive view of entity linking in conversational systems, we set out to analyze at least one dataset for each of the three main categories of conversational problems. 
To this end, we shortlisted datasets that resemble real conversations. That is, multi-domain and multi-turn  datasets, collected based on actual interactions between two humans. Datasets that are extracted from web services (e.g., Reddit and Stack Exchange) or created based on templates (e.g., bAbI~\citep{Sukhbaatar:2015:EEM}) were thus ignored.
To ensure that the selected datasets are sizable, they were required to contain at least 100 dialogues.
This list was further narrowed down by selecting relatively popular datasets based on citation counts and publication year.\footnote{While admittedly this is a loose measure, it helps to identify datasets that became widely accepted by the research community.}
By applying these criteria, nine datasets were shortlisted. 
In the final step, each dataset in our shortlist was closely examined, and at least one dataset was selected for each conversational problem; see Table~\ref{tbl:extracted-candidates} for an overview of the selected datasets. 
The reasoning behind our selections is detailed below.

\setlength{\tabcolsep}{2pt}
\begin{table}
\small
\caption{Overview of the selected conversational datasets for the entity annotation process. A sample of QuAC, MultiWOZ, and WoW, and all dialogues in TREC CAsT 2020 were used for generating the ConEL dataset.}
\label{tbl:extracted-candidates}
\resizebox{\columnwidth}{!}{%
\begin{tabular}{@{~}l||@{~~}l|@{~~}l|@{~}l@{~}}
\hline
\textbf{Dataset}	 & \textbf{Task} 				& \textbf{\#Convs~} 	& \textbf{Avg. Turns} \\ 
\Xhline{2pt}
QuAC~\citep{Choi:2018:QUA}& QA & 13.6K 				& 14.5 \\
MultiWOZ~\citep{Zang:2020:MWO}& Task-oriented 		& 8.4K 					& 13.5 \\
WoW~\citep{Dinan:2018:WWK} & Social chat & 22.3K 				& 9.1 \\
TREC CAsT 2020~\citep{Dalton:2020:CAT} & QA & 25 & 17.3 \\
\hline
\end{tabular}%
}
\end{table}

\medskip \noindent \textbf{QA.}
Among the QuAC~\citep{Choi:2018:QUA}, CoQA~\citep{Reddy:2019:CQA}, and QReCC~\citep{Anantha:2020:ODQ} datasets, we selected QuAC for QA dialogues. 
QuAC is a widely used dataset for conversational QA and contains 13.6K dialogues between two crowd workers.  
CoQA, on the other hand, is a machine reading comprehension dataset with provided source texts for every dialogue. 
Since these source texts are not necessarily available in real conversations, CoQA was left out. 
QReCC is built based on questions from other datasets, including QuAC and TREC CAsT, and is focused on question rewriting.  Because of the overlapping questions with other datasets, it was also ignored.

\medskip \noindent \textbf{Task-oriented.}
The MultiWOZ~\citep{Zang:2020:MWO} and KVRET~\citep{Eric:2017:KVR} datasets were examined for task-oriented dialogues.
MultiWOZ covers seven various goal-oriented domains: \emph{Attraction, Hospital, Police, Restaurant, Hotel, Taxi, and Train}. 
KVRET, on the other hand, deals with only three domains, all of which are in-car situations.
We, therefore, selected the MultiWOZ dataset, which also has more dialogues than KVRET (8.4K vs. 3K).
Note that MultiWOZ has several versions~\citep{Budzianowski:2018:MWO, Eric:2020:MWO, Zang:2020:MWO}; we used the latest version, MultiWOZ 2.2~\citep{Zang:2020:MWO}.

\medskip \noindent \textbf{Social chat.}
The Wizard of Wikipedia (WoW)~\citep{Dinan:2018:WWK}, Empathetic Dialogues~\citep{Rashkin:2019:TEO}, Persona-Chat~\citep{Zhang:2018:PDA}, and TaskMaster-1~\citep{Byrne:2019:TTR} datasets were shortlisted for social chat dialogues.
We excluded TaskMaster-1, as the majority of dialogues (7.7K) were collected by crowd workers who were instructed to write full conversations, i.e., played both the user and the system roles on their own.
Persona-Chat and Empathetic Dialogues are more focused on emotional and personal topics, while WoW is knowledge grounded and makes use of knowledge retrieved from Wikipedia.
We therefore chose WoW as a social chat dataset.

\medskip \noindent Additionally, we also included the TREC 2020 Conversational Assistance Track (CAsT)~\citep{Dalton:2020:CAT} dataset in our study. 
TREC CAsT~\citep{Dalton:2019:CAT}  is an important initiative by the IR community, and is focused on the information seeking aspect of conversations. 
Unlike other datasets, which represent dialogues as a sequence of user-system exchanges, TREC CAsT 2019 provides relevant passages that a system may return in response to a user utterance -- therefore, a unique conversation cannot be made for a given conversational trajectory. 
This has been changed in TREC CAsT 2020~\citep{Dalton:2020:CAT}, where a canonical response is given for each user utterance.  We generated conversations for our crowdsourcing experiments using these canonical responses.
In the remainder of this chapter we refer to TREC CAsT 2020 as CAsT.

\section{Entity Annotation Process}
\label{ch:conel:sec:annot}

This section describes the process of annotating dialogues from the selected conversational datasets. 
Our aim is to identify entities that can aid machine understanding of user utterances; this includes named entities, concepts, and mentions of personal entities.
Note that our focus is on user utterances, since the system is supposedly aware of the text it generates during a conversation.  The knowledge graph we use for annotations is Wikipedia (2019-07 dump).

The annotation process was performed via crowdsourcing using Amazon's Mechanical Turk (MTurk). 
In order to reduce the cognitive load for this complex task and to obtain the best annotation results, we ran a number of pilot experiments. 
In these experiments, we tested multiple task structures and interfaces using MTurk and compared the results with expert annotations of the same dialogues. 
The best task designs and interfaces were then used for the final annotations. 
Below, we describe the task design for annotating concepts, named entities, and personal entities (Sections~\ref{ch:conel:sec:annot:con_ens}  and~\ref{ch:conel:sec:annot:pe}), followed by the process of dialogue selection and annotation (Section~\ref{ch:conel:sec:annot:selection}).

\subsection{Concepts and Named Entities}
\label{ch:conel:sec:annot:con_ens}

We employ a two-step process for annotating explicitly mentioned entities, i.e., concepts and named entities.

\begin{figure}[t]
\fbox{\includegraphics[width=0.97\linewidth]{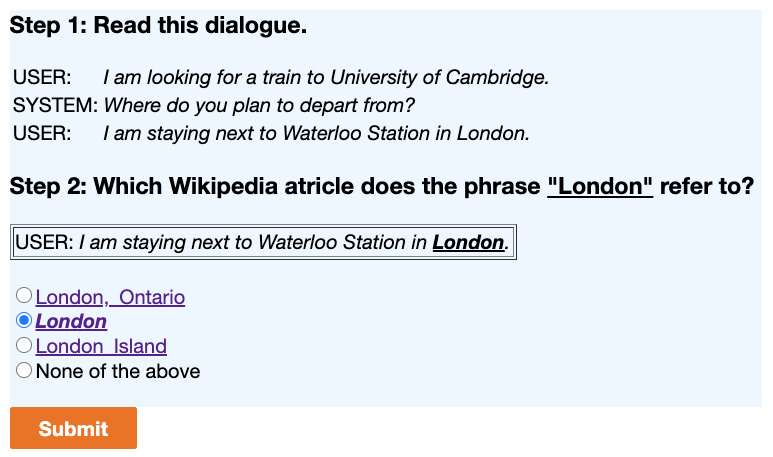}}
  \caption{Annotation interface for the entity-mention selection task (Stage 1). To keep the cognitive load low, only multiple-choice questions were used.  Possible answer entities are linked to their corresponding Wikipedia article.}
 
  \label{fig:stage1_example}

\end{figure}

\medskip \noindent \textbf{Stage 1: Selecting entity-mention pairs.}
First, we aim to map each mention to a single entity in the knowledge graph. 
Workers were presented with a dialogue, a mention from the latest user utterance, and a set of candidate entities or None. 
They were instructed (using a concise description and a couple of examples) to find the Wikipedia article that is referred to by the mention; Figure~\ref{fig:stage1_example} shows an excerpt from this task. 
The ``None of the above'' option is selected when the candidate pool does not contain the correct entity or the given mention is not appropriate.
These mentions were later examined by an expert annotator and assigned the correct entity or ignored.
To reduce the cognitive load on the workers, long conversations were trimmed; i.e., the middle turns in conversations with more than six turns were not presented.

\medskip \noindent \textbf{Stage 2: Finding the helpful entities.}
The mention-entity pairs obtained in the first stage are not necessarily important for machine understanding of user utterances; consider, for instance, the entity \textsc{College} in utterance ``I have wanted to travel to Amsterdam since college. What are the tourist attractions there?''
In Stage 2, we asked workers to filter the entity-mention pairs identified in Stage 1 by selecting only those pairs that can help the system to identify the user's intent. 
Specifically, we provided them with a conversation history and all mention-entity pairs from a user utterance, and gave them the following instruction:
\emph{``Imagine you are an AI agent (e.g., Siri or Google Now), having a dialogue with a person. You have access to Wikipedia articles (and some other information sources) to answer the person's questions. Select the Wikipedia articles that help you to find an answer to the person's question.''}
\fa{We presented mention-entity pairs two times to the users (all in one assignment): once they were asked to select named entities, and the other time they were asked to select all ``helpful'' entities.
Using this interface, we were able to identify named entities and further analyze the differences between concepts and named entities.}

\medskip \noindent \textbf{Generating annotation candidates.}
We employ a pooling approach to generate an extended set of candidate mentions and entities.
Three EL tools were used to annotate the dialogues: TagMe~\citep{Ferragina:2010:TAG}, WAT~\citep{Piccinno:2014:FTM}, and REL~\citep{vanHulst:2020:REL}.
Each tool was employed in two ways: (i) the \emph{turn} method, which annotates a single turn, irrespective of the conversation history, and (ii) the \emph{history} method, which annotates each turn given the conversation history up to that turn.  For the CAsT dataset, only user utterances were given to the EL tool, while for other datasets both system and user utterances were considered as conversation history. This  is due to relatively long system utterances in the CAsT dataset, which makes it infeasible for the EL tools to annotate the whole conversation history.  
To further improve the recall of our pool, we included the 
top-$10$ Wikipedia search results, using mentions as queries sent to the MediaWiki API.\footnote{\url{https://www.mediawiki.org/wiki/API:Main_page}}

\subsection{Personal Entities}
\label{ch:conel:sec:annot:pe}

\begin{figure}[t]
\fbox{\includegraphics[width=0.97\linewidth]{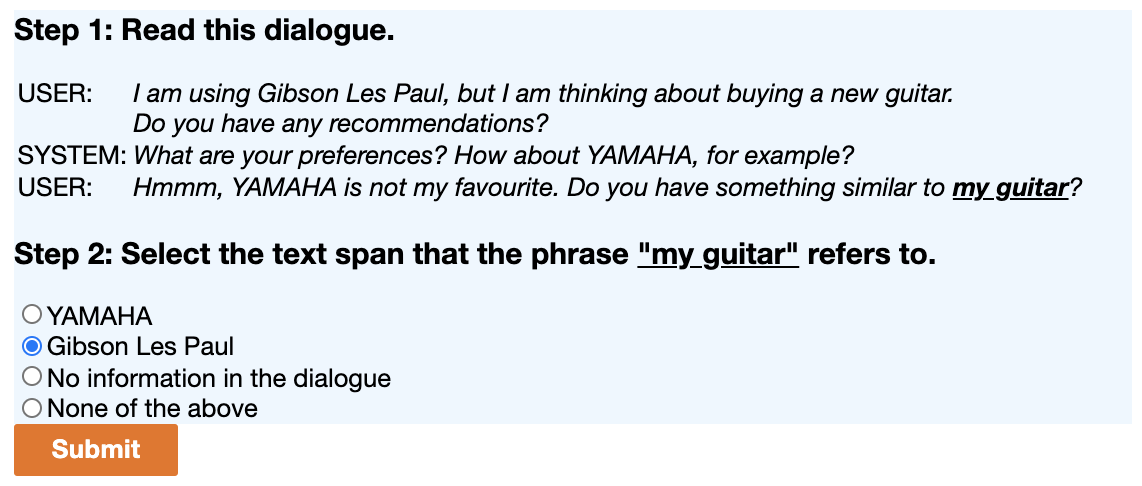}}
  \caption{Annotation interface for mapping a personal entity mention (``my guitar'') to the corresponding explicit entity mention in the conversation (``Gibson Les Paul'').}
  \label{fig:pe_ann_example}
\end{figure}

Annotating conversations with personal entities requires identifying  \emph{personal entity mentions} and mapping them to the corresponding \emph{explicit entity mentions} in the conversation history (if exists); e.g., mapping the personal entity mention ``my guitar'' to the explicit entity mention ``Gibson Les Paul.''
Once this mapping was done, mention-entity pairs can be identified as described in Stage 1 of Section~\ref{ch:conel:sec:annot:con_ens}.
We note that in some cases, explicit entity mentions are not present in the conversation history and the system needs to detect them from other information sources (e.g., previous conversations or user profile data).  
In this study, we confined ourselves to the cases where explicit entity mentions can be found in conversation history; i.e., personal entity mentions without explicit entity mentions in the conversation history were not mapped to any entity.

We designed a crowdsourcing experiment, where workers were given a conversation history along with the personal entity mention, and their task was to select the text span in the conversation history that the given personal entity mention refers to. 
Figure~\ref{fig:pe_ann_example} shows an example of this task.
``None of the above'' answers were resolved by an expert annotator.

\needspace{4\baselineskip}
\medskip \noindent \textbf{Generating annotation candidates.}
We used a simple yet effective method to find personal entity mentions.
Inspired by~\citet{Li:2014:PKG}, we detect all text spans starting with ``my'' followed by one or several adjectives, common nouns, proper nouns and/or numbers (using the SpaCy\footnote{\url{https://spacy.io/}} part-of-speech (POS) tagger).
We further allowed for the word \emph{``of''} to be part of the mention (e.g., ``my favourite forms of science fiction''). 
For each personal entity mention, we included all the candidate mentions that were identified by our EL methods (cf. Section~\ref{ch:conel:sec:annot:con_ens}).

\subsection{Dialogue Selection and Annotation}
\label{ch:conel:sec:annot:selection}

Annotating all dialogues in the selected datasets was infeasible for us, due to its high costs. 
We therefore selected a random sample of dialogues from a pool of presumably difficult dialogues from the QuAC, MultiWOZ, and WoW datasets. This pool contains dialogues with  at least one \emph{complex mention}, a personal entity mention, or a clarification question in user utterances.
By complex mention we refer to cases where the same mention is linked to different entities by the EL tools (i.e., REL, WAT, and TagMe). 
The clarification questions were identified based on the patterns stated by~\citet{Braslavski:2017:WDY}, and personal entity mentions were extracted as described in Section~\ref{ch:conel:sec:annot:pe}.
A total of $100$ samples were selected ($25$ for each dataset), amounting to $708$ user utterances.
Note that unlike the other datasets, CAsT contains only 25 dialogues, therefore all its dialogues were annotated.
Based on this selection, we are able to analyze the differences between the three main categories of conversational tasks.

The CAsT dataset comes with manually rewritten user queries, where each rewritten query can be answered independently of the conversation history. We also annotated the manually rewritten CAsT queries to allow for a comparison between raw and rewritten queries.

To extend our analysis on personal entity linking, we annotated another sample of dialogues from the WoW dataset. 
To generate this sample, we randomly selected $500$ dialogues that contain personal entity mentions and presented them to crowd workers to find their entity references in the dialogues (cf. Section~\ref{ch:conel:sec:annot:pe}).
Workers agreed that, in 180 dialogues of this sample, the references to the personal entity mentions are present in the dialogue. 
We then randomly selected 25 dialogues (containing 216 user utterances) out of these 180 dialogues and annotated their concepts, named entities, and personal entities.

To ensure high data quality, the annotation tasks were performed by top-rated MTurk workers, i.e., Mechanical Turk Masters.
Since the number of Masters is small, and they mainly select tasks with a high number of human intelligence tasks (HITs), the remaining 0.4\% of our annotation tasks were performed by high quality workers with a task approval rate of 99\% or higher.
We collected three judgments for each annotation and paid the workers \textcent6 for each annotation assignment, resulting in a final cost of around \$620.
Fleiss' Kappa inter-annotator agreement was 0.76, 0.30, 0.61 for Stage 1, Stage 2, and personal entity annotations, respectively.
Disagreements were resolved by an expert annotator.

\section{Annotation Results}

In this section, we describe our findings based on the analysis of the entity annotations obtained for the selected datasets. 
We also present baseline results for the entity-annotated conversations.
The results are shown in Tables~\ref{tbl:el_all}--\ref{tbl:el_pe}.
In these tables, the last character of each method, ``t'' or ``h,'' stands for ``turn'' or ``history,'' respectively (cf. Section~\ref{ch:conel:sec:annot:con_ens}).
Precision, recall, and F1 scores are micro-averaged and computed using the strong matching approach~\citep{Usbeck:2015:GGE}. 
To understand the frequency of personal entities in conversational datasets, we applied the method described in Section~\ref{ch:conel:sec:annot:pe} to identify all personal entity mentions in all the datasets.
We found that WoW contains more dialogues with personal entity mentions compared to other datasets; i.e., 33\% of dialogues in WoW vs. 0.3\%, 11\%, and 12\% of dialogues in QuAC, MultiWOZ, and TREC CAsT, respectively.
These results indicate that \textbf{personal entity mentions are mainly present in social chat conversations}.

Comparing concepts and named entities, we found that 43\% of linked entities in the ConEL dataset are marked as named entities (NEs) by crowd workers, which implies that the remaining 57\% entities are concepts.
This indicates that \textbf{in addition to named entities, concepts are also found useful for understanding the intents of user utterances.}

\begin{table}[t!]
\centering
\caption{Entity linking results on the ConEL dataset.} 
\label{tbl:el_all}
\resizebox{\textwidth}{!}{%
\begin{tabular}{l || rrr | rrr | rrr | rrr || rrr} 
   \hline
	\multirow{2}{*}{\textbf{}}
	& \multicolumn{3}{c |}{\textbf{QuAC}} & \multicolumn{3}{c |}{\textbf{MultiWOZ}} 
	& \multicolumn{3}{c |}{\textbf{WoW}} & \multicolumn{3}{c ||}{\textbf{CAsT}} 
	& \multicolumn{3}{c }{\textbf{All}}\\
	&
	\multicolumn{1}{c}{P} & \multicolumn{1}{c}{R} & \multicolumn{1}{c|}{F} & 
	\multicolumn{1}{c}{P} & \multicolumn{1}{c}{R} & \multicolumn{1}{c|}{F} & 
	\multicolumn{1}{c}{P} & \multicolumn{1}{c}{R} & \multicolumn{1}{c|}{F} & 
	\multicolumn{1}{c}{P} & \multicolumn{1}{c}{R} & \multicolumn{1}{c||}{F} & 
	\multicolumn{1}{c}{P} & \multicolumn{1}{c}{R} & \multicolumn{1}{c}{F} \\
   \Xhline{2pt}
	$\textrm{TagMe}_t$ 
	& 29.5	& 37.1	& 32.9
	& 18.5	& 23.9	& 20.9
	& 33.0	& 41.4	& \textbf{36.7}
	& 52.7	& \textbf{47.3}	& 49.8
	& 35.5	& 39.7	& 37.5  \\
	$\textrm{TagMe}_h$
	& 34.7	& 32.4	& 33.5
	& 18.7	& 23.9	& 21.0
	& 33.0	& 41.4	& \textbf{36.7}
	& 57.5	& 46.1	& \textbf{51.2}
	& 38.2	& 38.0	& \textbf{38.1} \\ \hline
	 $\textrm{WAT}_t$
	& 23.2	& 39.0	& 29.1
	& 19.5	& \textbf{36.6}	& \textbf{25.5}
	& 24.8	& 45.7	& 32.2
	& 46.6	& 41.3	& 43.8
	& 28.6	& 40.7	& 33.6 \\
	 $\textrm{WAT}_h$
	& 27.4	& \textbf{48.6}	& 35.1
	& 19.3	& 31.0	& 23.8
	& 23.5	& \textbf{55.7}	& 33.1
	& 44.7	& 45.5	& 45.1
	& 29.6	& \textbf{45.5}	& 35.8 \\ \hline
	 $\textrm{REL}_t$
	& 23.7	& 21.9	& 22.8
	& \textbf{39.3}	& 15.5	& 22.2
	& 43.8	& 10.0	& 16.3
	& 68.1	& 19.2	& 29.9
	& 38.8	& 17.7	& 24.3 \\
	 $\textrm{REL}_h$
	& \textbf{37.6}	& 33.3	& \textbf{35.4}
	& \textbf{39.3}	& 15.5	& 22.2
	& \textbf{52.9}	& 12.9	& 20.7
	& \textbf{70.2}	& 19.8	& 30.8
	& \textbf{47.6}	& 21.3	& 29.4 \\
   \hline
   \end{tabular}
}
\end{table}

Table~\ref{tbl:el_all} shows the results of different EL methods on the ConEL dataset. While TagMe achieves the highest F1 scores on WoW and CAsT, WAT and REL are the best performing tools (with respect to F1) on the MultiWOZ and QuAC datasets, respectively.
Comparing the ``turn'' and ``history'' methods, we observe that conversation history improves EL results for most datasets and tools. 
We also find that REL has higher precision but lower recall compared to TagMe and WAT. 
One might argue that high precision EL is preferred in a conversational setting, as incorrect results can lead to high user dissatisfaction. This claim, however, requires further investigation, and the effect of EL on end-to-end conversational system performance is yet to be evaluated.

\begin{table}[t!]
\centering
\caption{Breakdown of entity linking results for named entities ($\textrm{F}_{\textrm{NE}}$) and concepts ($\textrm{F}_{\textrm{C}}$).}
\label{tbl:el_breaksown}
\resizebox{\textwidth}{!}{%
\begin{tabular}{@{~}l||@{~}c@{~~}c|@{~}c@{~~}c|@{~}c@{~~}c|@{~}c@{~~}c||@{~}c@{~~}c@{~}}
	\hline
	\multirow{3}{*}{}
	& \multicolumn{2}{@{~}c|@{~}}{\textbf{QuAC}} & \multicolumn{2}{@{~}c|@{~}}{\textbf{Multi}} 
	& \multicolumn{2}{@{~}c |@{~}}{\textbf{WoW}} & \multicolumn{2}{@{~}c ||@{~}}{\textbf{CAsT}} 
	& \multicolumn{2}{@{~}c }{\textbf{All}} \\
	& &  & \multicolumn{2}{c |@{~}}{\textbf{WOZ}} & && && & \\
	& 
	$\textrm{F}_{\textrm{NE}}$ & $\textrm{F}_{\textrm{C}}$ &
	$\textrm{F}_{\textrm{NE}}$ & $\textrm{F}_{\textrm{C}}$ &
	$\textrm{F}_{\textrm{NE}}$ & $\textrm{F}_{\textrm{C}}$ &
	$\textrm{F}_{\textrm{NE}}$ & $\textrm{F}_{\textrm{C}}$ &
	$\textrm{F}_{\textrm{NE}}$ & $\textrm{F}_{\textrm{C}}$ \\
	\Xhline{2pt}
	$\textrm{TagMe}_t$
	& 32.2	& 6.3	& 17.3	& 9.4	& \textbf{34.6}	& 11.8	& 24.4	& 40.7	& 27.5	& 21.6 \\
	$\textrm{TagMe}_h$
	& 30.5	& \textbf{11.3}	& 15.9	& 11.0	& \textbf{34.6}	& 11.8	& 24.3	& \textbf{43.3}	& 26.5	& \textbf{24.4} \\\hline
	$\textrm{WAT}_t$
	& 25.0	& 8.9	& 15.5	& \textbf{15.4}	& 23.8	& 15.0	& 22.6	& 35.1	& 22.1	& 20.2 \\
	$\textrm{WAT}_h$
	& 31.7	& 8.5	& 14.8	& 14.7	& 22.4	& \textbf{16.2}	& 22.1	& 35.9	& 23.9	& 20.7 \\\hline
	$\textrm{REL}_t$
	& 26.1	& 0.0	& \textbf{31.7}	& 3.1	& 18.2	& 8.5	& 63.8	& 2.4	& 35.1	& 2.5 \\
	$\textrm{REL}_h$
	& \textbf{40.7}	& 0.0	& \textbf{31.7}	& 3.1	& 25.0	& 8.3	& \textbf{66.0}	& 2.4	& \textbf{43.1}	& 2.5  \\
\hline
\end{tabular}
}
\shrink
\end{table}

Table~\ref{tbl:el_breaksown} compares EL results for named entities and concepts separately, where F1 scores are computed based on only named entities $\textrm{F}_{\textrm{NE}}$ or concepts $\textrm{F}_{\textrm{C}}$.
We observe that REL is better at linking named entities, while TagMe is better at linking concepts. 
This shows that although it is important to achieve high performance for both named entities and concepts, there is no single EL tool that excels at both.
The results in Tables~\ref{tbl:el_breaksown} and~\ref{tbl:el_all} suggest that \emph{\textbf{all existing EL tools that we examined are suboptimal for EL in a conversational setting.}}

\begin{table}[t!]
\centering
\caption{Entity linking results on TREC CAsT raw and rewritten dialogues.}
\label{tbl:el_cast}
\begin{tabular}{l||  ccc | ccc } 
   \hline
	\multirow{2}{*}{\textbf{}}
	& \multicolumn{3}{c |}{\textbf{CAsT (raw)}} & \multicolumn{3}{c }{\textbf{CAsT (rewritten)}} \\
	&
	P & R & F & 
	P & R & F \\
   \Xhline{2pt}
	 $\textrm{TagMe}_t$
	& 52.7	& \textbf{47.3}	& 49.8
	& 64.1	& \textbf{66.4}	& \textbf{65.2}  \\
	 $\textrm{TagMe}_h$
	& 57.5	& 46.1	& \textbf{51.2}
	& 65.6	& 64.6	& 65.1 \\ \hline
	 $\textrm{WAT}_t$
	& 46.6	& 41.3	& 43.8
	& 52.9	& 45.3	& 48.8 \\
	 $\textrm{WAT}_h$
	& 44.7	& 45.5	& 45.1
	& 54.9	& 50.8	& 52.8 \\ \hline
	 $\textrm{REL}_t$
	& 68.1	& 19.2	& 29.9
	& 73.8	& 27.1	& 39.6 \\
	 $\textrm{REL}_h$
	& \textbf{70.2}	& 19.8	& 30.8
	& \textbf{76.4}	& 27.9	& 40.8 \\
   \hline
   \end{tabular}
 
 \end{table}

Table~\ref{tbl:el_cast} shows the EL results for all raw and rewritten CAsT queries.
Similar to Table~\ref{tbl:el_all}, we observe that there is a trade-off between precision and recall across the different EL tools. 
The results also show higher scores for rewritten queries compared to raw queries, which is due to resolved coreferences and richer context in the rewritten queries.

Table~\ref{tbl:el_pe} shows EL results on a sample of WoW dialogues, all containing references to personal entities (cf. Section~\ref{ch:conel:sec:annot}). This sample is annotated with concepts, named entities, and personal entities.
The left block shows the results of different EL methods in their original form, i.e., without annotating personal entity mentions.

The right block in Table~\ref{tbl:el_pe} represents a modified version of the same methods, where each method is extended to identify and link personal entity mentions, denoted with \emph{PE}. 
Considering a personal entity mention $m_{pe}$, and an entity $e$, the PE method computes the cosine similarity between the word embedding of $m_{pe}$ and the entity embedding of entity $e$.
For every $m_{pe}$, we compute this similarity with all the previously linked entities in the conversation and find the most similar entity.
Mention-entity pairs $\langle m_{pe}, e \rangle$ below a certain threshold $\tau$ are ignored. This threshold allows for filtering personal entity mentions that do not have the corresponding entities in the conversation history.
We used Wikipedia2Vec~\citep{Yamada:2020:W2V} word and entity embeddings released by \citet{Gerritse:2020:GEE}.
The threshold $\tau$ was set empirically by performing a sweep (on the range [0, 1] in steps of 0.1) using 5-fold cross-validation.

\begin{table}[t!]
\centering
\caption{Entity linking results on a sample of the WoW collection containing references to personal entities. The left block shows the results of the original EL methods and the right block represents the results of modified EL methods, where personal entities are also annotated.}
\label{tbl:el_pe}
\begin{tabular}{l ||@{~~}S l@{~~}S l@{~~}S l l } 
    \hline
	& 
	P & R & F \\
   \Xhline{2pt}
	$\textrm{TagMe}_t$ & 62.2	&	50.2	&	55.6 \\
	$\textrm{TagMe}_h$ & 62.4	&	49.6	&	55.3 \\ \hline
	$\textrm{WAT}_t$  & 56.0	&	63.1	&	59.4\\
	$\textrm{WAT}_h$  & 55.6	&	\textbf{67.1}	&	\textbf{60.8}\\ \hline
	$\textrm{REL}_t$ & 64.0	&	18.0	&	28.1\\
	$\textrm{REL}_h$  & \textbf{78.0}	&	22.0	&	34.3\\
   \hline
\end{tabular}
\,
\begin{tabular}{l ||@{~~}S l@{~~}S l@{~~}S l l } 
   \hline
	& 
	P & R & F \\
   \Xhline{2pt}
	$\textrm{TagMe}_t$+PE &	61.7	&	51.1	&	55.9 \\
	$\textrm{TagMe}_h$+PE  &	62.0	&	50.7	&	55.8 \\ \hline
	$\textrm{WAT}_t$+PE &	56.2	&	64.4	&	60.0	\\
	$\textrm{WAT}_h$+PE &	55.7	&	\textbf{68.8}	&	\textbf{61.6}	 \\ \hline
	$\textrm{REL}_t$+PE &	63.2	&	18.6	&	28.8	\\
	$\textrm{REL}_h$+PE &	\textbf{77.2}	&	22.6	&	34.9	\\
   \hline
\end{tabular}
\end{table}

Comparing the left and right parts of Table~\ref{tbl:el_pe}, we observe a slight (albeit often negligible) performance increase for the PE method. These results show that identifying personal entity mentions and their corresponding entities is a non-trivial task and cannot be resolved with a simple extension of current approaches. This reinforces our finding that all examined EL tools are suboptimal for EL in conversations.

\section{Conclusion}
In this chapter, we studied entity linking in a broad setting of conversational systems: QA, task-oriented, and social chat.
Using crowdsourcing, we analyzed existing conversational datasets and annotated them with concepts, named entities, and personal entities.
We found that while both concepts and named entities are useful for understanding the intent of user utterances, personal entities are mainly important in social chats.
Further, we compared the performance of different established EL methods in a conversational setting and concluded that none of the examined methods can handle this problem effectively, falling short in providing both high recall and precision, as well as annotating concepts, named entities, and personal entity mentions. Our annotated conversational dataset (ConEL) and interface designs are made publicly available. These resources come with detailed instructions on the procedure of collecting the annotations, which can be used for further extension of the collection. 
Following the insights from this study, developing conversational entity linking methods and employing them in various types of conversational systems are obvious future directions.

\chapter{Entity Linking for Personalization}
\label{ch:crel}

The analysis in Chapter~\ref{ch:conel} revealed that traditional Entity Linking (EL) methods are suboptimal for conversations, particularly struggling to identify the concepts and personal entities crucial for personalizing conversational systems. This finding necessitates the development of a new EL method tailored for conversational settings. Therefore, this chapter addresses the research question:

\vspace{1em}
\noindent
\hspace{2.8em}%
\begin{minipage}[t]{0.83\linewidth}
\textbf{RQ2:} \textit{How can we develop an Entity Linking method to accurately identify important entities for personalized conversational systems?}
\end{minipage}
\vspace{1em}

\noindent
To answer this question, this chapter introduces CREL, a conversational entity linking toolkit developed to identify named entities, concepts, and personal entities. Unlike existing EL methods, CREL utilizes coreference resolution to link personal references (e.g., ``my dog'') to explicit mentions within the conversation. CREL is trained on \mbox{ConEL-2}, an expansion of the original ConEL dataset (see Chapter~\ref{ch:conel}) with more conversations for training. The findings show that CREL significantly outperforms state-of-the-art baselines in linking entities needed for personalized interactions.

\section{Introduction}
\label{ch:crel:sec:introduction}

Understanding user utterances in conversational systems is fundamental to creating natural and informative interactions with users.
Entity Linking is a powerful text understanding technique that has been extensively explored in previous studies~\citep{Shang:2021:ERO} and Chapter~\ref{ch:conel}, and has been shown to be effective in question-answering and retrieval tasks~\citep{Yamada:2018:SOQ, Gerritse:2022:ETE, Dalton:2014:EQF, Hasibi:2016:EEL, Xiong:2017:WED, Dargahi:2018:QUE}. 
However, the EL approaches developed for documents prove to be suboptimal for conversations, to the extent that the state-of-the-art EL models are outperformed by the outdated models when evaluated on conversational EL datasets (Chapter~\ref{ch:conel}). The reasons are three-fold: 
	\textbf{(i)~Personal Entity Annotation}: Conversations often contain personal statements, referring to entities by their pronouns; e.g., ``my cars.'' Identifying personal entities is essential for personalization, including building personal knowledge graphs (PKGs)~\citep{Balog:2019:PKG, Li:2014:PKG, Tigunova:2019:LBL, Tigunova:2020:CHA} and generating personalized responses to users~\citep{Joshi:2017:PGD, Luo:2019:LPE}. Existing approaches for annotating documents, or even conversations~\citep{Laskar:2022:BEE, Xu:2022:OCR, Shang:2021:ERO}, do not identify and link personal entity mentions.
 \textbf{(ii)~Concept Annotation}: EL models developed for documents or short texts focus mainly on annotating named entities~\citep{vanHulst:2020:REL, Cao:2021:GENRE}. In conversations, however, both named entities and concepts contribute to machine understanding of text and should be annotated as discussed in Chapter~\ref{ch:conel}.
 \textbf{(iii)~Training Data}: Existing EL approaches are trained on well-written texts, while conversations have multiple (informal) turns with information spread over the turns.

In this chapter, we overcome the above challenges and provide a toolkit and a training dataset as two resources for EL in conversations. The challenges are addressed in the following ways. 

\begin{figure}[t]
  \includegraphics[width=1.0\linewidth]{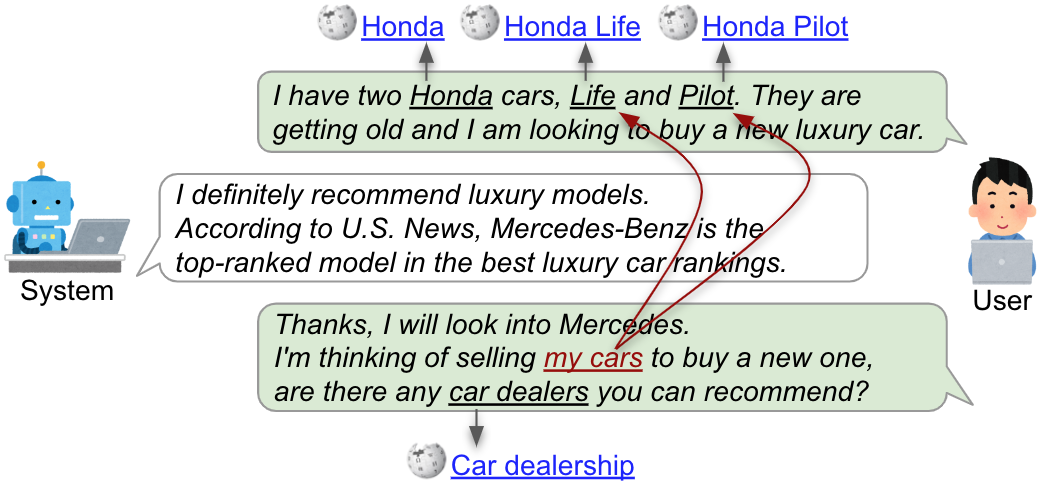}
    \caption{Example of entity linking in conversations.
      }
    \label{fig:example}
    \vspace{-1em}
\end{figure}

\medskip \noindent \textbf{Personal Entity Annotation.}
We detect \emph{personal entity mentions} (e.g., ``my cars''), find their corresponding \emph{explicit entity mentions} (e.g., ``Life'' and ``Pilot''), and link them to the corresponding entities in the knowledge graph (see Figure~\ref{fig:example}).
This task and coreference resolution both connect and stand in contrast to each other: they both aim to find mentions referring to the same entity~\citep{Singh:2011:LCD, Lee:2018:HOC, Joshi:2019:BFC, Kirstain:2021:CRS}, but personal entity mentions are sparse and different from common pronominal expressions that are resolved in coreference resolution.
This implies that state-of-the-art coreference resolution methods cannot be used out-of-the-box for resolving personal entities. We, therefore, identify personal entity mentions in a separate module and modify coreference resolution techniques to map personal entities to their explicit entity mentions.
	 
\medskip \noindent \textbf{Concept Annotation.} 
While open-domain knowledge graphs (e.g., Wikipedia) contain both named entities and concepts, conventional EL approaches are optimized for annotating only named entities, which hampers their performance for annotating conversations~\citep{vanHulst:2020:REL, Cao:2021:GENRE, Wu:2020:BLINK, Zhang:2022:ENT}. 
Specifically, the performance of named entity and concept linking in the widely used REL entity linker~\citep{vanHulst:2020:REL} varies dramatically (F-score of 43.1 and 2.5, respectively).
We mitigate this problem by training a mention detection model that can identify both concepts and named entity mentions in conversations and optimize the entity disambiguation model of REL~\citep{vanHulst:2020:REL} for all entity types. 
Our methods for personal entity, concept and named entity linking are integrated into our conversational EL tool, which is provided as a resource to the community. We show that this method outperforms state-of-the-art EL models.

\needspace{4\baselineskip}
\medskip \noindent \textbf{Training Data.}
A key to solving the aforementioned problems is training data. Existing studies on conversational entity linking either use private proprietary datasets~\citep{Shang:2021:ERO} or public datasets with an insufficient number of conversations for both training and evaluation purposes (Chapter~\ref{ch:conel}).
In this work, we build on the existing ConEL dataset (Chapter~\ref{ch:conel}) and address its shortcomings in our new dataset: (i) In ConEL, each personal entity mention is mapped to only a single explicit entity,
while in real conversations, personal entity mentions can have multiple corresponding entities (see Figure~\ref{fig:example}); (ii) ConEL contains a limited number of conversations with personal entities, as it synthesizes conversations mainly from question-answering and task-oriented benchmarks with a limited number of personal statements. Additionally, the ConEL-PE dataset, as in Chapter~\ref{ch:conel}, generated by annotating real social chats, contains only 25 annotated conversations, which cannot be used for training purposes.
Our newly developed dataset, referred to as \emph{ConEL-2}, consists of 1,327 utterances, annotated with personal entities, concepts, and named entities. This dataset has been used for training and evaluation of our conversational EL tool.

In summary, this work makes the following resources available to the community:\footnote{\url{https://github.com/informagi/conversational-entity-linking-2022}}
\begin{itemize}[leftmargin=0.9em]
  \item An open-source tool for entity linking in conversations, which can uniquely identify personal entities, concepts, and named entities. Our developed conversational EL methods outperform state-of-the-art baselines.
  \item A dataset containing EL annotations for conversations, which can be used for training and evaluation purposes. 
\end{itemize}

\section{Method}
\label{ch:crel:sec:method}

Our conversational entity linking approach consists of two components; 
(1) personal entity linking, and (2) concept and named entity linking.

\subsection{Personal Entity Linking}
\label{ch:crel:sec:method_pel}
Inspired by coreference resolution approaches~\citep{Joshi:2019:BFC, Kirstain:2021:CRS, Lee:2017:E2E, Lee:2018:HOC}, our model identifies personal entity mentions and their entity antecedents from conversation history. We disentangle mention detection from entity antecedent assignment and employ a BERT-based model (cf. Sect.~\ref{ch:crel:sec:method_cne}) for detecting entity mentions and a rule-based approach~\citep{Li:2014:PKG} for identifying personal entity mentions. Specifically, we  follow Chapter~\ref{ch:conel} to identify personal mentions, starting with ``my'' or ``our'' tokens followed by adjectives, common nouns, proper nouns, pronouns, numbers, particles, and articles. Also, ``of'' and ``in'' are allowed to be part of the mention. 
While admittedly this approach cannot capture all forms of personal entity mentions, we believe that it covers a reasonably large number of cases to be used for early experimentation in this direction. We leave the comprehensive research on different forms of personal entity mention for future work.

Similar to~\citet{Kirstain:2021:CRS}, we denote the start ($s$) and end token ($e$) representations of mentions as:
\begin{gather}
  m^{s} = GeLU(\mathbf{W}^{s}v) \quad m^{e} = GeLU(\mathbf{W}^{e}v),
\end{gather}
where $\mathbf{W}^{s}$ and $\mathbf{W}^{e}$ are trainable weights and $v$ is contextualized representation of a token.
We use LongFormer~\citep{Beltagy:2020:Longformer} to obtain contextualized embeddings as it can accommodate the long conversation history. The input to the LongFormer model is the current turn and conversational history. The hidden state in the output of the last layer of the model is used as $v$. 

Assuming mention $m_{bf}$ appears before mention $m_{af}$, the compatibility score between the two mentions is computed by:
\begin{align}
  score(m_{bf}, m_{af}) & =  m^{s}_{bf} \cdot B^{ss} \cdot m^{s}_{af}
        + m^{s}_{bf} \cdot B^{se} \cdot m^{e}_{af} \nonumber \\
        & + m^{e}_{bf} \cdot B^{es} \cdot m^{s}_{af}
        + m^{e}_{bf} \cdot B^{ee} \cdot m^{e}_{af},
 \label{eq:score}
\end{align}
where $B$ represents trainable weights and the superscripts $s$ and $e$ denote the start and end tokens of a mention.
Equation~\ref{eq:score} is computed for all pairs of personal entity mentions and explicit entity mentions.
During inference, personal and explicit entity mention pairs with the $score$ higher than threshold $\tau$ are selected.

\subsection{Concept and Named Entity Linking}
\label{ch:crel:sec:method_cne}
Entity linking typically involves two subtasks: (i) identifying mention boundaries and (ii) disambiguating mentions using a knowledge graph. 
In order to utilize existing advancements on these two subtasks, we need to adapt them to the conversational setup.
For mention detection (MD), this translates to adapting MD models to the conversational domain and identifying mentions of both named entities (NEs) and concepts. 
We employ the vanilla BERT~\citep{Devlin:2019:BER} architecture and fine-tune it to predict BIO labels, corresponding to begin, inside, and outside of mentions. The BERT model utilized in this work is pre-trained on conversational datasets such as DailyDialogues~\citep{Li:2017:DDM}, OpenSubtitles~\citep{Lison:2016:OSE}, Debates~\citep{Zhang:2016:CFO}, and Blogs~\citep{Schler:2006:EAG}.

For entity disambiguation (ED), we train REL~\citep{vanHulst:2020:REL} disambiguation approach on our ConEL-2 dataset.
REL is one of the state-of-the-art EL tools that is optimized for document annotation, but its disambiguation approach, considering both the local and global context of mentions, can be seamlessly adapted to conversations. Our entity linking method, consisting of the aforementioned MD and ED models is named CREL, which is an extension of REL tool for conversations.

\section{The ConEL-2 Collection}
\label{ch:crel:sec:annot}

To facilitate the development of EL approaches for conversations, we construct the ConEL-2 collection.
We draw the conversations from an existing conversational dataset. To select the dataset, we focus on social chat datasets, as personal entities are mainly found in social chat conversations (Chapter~\ref{ch:conel}).
We choose the Wizard of Wikipedia (WoW)~\citep{Dinan:2018:WWK} dataset, which contains multi-domain multi-turn conversations collected from actual interactions between two people. The selected dialogues are annotated in three stages: (i) identifying personal entity mentions, (ii) pairing personal and explicit entity mentions, and (iii) entity linking. Unless stated otherwise, each human intelligence task (HIT) is annotated by three turkers, and additional turkers are added when needed based on our quality check strategies.

\begin{figure}[t]
  \includegraphics[width=1.0\linewidth]{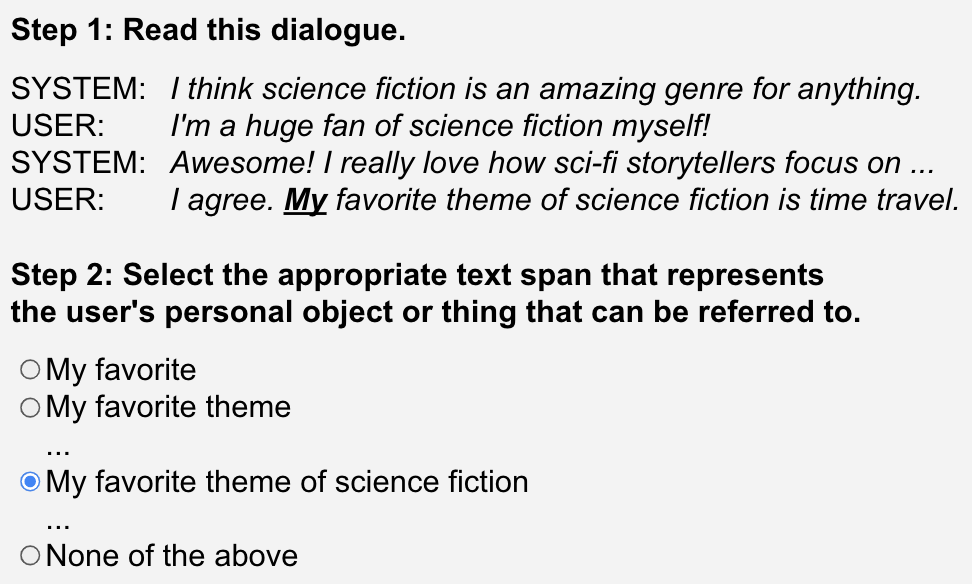}
    \caption{Annotation interface for identifying personal entity mentions (stage 1 of annotation process).}
    \label{fig:example_stage1}
\end{figure}

\medskip \noindent \textbf{Stage 1: Identifying personal entity mentions.}
In this stage, we ask turkers to identify text spans referring to personal entities (i.e., personal entity mentions).
First, we extract conversations that contain ``my'' or ``our'' in at least one user utterance, which amounts to 7,888 out of 22,311 conversations in the WoW dataset.
To make the annotation process technically and financially feasible, we randomly select 1,000 conversations from the extracted dialogues.
We then show the dialogue history and a set of candidate mentions to the turkers and ask them to find the best personal entity mention;  the interface is shown in Figure~\ref{fig:example_stage1}. Candidate mentions are text spans starting with ``my'' or ``our,'' followed by a maximum of 10 subsequent words. Our statistics on the collected annotations show that 98.9\% of the HITs were agreed upon by two or more turkers, with a Fleiss' kappa of 0.864.
We move forward with the 985 conversations, where at least two turkers agreed on all personal entity mentions of the conversation.

\medskip \noindent \textbf{Stage 2: Pairing personal and explicit entity mentions.}
In this stage, we ask turkers to find the corresponding explicit entity mentions for a given personal entity mention. For example, in Figure~\ref{fig:example}, the personal entity mention ``my cars'' is shown to turkers, and they are asked to find the corresponding explicit entity mentions ``Life'' and ``Pilot.''
We create a pool of candidate explicit entity mentions by using existing EL tools: TagMe~\citep{Ferragina:2010:TAG}, WAT~\citep{Piccinno:2014:FTM}, REL~\citep{vanHulst:2020:REL}, and GENRE~\citep{Cao:2021:GENRE}.
The detected mentions are manually processed and meaningless duplicates such as ``a restaurant'' and ``restaurant'' are resolved.
Turkers are then given the option to select one corresponding explicit entity mention or the  ``not in dialogue'' option if the corresponding entity mention is not found in the dialogue.
Fleiss' kappa for the inter-annotator agreement is 0.434, which is considered moderate. 
To improve the annotation quality, two extra annotations are collected for HITs where two turkers agreed on one option. We then select all conversations with equal or more than three turkers agreed on one option (either an explicit entity mention or the  ``not in dialogue'' option). This selection amounts to 290 conversations, which are passed to the next step.

We note that one of the shortcomings of the existing ConEL dataset is identifying only one explicit entity mention for each personal mention (cf. Section~\ref{ch:crel:sec:introduction}). 
In ConEL-2, we address this issue and ask turkers to find another corresponding explicit entity mention for all extracted conversations if any.

\medskip \noindent \textbf{Stage 3: Entity linking.}
In this stage, we ask turkers to select the corresponding Wikipedia article to the given entity mention, 
e.g., mapping the mention ``car dealers'' to the entity \emph{Car dealership} in Figure~\ref{fig:example}.
We first use the EL tools described in the stage 2 to create the pool of candidate mention-entity pairs. 
Noisy mentions such as ``please'' and ``I am'' are manually removed from the pool.
We also include top-$10$ Wikipedia search\footnote{\url{https://www.mediawiki.org/wiki/API:Main_page}} results using mentions as queries.
Turkers are then asked to select one of the candidate entities for each entity mention if any.

\begin{table}[t!]
  \centering
  \caption{Statistics of the ConEL-2 collection. 
    Personal entity annotations represent the number of annotations for personal entity mentions, with or without corresponding explicit entity mentions in the dialogue.
    }
  \label{tbl:annotation_statistics}
  \begin{tabular}{@{~}l|@{~~}l|@{~~}l|@{~~}l}
    \hline
    & Train & Val & Test \\
    \Xhline{1.5pt}
    Conversations & 174 & 58 & 58 \\
    User utterances & 800 & 267 & 260 \\
    NE and concept annotations & 1428 & 523 & 452 \\
    Personal entity annotations & 268 & 89 & 73 \\
    \hline
  \end{tabular}
\end{table}

The statistics show 98.7\% of the HITs were agreed by two or more turkers (Fleiss' kappa of 0.839), which highlights the high quality of our annotation process.
We therefore include all annotation results which are agreed by at least two turkers.
For the remaining 1.3\% of the non-agreed HITs, we recollected two extra annotations for each and select the entity that is selected by the majority of turkers.
The annotation results are shown in Table~\ref{tbl:annotation_statistics}. We split the dataset into train, validation, and test set with the 60\%-20\%-20\% ratio.

\section{Results}
\label{ch:crel:sec:results}

\subsection{Personal Entity Linking}
\label{ch:crel:sec:results_pe}

\medskip \noindent \textbf{Experimental Setup.}
We take the LongFormer-large model\footnote{\url{https://huggingface.co/allenai/longformer-large-4096}}~\citep{Beltagy:2020:Longformer} pre-trained on OntoNote Release 5.0~\citep{Hovy:2006:ON9}, and fine-tune it on the training set of the ConEL-2 dataset. This model is referred to as \seconv{}. 
We perform training with the batch size of 8 and learning rate of $10^{-5}$ using AdamW optimizer, and the validation set of ConEL-2 is used for early stopping.
For evaluation, we use validation and test sets of the ConEL-2 datasets as well as the ConEL-PE dataset (Chapter~\ref{ch:conel}). ConEL-PE contains 25 dialogues from the WoW dataset. We only use personal entity annotations for this evaluation.

\medskip \noindent \textbf{Baselines.}
We compare our model with the Wikipedia2Vec-based method (W2V) as in Chapter~\ref{ch:conel}, where antecedent assignment is based on similarity between Wikipedia2vec~\citep{Yamada:2020:W2V} embeddings of  entities and personal entity mentions. 
We also report on the official S2E  model\footnote{\url{https://github.com/yuvalkirstain/s2e-coref}}~\citep{Kirstain:2021:CRS} and its variation \seconvminus{}, where fine-tuning is performed only on our conversational dataset (without using OntoNote dataset).

\begin{table}[t!]
  \addtolength{\tabcolsep}{-1.9pt}
  \centering
  \miniskip
  \caption{Personal entity linking results. Micro-averaged precision, recall, and F1 scores are computed.}
  \label{tbl:pe}
  \resizebox{\textwidth}{!}{
    \begin{tabular}{l|ccc|ccc|ccc}
      \hline
      \multicolumn{1}{l|}{} 
      & \multicolumn{3}{c|}{ConEL-2 (Val)} & \multicolumn{3}{c|}{ConEL-2 (Test)} & \multicolumn{3}{c}{ConEL-PE} \\
      & P & R & F
      & P & R & F
      & P & R & F \\
      \Xhline{2pt}
      S2E~\citep{Kirstain:2021:CRS} & .375 & .415 & .394 & .292 & .302 & .297 & .317 & .481 & .382 \\
      W2V (Chapter~\ref{ch:conel}) & .324 & .508 & .395 & .458 & .429 & .443 & .436 & .630 & .515 \\ 
      S2E$^{-}_{\text{conv}}$ & .494 & \textbf{.615} & .548 & .450 & \textbf{.571} & .503 & .400 & \textbf{.741} & .519 \\
      S2E\textsubscript{conv}& \textbf{.714} & .538 & \textbf{.614} & \textbf{.673} & .524 & \textbf{.589} & \textbf{.613} & .704 & \textbf{.655} \\
      \hline
    \end{tabular}
  }
\end{table}

\medskip \noindent \textbf{Results.}
We measured the performance of our rule-based personal entity mention detection approach and obtained micro-averaged precision, recall, and F1 of 0.908, 0.898, and 0.903, respectively. This indicates that identifying personal entity mentions can be handled effectively by a simple approach. Assigning entity antecedents, on the hand, is a challenging task.

Table~\ref{tbl:pe} shows the results for personal entity linking. We observe that \seconv{} fine-tuned on both OntoNote and conversational datasets outperforms all baselines. It also indicates that, fine-tuning on conversational data substantially improves personal entity linking performance (S2E vs. \seconv{} results). 

\subsection{Concept and Named Entity Linking}
\label{ch:crel:sec:results_cne}

\medskip \noindent \textbf{Experimental Setup.}
For mention detection, we fine-tune bert-base-cased-conversational\footnote{\url{https://huggingface.co/DeepPavlov/bert-base-cased-conversational}} on our training data with the batch size of 16, learning rate of $5 \times 10^{-5}$, and AdamW optimizer, with the validation set of ConEL-2 for early stopping. For ED, we train the REL ED model, following the same training procedure as~\citet{vanHulst:2020:REL}.
As input for these models, we use only current turns for efficiency; although using the entire conversation history has shown slightly better performance (Chapter~\ref{ch:conel}), it doesn't change our conclusion, as it mainly benefits REL, with marginal benefits for WAT and TagMe.
For evaluation, we use the ConEL-2 and ConEL datasets. ConEL consists of 100 conversations that are collected from three different conversational tasks (question answering, task-oriented, and social chat) and annotated with concepts and named entities.
Only concept and named entity annotations are used for this evaluation.

\medskip \noindent \textbf{Baselines.}
REL~\citep{vanHulst:2020:REL}, GENRE~\citep{Cao:2021:GENRE}, TagMe~\citep{Ferragina:2010:TAG}, and WAT~\citep{Piccinno:2014:FTM} are strong publicly available EL tools that are used as our baselines. 
Additionally, we train three BERT-based MD models:
	(i) \emph{BERT$_{\text{conv}}$}: The bert-base-uncased model\footnote{\url{https://huggingface.co/bert-base-uncased}}, fine-tuned on our training data,
	(ii) BERT$_{\text{WP-conv}}$: Similar to BERT$_{\text{conv}}$, but fine-tuned on the Wikipedia anchor links from~\citet{Cao:2021:GENRE} before fine-tuning on ConEL-2 training data, and
		(iii) BERT$_{\text{NER-conv}}$: The bert-base-NER model which is originally trained for entity recognition (NER)\footnote{\url{https://huggingface.co/dslim/bert-base-NER}}, fine-tuned on our training data. 
The fine-tuning procedures are the same as CREL. We combine these models with the REL ED model trained on conversations (denoted as REL*) and report on their results.

\begin{table}[t!]
  \centering
  \caption{Concept and named entity linking results.
  F\textsubscript{MD} and F\textsubscript{EL} represent the micro-averaged F1 scores for mention detection and entity linking, respectively.}
  \label{tbl:md}
  \resizebox{\textwidth}{!}{
  \begin{tabular}{l|cc|cc|cc}
    \hline
    & \multicolumn{2}{c|}{ConEL-2 \footnotesize{(Val)}} & \multicolumn{2}{c|}{ConEL-2 \footnotesize{(Test)}} & 
    \multicolumn{2}{c}{ConEL} \\
    & F\textsubscript{MD} & F\textsubscript{EL} & F\textsubscript{MD} & F\textsubscript{EL} & 
    F\textsubscript{MD} & F\textsubscript{EL} \\
    \Xhline{2pt}
    GENRE \citep{Cao:2021:GENRE} & .290 & .252 & .320 & .299  & .350 & .211 \\
    REL \citep{vanHulst:2020:REL} & .304 & .244 & .279 & .231 & .462 & .245 \\
    TagMe \citep{Ferragina:2010:TAG} & .559 & .478 & .611 & .504 & .510 & .375 \\
    WAT \citep{Piccinno:2014:FTM} & .616 & .539 & .613 & .519 & .416 & .336 \\
    \hline
    REL*(BERT$_{\text{conv}}$) & \textbf{.748} & \textbf{.652} & .716 & .587  & .551 & .424 \\
    REL*(BERT$_{\text{WP-conv}}$) & .697 & .624 & .708 & .592 & .508 & .399 \\
    REL*(BERT$_{\text{NER-conv}}$) & .723 & .635 & .693 & .576 & .548 & .407 \\
    CREL & .742 & .651 & \textbf{.729} & \textbf{.597} & \textbf{.559} & \textbf{.429} \\
    \hline
    \end{tabular}
  }
\end{table}

\medskip \noindent \textbf{Results.}
The results are reported in Table~\ref{tbl:md}.
F\textsubscript{MD} and F\textsubscript{EL} are the F1 scores for MD and EL, respectively.
The results show that the REL and GENRE, while being state-of-the-art EL toolkits, perform worse than TagMe and WAT. 
This is because REL and GENRE are intended to detect only named entities, while TagMe and WAT are able to identify concepts as well.
Comparing CREL with REL* methods, we observe that adaptation of BERT for conversational domain plays an important role in conversational entity linking, reflected by the highest score for ConEL-2 (test) and ConEL datasets. 

\begin{table}[t!]
  \centering
  \caption{Overall entity linking results, annotating both personal entities (PE) and concepts and named entities (EL).}
  \label{tbl:overall}
  \scalebox{0.95}[0.95]{
    \begin{tabular}{ll|ccc|ccc|ccc}
      \hline
      \multicolumn{1}{l}{} & \multicolumn{1}{l|}{}
      & \multicolumn{3}{c|}{ConEL-2 \footnotesize{(Val)}} & \multicolumn{3}{c|}{ConEL-2 \footnotesize{(Test)}} & \multicolumn{3}{c}{ConEL-PE} \\
     EL & PE & \multicolumn{1}{c}{P} & \multicolumn{1}{c}{R} & \multicolumn{1}{c|}{F}
      & \multicolumn{1}{c}{P} & \multicolumn{1}{c}{R} & \multicolumn{1}{c|}{F}
      & \multicolumn{1}{c}{P} & \multicolumn{1}{c}{R} & \multicolumn{1}{c}{F} \\
      \Xhline{2pt}
      REL & W2V & .523 & .159 & .244 & .516 & .159 & .243 & .488 & .206 & \underline{.290} \\
      REL & S2E\textsubscript{conv} & .523 & .162 & .248 & .522 & .165 & \underline{.250} & .466 & .206 & .286 \\ 
      \hline
      GENRE & W2V & .468 & .180 & .260 & .569 & .223 & .320 & .537 & .256 & .347 \\
      GENRE & S2E\textsubscript{conv} & .589 & .169 & .263 & \textbf{.675} & .217 & \underline{.328} & \textbf{.646} & .256 & \underline{.367} \\ 
      \hline
      TagMe & W2V & .504 & .407 & .451 & .530 & .440 & .481 & .528 & .518 & .523 \\
      TagMe & S2E\textsubscript{conv} & .612 & .395 & .480 & .621 & .434 & \underline{.511} & .636 & .518 & \underline{.571} \\ 
      \hline
      WAT & W2V & .506 & .489 & .497 & .545 & .464 & .501 & .536 & .628 & .579 \\
      WAT & S2E\textsubscript{conv} & .585 & .490 & .534 & .582 & .464 & \underline{.516} & .571 & .648 & \underline{.607} \\ 
      \hline   
      CREL & W2V & .636 & .589 & .612 & .607 & .554 & .579 & .573 & .714 & .635 \\
      CREL & S2E\textsubscript{conv} & \textbf{.712} & \textbf{.593} & \textbf{.647} & .634 & \textbf{.566} & \underline{\textbf{.598}} & .600 & \textbf{.724} & \underline{\textbf{.656}} \\
      \hline

  \end{tabular}
  }
\end{table}

\subsection{Overall Performance}
\label{ch:crel:sec:results_overall}

Putting all the pieces together, we report on the overall performance of our model for linking personal entities, concepts and named entities.
For this experiment, we report only on the ConEL-2 and ConEL-PE datasets, as ConEL does not have annotations for personal entities. The results are shown in Table~\ref{tbl:overall}.
We notice at first glance that our proposed  personal entity linking extension (S2E\textsubscript{conv}) leads to improved F-scores compared to W2V in most cases. The results also show that our EL tool (CREL with S2E\textsubscript{conv}) achieves the highest F-score for all sets, reinforcing the importance of adapting EL approaches for conversations.

\section{Conclusion}

In this chapter, we tackled the problem of entity linking in conversational settings, where not only named entities, but also concepts and personal entities are important.
To this end, we collected conversational entity annotations, which allows us to train and evaluate entity linking in conversations.
Additionally, based on the collected annotations, we develop a conversational EL tool.
Our empirical results show that our EL tool achieves the highest performance over conventional EL tools for documents and short texts.
The collected annotations and our EL tool with detailed instructions are publicly available as a research resource for further study in conversations.
Future work includes incorporating coreference resolution methods to identify more diverse personal entity mentions.

\chapter{Personalized Response Generation}
\label{ch:laps}

The previous chapters developed methods for personal information extraction, yet effectively utilizing extracted personal context for personalized response generation poses a different challenge. 
The central research question this chapter addresses is:

\vspace{1em}
\noindent
\hspace{2.8em}%
\begin{minipage}[t]{0.83\linewidth}
\textbf{RQ3:} \emph{How can users' personal preferences be utilized to generate personalized responses?}
\end{minipage}
\vspace{1em}

\noindent
The difficulty in addressing this question stems from the lack of realistic large-scale dialogue datasets for training and evaluating personalized response generation models.
The existing datasets often lack the multi-session nature and real-world user preferences, while previous collection approaches that rely on experts are difficult to scale.

In this chapter, we first introduce LAPS, a method that leverages large language models (LLMs) to guide human workers in efficiently generating personalized dialogues to answer this research question. 
The LAPS approach collects large-scale, human-written, multi-session, and multi-domain conversations with real user preferences.

Building upon LAPS, to answer RQ3, we then train a preference extraction model and perform personalized response generation using a semi-structured \textit{preference memory} constructed from user's preferences.
The experiments demonstrate that preference memory improves accuracy, provides explanations, and mitigates LLMs' recall issues in response generation.
\section{Introduction}
\label{ch:laps:sec:introduction}

\begin{figure}[t]
\centering
\includegraphics[width=0.99\linewidth]{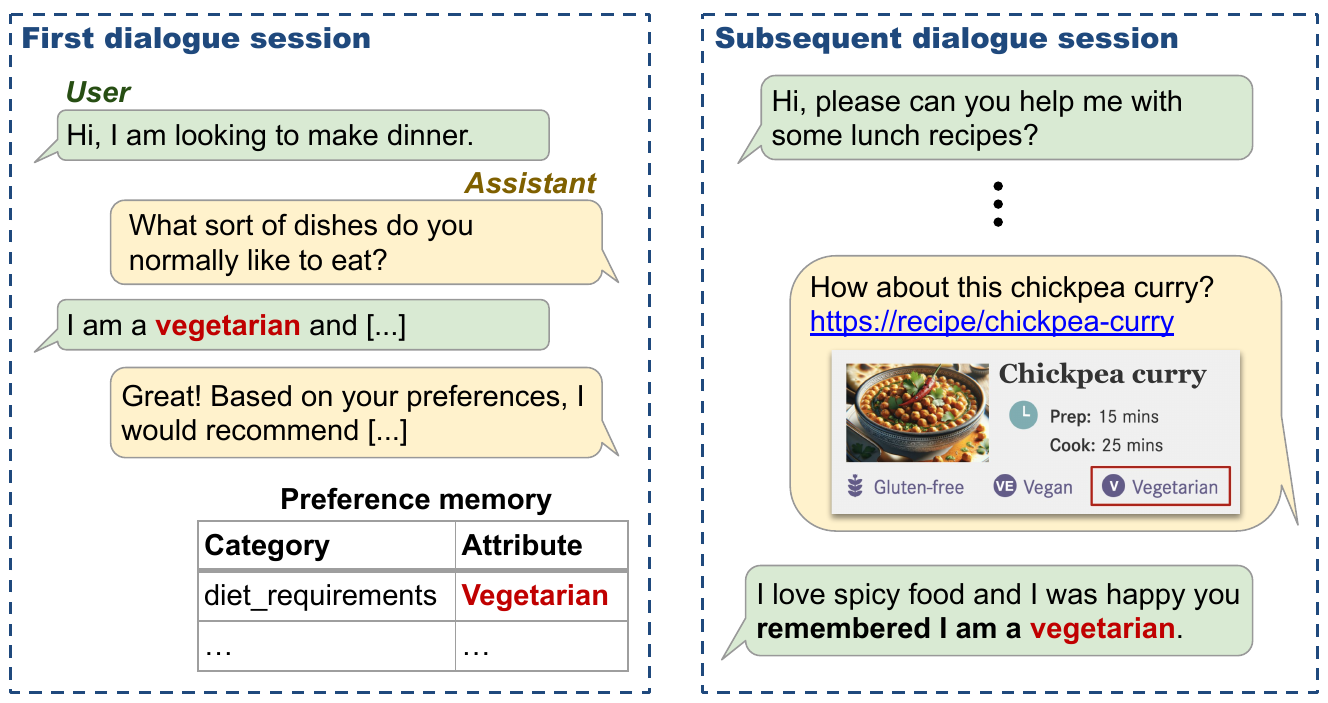}
\caption{A snippet from a multi-session dialogue.
User preferences are extracted and stored in memory to generate personalized recommendations in subsequent sessions.
}
\label{fig:dialogue-illustration}
\miniskip
\end{figure}

Personalization is paramount for conversational search and recommendation~\citep{Luo:2019:LPE, Brown:1987:PSU}. 
Conversational agents need to meet users' expectations and provide them with individually tailored responses. 
In real-world scenarios, where users interact with dialogue agents across multiple sessions, conversational systems need to accurately understand, extract, and store user preferences and articulate personalized recommendations based on the stored user profile; see Figure~\ref{fig:dialogue-illustration}. 
The Information Retrieval (IR) community has studied various aspects of personalization in conversational systems. For instance, the concept of Personal Knowledge Graph (PKG)~\citep{Balog:2019:PKG} is introduced to enable personalization of (conversational) search systems. Similarly, TREC Interactive Knowledge Assistance Track (iKAT) utilizes Personal Text Knowledge Base (PTKB)~\citep{Aliannejadi:2024:IKAT} for persona-based conversational search.
Recent years have also witnessed tremendous progress in LLMs~\citep{Brown:2020:LMF,Ouyang:2022:TLM}. Yet LLM-based conversational agents do not effectively handle user preferences, delivering generalized recommendations that fail to capture the nuanced interests of individual users.

A major hindrance in developing a personal conversational system is the unavailability of \emph{large-scale human-written} conversational datasets.
These are needed for training new models and understanding user behavior, in expressing their preferences over multiple dialogue sessions~\citep{Radlinski:2019:CCP, Joho:2017:FIW}. However, constructing such datasets has proven a daunting task~\citep{Bernard:2023:MGS, Radlinski:2019:CCP}. 
Preference elicitation in conversations is complex, and crowd workers engage poorly with the task.
While human experts deliver quality results, recruiting them as intermediary coaches or as human agents to generate conversations does not scale.
The recently-emerged paradigm of collecting synthetic conversational data through LLMs~\citep{Gilardi:2023:COC,Yunxiang:2023:CMC,Lee:2022:GPD,Xu:2023:BOC,Chen:2023:PLM,Kim:2023:SMD,Kim:2022:BTF,Jandaghi:2023:FPC,Cheng:2025:BMC,Askari:2025:SSM} raises concerns about dialogue diversity~\citep{Reif:2023:VLD,Chung:2023:IDM,Yu:2023:LLM,DellAcqua:2023:NAJ,Park:2023:DDS}.
Crucially, LLM-generated conversations do not represent actual user preferences and interactions, undermining the credibility for developing \emph{future} personal conversational systems.

The critical question that arises here is:

\vspace{1em}
\noindent
\hspace{2.8em}%
\begin{minipage}[t]{0.83\linewidth}
    \textbf{RQ3a:}  \textit{How can we collect cost-effective yet large-scale multi-session human-written conversational datasets for personalized response generation?}
\end{minipage}
\vspace{1em}

\noindent We address this question by proposing LAPS, an \underline{\textbf{L}}LM-\underline{\textbf{A}}ugmented \underline{\textbf{P}}ersonalized \underline{\textbf{S}}elf-Dialogue method to collect large-scale personal conversations.
LAPS employs an LLM to dynamically generate personal \textit{guidance} for crowd workers, playing both user and assistant roles. 
The guidance is generated based on the previously elicited user preferences and the current state of the dialogue, determined by a dialogue act classifier. After each dialogue session, LAPS extracts preferences from the dialogue and stores them in a \textit{preference memory}. This memory is a key-value store of user preferences, analogous to the PKG~\citep{Balog:2019:PKG} and PTKB~\citep{Aliannejadi:2024:IKAT} concepts. Using LAPS, we collect 1,406 multi-domain multi-session dialogues, paired with 11,215 preferences.

Our follow-up research question concerns the quality of LAPS-generated datasets:

\vspace{1em}
\noindent
\hspace{2.8em}%
\begin{minipage}[t]{0.83\linewidth}
    \textbf{RQ3b:}  \textit{How do the LLM-augmented self-dialogues compare to human- and synthetically-generated conversations?}
\end{minipage}
\vspace{1em}

\noindent We compare our conversations with a wide range of widely used conversational datasets and show that LAPS-generated conversations score higher in diversity (based on Dist-n, Ent-n, and SELF-BLEU metrics) and quality (based on UniEval~\citep{Zhong:2022:TUM}). We further compare LAPS- and LLM-generated dialogues using GPT-3.5 and GPT-4 and show that LLM-generated dialogues are less diverse than those involving humans, even with temperature tuning.

Although LAPS extracts preferences in a semi-structured format and stores them in a preference memory, one could wonder whether such memory is needed, given LLMs' abilities in handling long context from the previous sessions.
This leads us to the second follow-up research question:

\vspace{1em}
\noindent
\hspace{2.8em}%
\begin{minipage}[t]{0.83\linewidth}
    \textbf{RQ3c:}  \textit{How can preference memory enhance the effective utilization of user preferences in recommendations?}
\end{minipage}
\vspace{1em}

\noindent To address this question, 
we train a preference extraction model on our dataset and use it to build a preference memory from previous sessions.
These preferences are then incorporated into the LLM's prompt for generating personalized recommendations. 
We compare this approach to the baseline prompting method, where dialogue histories of all sessions are appended to the prompt.
Our experiments show that by incorporating preference memory, the model can more accurately utilize the users' disclosed preferences for recommendations than the baseline method.
The notable advantage of preference memory is that it contributes to more explainable recommendations.
Finally, we found that when using the baseline method, the LLM struggles with recalling user preferences; likely due to lengthy prompts. 

\medskip 
\noindent\textbf{Contributions} of this work are as follows:%
\begin{itemize}[leftmargin=2em]
  \item We introduce LAPS method for collecting scalable multi-session personalized dialogues with actual user preferences.
   
  \item We analyze and compare various dialogue collection methods, demonstrating that LAPS collects lexically diverse and high-quality dialogues, uncovering the diversity issue of generating fully-synthetic dialogues with LLMs.
  \item We study the benefits of storing and using user preferences in a semi-structured format (preference memory) and show that it helps an LLM in recalling previously disclosed preferences when generating personalized recommendations.
   \item Enabled by LAPS, we release a unique conversational dataset that is multi-session, human-written, large-scale, and contains users' personal preferences.\footnote{The code and dataset is available at \url{https://github.com/informagi/laps}}

\end{itemize}

\section{Related Work}
\label{ch:laps:sec:related}

\subsection{Dialogue Collection Methods}
\label{ch:laps:sec:related:collection}

\textbf{Human-Human} interactions are arguably the optimal strategy for collecting natural dialogues. MultiWOZ \citep{Budzianowski:2018:MWO} and PersonaChat \citep{Zhang:2018:PDA} are notable examples, offering task-oriented and chit-chat datasets, respectively. These datasets, however, focus less on real user preferences, relying instead on predefined tasks or personas.
Addressing this limitation, \citet{Radlinski:2019:CCP} introduced CCPE-M, a dataset emphasizing actual user preferences. Similarly, \citet{Bernard:2023:MGS} introduced MG-ShopDial, an e-commerce conversational dataset with genuine preferences. Despite their quality, the small size of these datasets (502 dialogues for CCPE-M and 64 for MG-ShopDial) limits their utility for training large models, and they overlook the multi-session aspect of real-world dialogues.

A quality-quantity trade-off in dialogue collection is highlighted by \citet{Bernard:2023:MGS}.
Initially attempting crowdsourcing, \citet{Bernard:2023:MGS} shifted to a volunteer-based collection due to low engagement, resulting in higher quality but fewer dialogues. This underscores the challenge in gathering large, high-quality datasets representing true user preferences.

\medskip \noindent \textbf{Self-Dialogue}, where a single worker simulates both roles, is an effective approach for large-scale data collection. Introduced by \citet{Krause:2017:EBO}, this technique has been shown to produce high-quality data, with \citet{Fainberg:2018:TMS} noting its increased coherence and reduced errors compared to human-human dialogue. \citet{Byrne:2019:TTR} further validated this, emphasizing the superior linguistic diversity and fewer mistakes in self-dialogue data.
However, self-dialogue has limitations, especially in managing complex conversations, such as those involving preference elicitation, while acting in two distinct roles.
Additionally, self-dialogue cannot authentically capture unknown user preferences as the same person plays both user and assistant roles, leading to fewer clarification scenarios than in human-human interactions~\citep{Fainberg:2018:TMS}.
Our approach introduces LLM-augmented personalized self-dialogues, leveraging LLMs to ease the cognitive load on workers and addressing the limitation of capturing unknown preferences by involving an LLM, which doesn't have pre-existing knowledge of user preferences.

\medskip \noindent \textbf{(Semi)-Synthetic Dialogue Generation} offers an alternative to relying on crowd workers \citep{Soudani:DAC:2023, Lee:2022:GPD,Xu:2023:BOC,Chen:2023:PLM,Kim:2023:SMD,Kim:2022:BTF,Jandaghi:2023:FPC,Leszczynski:2023:TWS}. \citet{Lee:2022:GPD} developed PERSONACHATGEN using two LLMs for dialogues between personas, requiring multiple model calls for one dialogue. \citet{Chen:2023:PLM} optimized this by using a single model call with a carefully crafted prompt. \citet{Leszczynski:2023:TWS} took a different approach, generating synthetic dialogues by combining user-generated content and metadata.
However, concerns exist about the diversity in LLM-generated texts. \citet{Reif:2023:VLD} noted lexical and syntactic repetition, developing LinguisticLens for syntactic diversity analysis. \citet{Chung:2023:IDM} emphasized the need for human intervention to enhance diversity, and \citet{Yu:2023:LLM} pointed out the uniformity in LLM outputs from simple prompts, suggesting diverse prompts for more varied data. The diversity issue is also evident in fields like social science and business \citep{DellAcqua:2023:NAJ,Park:2023:DDS}.

To address this, semi-synthetic data collection methods like having crowd workers edit LLM-generated texts have been effective \citep{Shah:2018:BNC,Rastogi:2020:TSM,Bae:2022:BRS}. \citet{Shah:2018:BNC} introduced M2M, a framework using templates for initial dialogue generation and crowd workers for rewriting. Similarly, \citet{Rastogi:2020:TSM} used a schema-guided method. 
Our research improves upon these by using LLMs for providing workers with guidance for response composition, promoting diversity and reducing the influence of generated drafts.

\subsection{Personalized Conversational Systems}
Problem-driven conversational systems, especially those for search and recommendation, benefit from personalization.
Shifting from traditional approaches such as collaborative filtering, recent personalization focuses on more interactive approaches, such as preference elicitation~\citep{Radlinski:2019:CCP,Li:2023:EHP,Christakopoulou:2016:TCR,Sun:2018:CRS,Zhao:2022:KCP,Kostric:2021:SUP,Kostric:2023:GUQ}.
These elicited preferences can be stored as a knowledge graph~\citep{Balog:2019:PKG} through entity linking~\citep{vanHulst:2020:REL} or in text format~\citep{Aliannejadi:2024:IKAT}, and later used for personalization. 
Storing user preferences in a (semi-)structured format enables conversational agents to better satisfy users' information needs and can be also useful to mitigate bias~\citep{Gerritse:2020:BCS}. 
Following this line, our method elicits user preferences and stores them in a semi-structured preference memory.

\medskip \noindent \textbf{Memory and feedback} also offer the promise of making systems better aligned with user needs.
This approach, tracing back to ALFRED~\citep{Riesbeck:1981:FDR}, involves storing past failures to improve future interactions. Recently, \citet{Madaan:2023:SIR} advanced this concept with SELF-REFINE, enabling LLMs to refine their outputs iteratively using their own feedback.
For tasks requiring deeper personalization, human feedback is invaluable.
\citet{Madaan:2022:MPE} enhanced LLM response quality by adding a memory module that remembers information from the user's past sessions.
Aligned with \citet{Madaan:2022:MPE}, our approach includes a memory module to store user preferences from past interactions, enabling enhanced personalization.

\section{Dialogue Collection Method}
\label{ch:laps:sec:method}

\begin{figure}[t]
  \centering
  \includegraphics[width=\columnwidth]{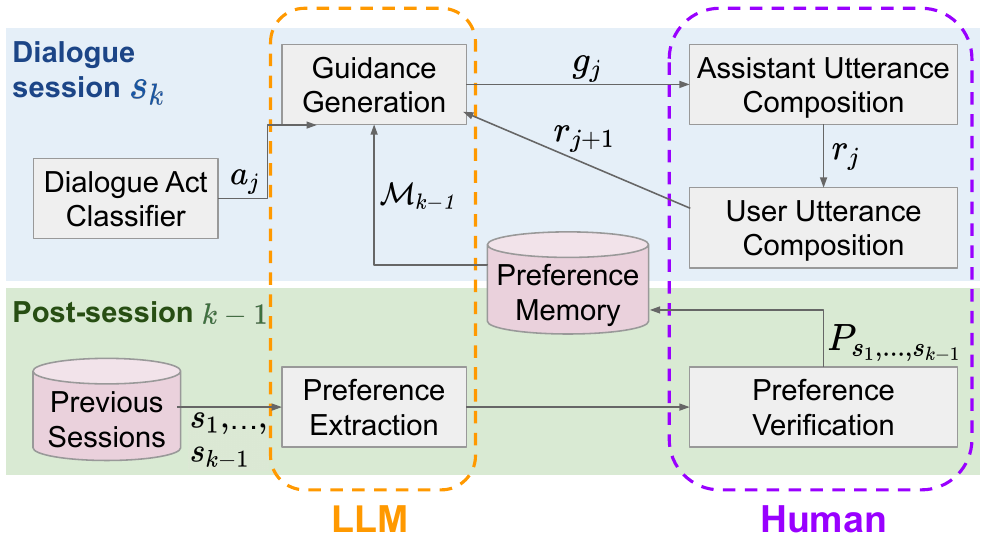}
  \caption{Overview of our dialogue collection method (LAPS).} 
  \label{fig:block-diagram}
\end{figure}

We propose LAPS, an LLM-Augmented Personalized Self-Dialogue construction method, capable of collecting large-scale, human-written, multi-session, and multi-domain conversations, paired with extracted user preferences. The method consists of four key elements (cf. Fig.~\ref{fig:block-diagram}): (i) dialogue act classification, (ii) guidance generation, (iii) utterance composition, and (iv) preference extraction.

The dialogue act classifier determines the next action that the assistant should take; e.g., recommend.  
Based on the dialogue act, the LLM generates guidance considering the dialogue history and the previously extracted preferences.
The human agent then composes the assistant response based on the LLM-generated guidance, and then switches to the role of a user, providing a response to the previous utterance. The process continues until a relevant recommendation is made and the dialogue session is completed.
Upon completion of a session, the preferences are extracted from the dialogue using an LLM and checked by the human agent.
These preferences, once confirmed by the same human agent, are stored in the \textit{preference memory} and used in subsequent sessions for generating personalized guidance. The human agent is then encouraged to initiate a new dialogue session for another scenario in the given domain.  This process continues until the human agent exits the job or reaches the end of all pre-defined session scenarios. 

\subsection{Task Formulation}
\label{ch:laps:sec:method:formulation}
The objective of this task is twofold: firstly, to create a large-scale collection of multi-session dialogues written by humans, focusing on user preferences; and secondly, to extract and compile the specific user preferences mentioned within these dialogue sessions.
Formally, the dialogue collection method $F$ is defined as a mapping from a set of task descriptions $\mathbb{T}$ and human agents $\mathbb{H}$ to a set of dialogue sessions and their extracted preferences $\mathbb{S}$:
\begin{align*}
    &F : \mathbb{T}, \mathbb{H} \rightarrow \mathbb{S}\\
    &F(t,h) = [(s_1, P_{s_1}), \ldots, (s_n, P_{s_n})],
\end{align*}
where $t$ is a task description for a topic,  $h$ is a human agent with identical preferences for a given topic, and $n$ represents the total number of sessions. The dialogue session $s_i$ and its corresponding preference set $P_{s_i}$ are defined as:
\begin{align}
    &s_i = [u_1^i, u_2^i, \ldots, u_m^i], \nonumber \\
    &P_{s_i} = \{(c, \wp_j)   \, | \, c \in C, j \in \mathbb{N}\}, \label{eq:pref}
\end{align}
where the dialogue session $s_i$ composed of a sequence of utterances $u$, and the preference set $P_{s_i}$ is a set of  category-preference pairs $(c, \wp_j)$, where the category $c$ belongs to the set of categories $C$; e.g., \{\texttt{allergy}, \texttt{cuisine}, \texttt{diet}\}.
We note that the preference set $P_{s_i}$ is generated only after the completion of session $s_i$, by extracting user preferences from the dialogue post-session and validating them with the same worker. The extracted preferences of a human agent are stored in a memory component $\mathcal{M}$, defined as: 
\begin{equation*}
\mathcal{M}_k = \bigcup_{\substack{1 < i \le k \\ k \le n}} P_{s_i},
\end{equation*}
where session $s_k$ is the last completed session by the human agent. 

Here we draw an analogy between the preference memory $\mathcal{M}$ and PKGs~\citep{Balog:2019:PKG}, where the personal information of users is stored according to an ontology. The \texttt{<subject, predicate, object>} triplets in PKGs correspond to human ($h$), category ($c$), and preference ($\wp$) in preference memory, respectively. We note that unlike a PKG that is built based on a pre-defined ontology, preference memory uses a more relaxed version of categories that are extracted on-the-fly from user utterances.  Preference memory can be also viewed  as a semi-structured form of PTKB, where free-form sentences about a user's persona are transformed in a key-value format.

\subsection{Guidance Generation}
Generating guidance for human agents is central to collecting large-scale, high-quality,  human-written utterances in LAPS.
Large-scale construction of conversational data requires recruiting crowd workers. However, due to the complexity of the task and high cognitive load of generating conversational utterances, crowd workers show poor engagement~\citep{Bernard:2023:MGS}. The challenge is even more intense for the preference elicitation task, where we need to coach the human agent simulating the system to ask engaging questions to reveal user preferences~\citep{Radlinski:2019:CCP}. A remedy could be utilizing LLMs to generate system utterances. This, however, results in less diverse conversations and is not in line with our aim of generating training data for \emph{future} personalized conversational search and recommendation systems. Even by instructing crowd workers to re-write LLM-generated utterances, we observed (in our pilot studies) that crowd workers become less creative in generating their own utterances and tend to replicate pre-generated utterances.

To alleviate the aforementioned problems, we propose to coach crowd workers throughout the conversation using automatically generated personalized guidance, and let the workers compose their own utterances via dialogue self-play.
Using this approach, we reduce the cognitive load of a highly complex task to a minimum, allowing workers to focus on a simple sub-task at a time and generate high-quality and engaging conversations. Formally, to compose the assistance utterances $u_j$, the human agent receives personalized guidance $g_j$ generated by an LLM. The guidance generation function $\mathcal{G}$ is defined as:
\begin{equation*}
\mathcal{G}\left(u_1, \ldots, u_{j-1}, a_j, \mathcal{M}_k\right) = g_{j},
\end{equation*}
where $\mathcal{M}_k$ denotes the preference memory extracted from sessions $(s_1, \ldots, s_k)$, utterances $u_1, \ldots, u_{j-1}$ represents conversations history up until turn $j$, and $a_j$ is the action that needs to be taken for turn $j$, obtained from the dialogue act classifier (cf. Subsection~\ref{ch:laps:sec:method:act}).

\medskip \noindent \textbf{Instantiation.}
As an instantiation of this function, we prompt GPT-3.5 \texttt{turbo} to generate personalized guidance. The guidance prompt template takes the dialogue history $u_1, \ldots, u_{j-1}$, preference memory $\mathcal{M}_k$, and instructions for the current dialogue act $a_j$ as inputs. The prompts include detailed step-by-step instructions for chain-of-thought prompting~\citep{Wei:2022:CTP}.
For instance, the prompt for the \emph{preference elicitation} act in a given \texttt{DOMAIN} is as follows:
\vspace{0.5em}
\begin{mdframed}[linecolor=black, linewidth=0.5pt, roundcorner=5pt, backgroundcolor=gray!10, font=\small]
    \textit{You are an advisor, who supports} \textbf{\$\texttt{DOMAIN}} \emph{recommendation assistants to compose responses to users.
    In this step, the assistant \textbf{collects information about user's preferences}} ([$a_j$]).
    
    \vspace{2mm}
    
    \noindent Preference memory: [$\mathcal{M}_k$]\\
    \noindent Dialogue history: [$u_1, \ldots, u_{j-1}$]
    
    \vspace{2mm}
    \noindent \textit{Step 1: Identify the last user turn.\\
    Step 2: Explain the intent of the last user utterance.\\
    Step 3: \textbf{Which preference(s) should the assistant ask next} given the dialogue history?\\
    Step 4: \textbf{Compose very short guidance for the human assistant on how to write a response to the user.}\\
    Step 5: Output in JSON format following [...]\\
    Ensure that you distinctly label and delineate Steps 1, 2, 3, 4, and 5. Let's think step by step:
    }
\end{mdframed}
\vspace{0.5em}
The guidance prompt for \emph{recommendation} act is similar to the \emph{preference elicitation} act, except that the assistant is instructed to recommend an item based on the user's personal preferences with URLs. The guidance also includes ``\textit{When making recommendations, if necessary, effectively utilize the user's preferences, such as [...]}'' to encourage the assistant to use the user's preferences disclosed in the previous sessions if necessary.

\subsection{Dialogue Act Classification}
\label{ch:laps:sec:method:act}
Dialogue act is an action that can change the (mental) state of conversation and guide the system to generate the next utterance~\citep{Gao:2019:NAC}. Dialogue act is used as an input to the guidance generation function and is obtained by the dialogue act classifier $\mathcal{A}$:
\begin{equation*}
    \mathcal{A}(u_{1}, \ldots, u_{j-1}, a_{j-1}) = a_{j}
\end{equation*}
which determines action $a_j$ based on the dialogue history $u_1, \ldots, u_{j-1}$ and the previous dialogue act $a_{j-1}$.

\medskip \noindent \textbf{Instantiation.}
We instantiate the act classifier function by prompting GPT-3.5 \texttt{Turbo}.
A series of dialogue acts are defined, outlining the specific actions to be taken sequentially. The primary dialogue acts in our setup are: (1) \textit{greeting}, (2) \textit{preference elicitation}, (3) \textit{recommendation}, (4) \textit{follow-up questions}, and (5) \textit{goodbye}. A dialogue act is selected upon completion of the previous act, which is determined by the LLM using detailed chain-of-thought instruction prompts; e.g., the \textit{recommendation} act is selected when the LLM produces the response ``true'' for the instruction prompt ``\textit{[...] Has the user shared any of the preferences listed above? Has the assistant collected why the user has the preference? If both are true, return true.}''

When collecting preferences for the first session, the \textit{preference elicitation} act is further divided into three sub-actions for collecting (i) \emph{must-have}, (ii) \emph{should-have}, and (iii) \emph{could-have} preferences. Once the \emph{must-have} and \emph{should-have} preferences are collected, the subsequent sessions only collect \emph{could-have} preferences, using the preference memory for other preferences.  

\subsection{Utterance Composition}
\label{ch:laps:sec:method:utterance-composition}
Conversation utterances are composed by the human agent for both user and assistant roles via dialogue self-play.
For the assistant role, the composition process is supported by the LLM-generated \textit{guidance}, which enables generating diverse human-written system utterances. For the user role, the human agent mainly needs to elaborate his preferences and state his opinion about the recommendations. The system and user utterance generation functions are defined as:
\begin{align*}
 & \mathcal{W}_s\left(S_k, g_{j}, u_1, \dots, u_{j-1}\right) =  u_j, \\
 & \mathcal{W}_u\left(S_k, u_1, \dots, u_j\right) = u_{j+1}, 
\end{align*}
where $\mathcal{W}_s$ and  $\mathcal{W}_u$  represent the function for writing system and user responses, respectively.
Here, $g_j$ denotes the guidance for generating utterance $u_j$, and  $S_k = \{s_1, \dots, s_k\}$ represents the session history, comprising all previous dialogue sessions up until, but not including, the current session.%

\medskip \noindent \textbf{Instantiation.}
These functions are instantiated by recruiting crowd workers and instructing them to write a self-dialogue for our task: 
\textit{``Your task is to chat with yourself both as a user (seeking a cooking recipe or movie) and an assistant (offering recipe or movie recommendations). You need to discuss preferences and receive suggestions while playing both roles.''}
This instruction is followed by brief descriptions of roles, session settings, and chat interface instructions, as well as general information about the payment and rejection policy.

For the user role, workers are instructed to be themselves, provide their preferences, review the recommendations, and offer feedback.
In contrast, the assistant role entails more tasks. Assistants must elicit preferences to inform their recommendations and provide URLs for the recommended items. Following user feedback, assistants confirm why the user likes or dislikes the recommendation to obtain a reusable preference for the next session.

\subsection{Preference Extraction}
Upon completion of a dialogue session, user preferences are extracted from the dialogue. Due to the inherent ambiguity and complexity of human-written conversations, we collect these preferences from the same human agent that generates the conversation.
Note that here we only collect user preferences that are mentioned in the course of previous conversation sessions and not general user preferences. Formally, given the dialogue session $s_i$, the preference extraction function $\mathcal{E}$ is defined as:
\begin{equation}
\label{eq:preference-extraction}
\mathcal{E}(s_i, C) =  P_{s_i},
\end{equation}
where $C$ denotes the set of preference categories, and $P_{s_i}$ is the set of category-preference pairs extracted from the dialogue session.

\medskip \noindent \textbf{Instantiation.}
Extraction of preferences is performed using a semi-automated approach, where the initial set of preferences is extracted by an LLM and then validated by the human agent. We prompt GPT4 with chain-of-thought instructions to generate preference attributes for each preference category $c \in C$, given the dialogue session $s_i$ and preference categories $C$.

Once an initial set of preferences is extracted, the worker is instructed to confirm the correctness of each extracted preference and verify all disclosed preferences during the conversations are extracted.
The worker is then directed to start a new conversation session or terminate the task.

\subsection{ Evaluation}
\label{ch:laps:sec:method:dialogue-eval}

{\renewcommand{\arraystretch}{1.3}
\begin{sidewaystable}
\centering
\footnotesize
\setlength{\tabcolsep}{3pt}

\caption{Overview of the selected baseline datasets and ours. EXP and CW denote an expert and a crowd worker, respectively.} 

\label{tab:dataset-comparison}
\resizebox{\textwidth}{!}{
\begin{tabular}{l || l | c | l | l | c | c | c | c | c }
\hline
\multirow{2}{*}{\textbf{Dataset}} & \multicolumn{1}{l|}{\textbf{Collection}} & \multirow{2}{*}{\textbf{\#Dial}} & \multirow{2}{*}{\textbf{Tasks}} & \multirow{2}{*}{\textbf{Domains}} & \multirow{2}{*}{\textbf{Scalability}} & \multicolumn{1}{c|}{\textbf{Actual User}} & \multicolumn{1}{c|}{\textbf{Preference}} & \multicolumn{1}{c|}{\textbf{Preference}} & \multicolumn{1}{c}{\textbf{Multi-}} \\ 
& \multicolumn{1}{l|}{\textbf{Method}} & & & & \multicolumn{1}{c|}{\textbf{}} & \multicolumn{1}{c|}{\textbf{Preferences}} & \multicolumn{1}{c|}{\textbf{Elicitation}} & \multicolumn{1}{c|}{\textbf{Extraction}} & \multicolumn{1}{c}{\textbf{Session}} \\
\Xhline{2pt}
\makecell[l]{SGD} & \makecell[l]{Semi-synthetic} & 16,142 & \makecell[l]{Booking, rec., etc.} & \makecell[l]{20 domains, inc. restaurants} & \ding{51} & \texttimes & \texttimes & \texttimes & \texttimes \\
\grayline
\makecell[l]{M2M}  & \makecell[l]{Semi-synthetic} & 3,008 & \makecell[l]{Booking} & \makecell[l]{Restaurants, movies} & \ding{51} & \texttimes & \texttimes & \texttimes & \texttimes \\
\grayline
\makecell[l]{PersonaChatGen} & \makecell[l]{Fully-synthetic} & 1,649 & \makecell[l]{Personal chit-chat} & \makecell[l]{Open domain} & \ding{51} & \texttimes & \texttimes & \texttimes & \texttimes \\
\grayline
\makecell[l]{Taskmaster-1} & \makecell[l]{Self-dialogue} & 7,708\textsuperscript{\dag} & \makecell[l]{Booking, ordering} & \makecell[l]{6 domains, inc. restaurants}  & \ding{51} & \texttimes & \texttimes & \texttimes & \texttimes \\
\grayline
\makecell[l]{MultiWOZ}  & \makecell[l]{Human-human (CW)} & 8,438 & \makecell[l]{Booking, rec., etc.} & \makecell[l]{7 domains, inc. restaurants} & \ding{51} & \texttimes & \texttimes & \texttimes & \texttimes \\
\grayline
\makecell[l]{CCPE-M}  & \makecell[l]{Human-human (EXP)} & 502 & \makecell[l]{Rec.} & \makecell[l]{Movies} & \texttimes & \ding{51} & \ding{51} & $\triangle$ & \texttimes \\
\grayline
\makecell[l]{MG-ShopDial} & \makecell[l]{Human-human (EXP)} & 64 & \makecell[l]{Rec., QA, etc.} & \makecell[l]{E-commerce} & \texttimes & \ding{51} & \ding{51} & \texttimes & \texttimes \\
\hline
\makecell[l]{LAPS} & \makecell[l]{LAPS} & 1,406 & \makecell[l]{Rec.} & \makecell[l]{Recipes, movies} & \ding{51} & \ding{51} & \ding{51} & \ding{51} & \ding{51} \\
\hline
\end{tabular}
}
\tiny
\begin{flushleft}
\hspace{1.0em} \textsuperscript{\dag} Includes only self-dialogues.\\
\hspace{1.0em} $\triangle$ Annotations are provided but no entity-relation pair extraction with like/dislike distinction.
\end{flushleft}
\end{sidewaystable}}

\medskip \noindent \textbf{Baseline Datasets.}
We evaluate our LAPS method by comparing existing conversational datasets with the conversations generated by LAPS. 
We carefully select the baseline datasets by examining 170 datasets listed in Chapter~\ref{ch:conel}, and narrowing down our selection by the following criteria: (i) inclusion of preference elicitation, (ii) utilization of preferences, and (iii) focus on task-oriented dialogues. We prioritize datasets that are both published in peer-reviewed venues and publicly available. 
The selected datasets are summarized in Table~\ref{tab:dataset-comparison}.
For a fair comparison, when possible, we select a single domain from each baseline dataset that overlaps with the domains of the dataset created by LAPS, namely recipe and movie. For open-domain datasets, e.g., PersonChatGen~\citep{Lee:2022:GPD}, we select dialogues with food-related personas, categorized under ``Food'' and ``Drink.''

\medskip \noindent \textbf{Baseline Methods.}
We tried three other human-based dialogue collection methods: \emph{Human-Human}, \emph{self-dialogue}, and \emph{LLM-human}. These methods do not scale and cannot generate quality conversations. In the \emph{Human-Human} method, we encountered difficulties in pairing up two workers simultaneously due to the high dropout rates, caused by the complex nature of multi-session preference elicitation. For the \emph{self-dialogue} method, we followed~\citet{Speggiorin:2022:TPM} and simplified the assistant role by providing users with a predefined response. This, however, led to homogeneous dialogues, even though the workers were instructed to rewrite the response. In the \emph{LLM-human} method, we generated assistant responses using GPT-3.5 \texttt{turbo} and asked workers to rewrite the utterances. However, even after experimenting with multiple temperature settings, the dialogue remained homogeneous (a common issue with LLMs reported also~\citep{Reif:2023:VLD, Chung:2023:IDM, Yu:2023:LLM, DellAcqua:2023:NAJ, Park:2023:DDS}) and workers often accepted the grammatically correct responses without re-writing them.

While we focus on collecting dialogues with actual user preferences, one might wonder whether an LLM could also play the role of a human agent. To address this question, we prompt the LLM to act both as a user and an agent, using the same input and output as human agents in LAPS (cf.\S~\ref{ch:laps:sec:method:utterance-composition}). We report on the results obtained from GPT-3.5 \texttt{turbo} and GPT-4-1106 \texttt{preview} with a temperature parameter of 1.0. These dialogues are 3 sessions long.

\medskip \noindent \textbf{Dialogue Diversity.}
We use three metrics to measure the lexical diversity of the collected conversations: (i) Dist-n~\citep{Li:2016:DPO}, (ii) Ent-n~\citep{Zhang:2018:GID}, and (iii) Self-BLEU~\citep{Zhu:2018:TBP}.
\textit{\textbf{Dist-n}} measures the response diversity by computing the ratio of distinct n-grams to all n-grams in the given collection. Following~\citet{Li:2016:DPO}, we use Dist-1 and -2 to measure the lexical diversity of conversation utterances.
\textit{\textbf{Ent-n}} aims to enhance the Dist-n measure by incorporating the frequency differences of n-grams into consideration, leveraging the entropy of the n-gram distribution. We report on Ent-4, following~\citet{Zhang:2018:GID}.
\textit{\textbf{Self-BLEU}} considers one utterance as a hypothesis and the rest of the utterances in the collection as references and calculates the BLEU score for each utterance. Following the NLTK's default setting~\citep{Bird:2009:NLP} and~\citet{Zhu:2018:TBP}, we compute the BLEU scores for $n=1,2,3,4$ for each hypothesis-reference pair and take the mean over all computed BLEU scores.

\textit{Normalization.}
A frequently overlooked pitfall of diversity metrics is their dependency on the total number of words in the dialogues.
For instance, Dist-n scores are typically higher for datasets with fewer total number of words, as they are less likely to have repeated words.
To address this, we set a word cutoff for all datasets and randomly sample dialogues until the total word count reaches this cutoff.
The cutoff is set at 7,012 words, which is the minimum number of total words for user/system utterances across all datasets. 
We perform 100 random sampling per dataset and report the average scores, as well as two-tailed independent t-test results.
Additionally, since M2M~\citep{Shah:2018:BNC} dataset is lowercased, we lowercase all other datasets for a fair comparison.

\medskip \noindent \textbf{Dialogue Quality.}
Dialogue quality evaluation is inherently challenging due to its subjective nature. 
Human evaluation, often considered the gold standard, is known to be highly sensitive to task design and instructions. Even with much care, it still suffers from differing bias and high variance per annotator, especially in crowdsourced environments~\citep{Deriu:2021:SEM,Smith:2022:HEC,Finch:2023:DFY,Li:2019:AID,Zhang:2023:BLL}.
Automatic evaluation, while capable of mitigating the aforementioned issues, is also known to have its own limitations, including a bias towards machine-generated responses~\citep{Liu:2023:GNE}.
Aware of these limitations, we opt for automatic evaluation for two reasons: (1) our aim is to ensure our dialogue quality aligns with other high-quality datasets, rather than attaining state of the art, and (2) we use human workers to compose responses in their own words, which is less likely to be overestimated by metrics biased towards machine-generated responses.

After examining four reference-free automatic evaluation methods~\citep{Mehri:2020:USR,Lin:2023:UMA,Liu:2023:GNE,Zhong:2022:TUM}, we select \textbf{UniEval}~\citep{Zhong:2022:TUM} as our automatic evaluation metric considering availability, cost, and performance.
For evaluation aspects, we choose naturalness, understandability, and coherence from~\citet{Zhong:2022:TUM}.
The aspects that require the conditioning fact as an input are not used, as the factuality of user preferences falls outside the scope of our study.
We use the official implementation of UniEval\footnote{\url{https://github.com/maszhongming/UniEval}} and its default settings. 
To ensure that the evaluation is computationally feasible using our available computational resources, we randomly select 100 dialogues (consisting of 1.8K responses on average) from each dataset.
For significance testing, we use two-tailed independent t-tests ($p < 0.05$).

\subsection{Experimental Setup}
{\medskip \noindent \textbf{Domains.}
Our domains are \textit{recipe} and \textit{movie}.
The recipe domain involves planning for the next dinner (session 1), breakfast (session 2), and lunch (session 3).
The movie domain involves planning to watch a movie with family, friends, or alone (session 1), exploring another movie by the same director or actress/actor as the previous recommendation (session 2), and watching with different people or a different occasion (session 3). For each domain, we curated a list of categories that are relevant to the task and categorized them into categories of  \textit{must-have}, \textit{should-have}, and  \textit{could-have}. These categories are detailed in our online repository.
}

\medskip \noindent \textbf{Participants and Quality Control.}
We recruited Prolific\footnote{\url{https://www.prolific.com/}} workers from English-speaking countries,
having $ \geq 98\% $ approval rate and $ \geq 1000 $ previous submissions.
For the movie domain, we only invited workers that accurately performed that task for the recipe domain. Throughout the experiments, we actively communicated with workers, answering over 250 questions and incorporating their feedback to clarify instructions.
£14 was paid for completing three sessions, which took \textasciitilde65 minutes.
Workers were also allowed to terminate the task at an earlier session and receive partial payment.
This allowed us to collect high-quality multi-session dialogues, as workers completing all sessions tend to be more engaged in the task.
After crowdsourcing, we manually reviewed the dialogues to ensure data quality, making corrections or deletions as necessary.
Common errors (aside from malicious behavior of not providing meaningful responses) include failing to include URLs in recommendations and misunderstanding their current role in the conversation.

\medskip \noindent \textbf{Chat Interface.}
We collect conversations using TaskMAD~\citep{Speggiorin:2022:TPM} and further develop it to support new features for our task.
The human agent interacts with two chat interfaces for system and user roles, and each interface consists of a text box for composing responses and an instruction box to clarify role responsibilities.

\section{Personal Recommendation Method}
\label{ch:laps:sec:recommendation}

The end goal for large-scale preference elicitation conversational datasets is to enable personal conversational search and recommendation. Based on LAPS, we collect a large-scale dataset for the movie and recipe domains and use it to train a model for extracting personal preferences from conversations. The preferences are then used to generate personalized recommendations. 

\subsection{Preference Extraction}
\label{ch:laps:sec:rec:extraction}
In this task, we aim to automatically extract user preferences from a dialogue session (cf.\ Eq.~\ref{eq:preference-extraction}).
We cast this task as a seq-to-seq QA~\citep{Chung:2022:SIL} and fine-tune an LLM with instruction prompts eliciting user preference regarding category and conversation session. Formally, our method decomposes Eq.~\ref{eq:preference-extraction} as $ \mathcal{E}(s, \mathcal{C}) = \bigcup_{c \in \mathcal{C}} \mathcal{E}'(s, c) $, where $\mathcal{E}'$ is the preference extraction function for individual category $ c $.

We use FlanT5~\citep{Chung:2022:SIL} as the base model and use instruct prompts to read the given session and answer questions about the user's preferences, such as \textit{``What cuisine does the user like?''}
For the recipe domain, we use FlanT5 Small (80M parameters), Base (250M), and Large (780M) models and fine-tune them on our recipe dataset. For the movie domain, we explore a domain adaptation approach, where the initially fine-tuned model on the recipe domain (FlanT5-Large) is further fine-tuned on the movie domain.

\subsection{Personalized Recommendation}
With the personalized recommendation task, we aim to generate a recommendation response based on user's personal preferences.
Personal preferences are stated in the conversation history $h$ and all previously completed sessions $S_k$. By utilizing the preference extraction method (cf. \S~\ref{ch:laps:sec:rec:extraction}), we construct the preference memory $\mathcal{M}_k$ based on session history $S_k$ and use it for recommendation. Formally, recommendation utterance $ u^{pred} $ is generated by the recommendation generation function $R$, defined as $R\left(h, \mathcal{M}_{k}\right) = u^{pred} $.
Recommendation responses are generated by an LLM (LLaMA-2-7B~\citep{Touvron:2023:LOF}) using zero-shot prompting. The prompt contains instructions to generate a personalized recommendation appended with the preference memory and history of the current conversation. 

\subsection{Evaluation}
\label{ch:laps:sec:downstream:eval-metric}

\medskip \noindent \textbf{Recommendation Baseline.}
 For our personalized recommendation method, we consider a baseline method, where the preference memory $\mathcal{M}_k$ is replaced with raw utterances from session history $S_K$. This baseline allows us to measure the effectiveness of using semi-structured fine-grained preferences over raw conversational utterances.

\medskip \noindent \textbf{Recommendation Human Evaluation.}
Two human experts are instructed to evaluate the \emph{rationale} and \emph{relevance} aspects of recommendations: ``Which assistant's rationale is more in line with the user's preferences?'', and ``Which assistant's recommendation item meets user preferences?''
Following~\citet{Li:2019:AID}, we conduct pairwise comparisons between responses from our method and a baseline. For each aspect, the annotators choose win, lose, or tie options, and disagreements are resolved through discussion.
For evaluation, we randomly select 50 and 10 samples from the recipe and movie domains, respectively.

 \medskip \noindent \textbf{Recommendation Automatic Evaluation.}
Automatic machine translation measures such as ROUGE and BLEU assume significant overlap between ground truth and valid responses. This strong assumption does not hold for dialogue systems and in particular for personal recommendations, where valid responses represent high level of diversity. The lack of correlation between human evaluation is reported in previous study~\citep{Liu:2016:HNE} and has been observed in our study as well.
Addressing this challenge, we introduce an automatic reference-based evaluation metric, \textbf{Preference Utilization (PU)}, which aims to measure the utilization of user preferences in the recommendation response.
Let $ \mathcal{P} = \{\wp_1, \wp_2, \ldots\} $ denote the set of preferences (without their corresponding categories as defined in Eq.~\ref{eq:pref}).
The subsets $ \mathcal{P}^{pred} \subseteq \mathcal{P} $ and $ \mathcal{P}^{ref} \subseteq \mathcal{P} $ denote preferences $\wp_i \in \mathcal{P}$ that appear in predicted and reference recommendation responses, respectively. Preference utilization precision $P_{PU}$ and recall $R_{PU}$ are computed as: 
\begin{align*}
    P_{PU} = \frac{|\mathcal{P}^{pred} \cap \mathcal{P}^{ref}|}{|\mathcal{P}^{pred}|}, 
    \quad
    R_{PU} = \frac{|\mathcal{P}^{pred} \cap \mathcal{P}^{ref}|}{|\mathcal{P}^{ref}|}.
\end{align*}
In our experiments, we perform exact string matching to generate $\mathcal{P}^{pred}$ and $\mathcal{P}^{ref}$.
When reference utterance $ u^{ref} $ includes URLs, we extract the content from the corresponding web page and use it in string matching.
For our ground truth, we only use the assistant responses about recommendations which are accepted by the user.

\medskip \noindent \textbf{Preference Extraction Evaluation.}
We evaluate preference extraction using exact match and BERTScore~\citep{Zhang:2020:BET}.
Exact match assesses the case-insensitive string match of preferences between predictions and ground truth, while BERTScore accounts for different expressions of identical preferences.
Here, preferences within each category are flattened into a single string with commas as delimiters and used for computation of BERTScore.

\subsection{Experimental Setup}

For preference extraction, we fine-tuned Flan-T5 models on our training set using the HuggingFace Transformers library~\citep{Wolf:2020:TSN}.
For both domains, the batch size is set to 8, the learning rate to $5 \times 10^{-5}$, and the AdamW optimizer is used.
Training is conducted for 10 epochs, with the best checkpoint selected based on the validation set. We ran all our experiments on a single GPU (NVIDIA A100 40GB). For a personalized recommendation, we use Llama-2-7B~\citep{Touvron:2023:LOF} due to its ability to handle a sufficiently long context, while being capable of running on a single GPU. We run the model (from the HuggingFace model hub) 10 times for each dialogue and report the average score.

\section{Results}
\label{ch:laps:sec:results}

This section presents the results of the proposed methods for dialogue collection (\S\ref{ch:laps:sec:res:dialogue}) and personal recommendation (\S\ref{subsec:Personal Recommendation Evaluation})

\subsection{Dialogue Collection Evaluation}
\label{ch:laps:sec:res:dialogue}

\begin{table}[t!]
\centering
\caption{Statistics of LAPS dataset.}
\label{tab:dataset-overview}
\setlength{\tabcolsep}{3.5pt}
\begin{tabular}{l | l || c | c | c | c | c | c }
\hline
\multirow{3}{*}{\textbf{Domain}} & \multirow{3}{*}{\textbf{Split}} & \multicolumn{3}{c|}{\textbf{\#Dialogue Sets}} & \multirow{3}{*}{\textbf{\#Pref}} & \multirow{3}{*}{\textbf{\#Utt}} & \multirow{3}{*}{\textbf{\#Dial}} \\ 
\cline{3-5}
& & \textbf{\footnotesize Single-} & \textbf{\footnotesize Two-} & \textbf{\footnotesize Three-} & & & \\[-2.5pt]
& & \textbf{\footnotesize Session} & \textbf{\footnotesize Session} & \textbf{\footnotesize Session} & & & \\[-1pt]
\Xhline{2pt}
\multirow{4}{*}{Recipe} & Train & 163 & 24 & 160 & 5,538 & 9,342 & 691 \\
& Val & 24 & 5 & 21 & 772 & 1,333 & 97 \\
& Test & 48 & 10 & 41 & 1,600 & 2,610 & 191 \\
& \textbf{Total} & \textbf{235} & \textbf{39} & \textbf{222} & \textbf{7,910} & \textbf{13,285} & \textbf{979} \\
\hline
\hline
\multirow{4}{*}{Movie} & Train & 46 & 14 & 72 & 2,225 & 3,974 & 290 \\
& Val & 5 & 1 & 13 & 351 & 642 & 46 \\
& Test & 11 & 4 & 24 & 729 & 1,220 & 91 \\
& \textbf{Total} & \textbf{62} & \textbf{19} & \textbf{109} & \textbf{3,305} & \textbf{5,836} & \textbf{427} \\
\hline
\end{tabular}
\end{table}

Using LAPS, we collect a large-scale multi-session personalized dataset, as shown in Table~\ref{tab:dataset-overview}.
The number of preferences is the total number of $(s, p, o)$ triples, where $s$ is the user, $p$ is the preference category, and $o$ is the preference attribute. Using human verification, we identified  4.5\% error rate in preferences extracted by GPT-4, highlighting the need for human involvement for accurate preference extraction.

\begin{table}[t!]
\centering
\caption{Lexical diversity scores.
    Significance against all baselines is marked by \textsuperscript{+}.
}
\label{tab:eval-baselines}
\setlength{\tabcolsep}{7pt}
\begin{tabular}{l || ccc }
\hline
\multirow{2}{*}{\textbf{Dataset}} & \multirow{2}{*}{\textbf{Dist-1/2}} & \multirow{2}{*}{\textbf{Ent-4}} & \multicolumn{1}{c}{\textbf{Self-}} \\
& & & \multicolumn{1}{c}{\textbf{BLEU\textsuperscript{$\blacktriangledown$}}} \\
\Xhline{2pt}
SGD & 0.179 / 0.538 & 8.311 & 0.964 \\
M2M & 0.057 / 0.290 & 7.922 & 0.955 \\
PersonaChatGen & 0.165 / 0.523 & 8.261 & 0.970 \\
Taskmaster-1 & 0.207 / 0.644 & 8.384 & \underline{0.949} \\
MultiWOZ & 0.158 / 0.505 & 8.345 & 0.966 \\
CCPE-M & 0.175 / 0.571 & 8.414 & 0.961 \\
MG-ShopDial & \textbf{0.234} / \underline{0.653} & 8.199 & \textbf{0.935} \\
\hline
LAPS-Recipe & 0.207 / 0.650 & \underline{8.563}\textsuperscript{+} & 0.955 \\
LAPS-Movie & \underline{0.222}\textsuperscript{+} / \textbf{0.666}\textsuperscript{+} & \textbf{8.593}\textsuperscript{+} & 0.954 \\
\hline
\end{tabular}
\footnotesize
\begin{flushleft}
\hspace{1.5em} \textsuperscript{$\blacktriangledown$} Lower is better.
\end{flushleft}
\end{table}

\medskip \noindent \textbf{Lexical Diversity.}
Table \ref{tab:eval-baselines} shows the results of lexical diversity evaluation,
indicating that LAPS-Movie and -Recipe achieve the highest diversity scores with respect to Dist-2 and Ent-4. Considering all metrics, LAPS-Movie is on par with MG-ShopDial, a dataset of human-human dialogues involving trained volunteers as an assistant role.
These results demonstrate that LAPS collects lexically diverse dialogues as effectively as human-human dialogue collection methods.

\begin{figure}[t]
\miniskip
\centering
\includegraphics[width=0.9\linewidth]{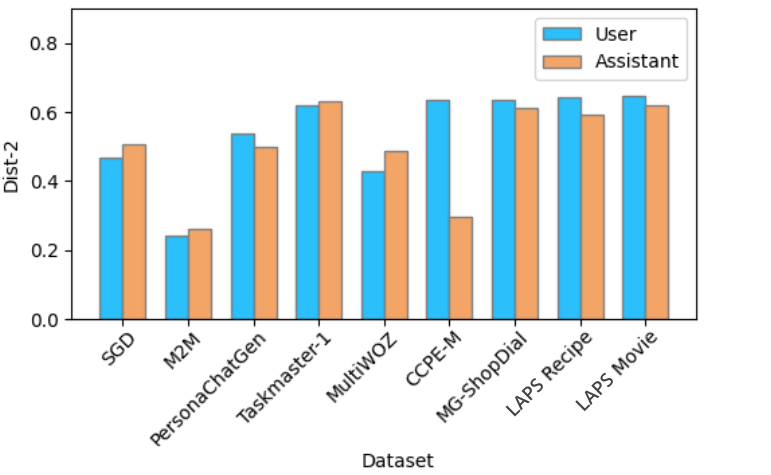}
\caption{Lexical Diversity of user and assistant utterances.}
\label{fig:diversity-user-asst}
\end{figure}

Comparing the lexical diversity of user and system utterances in Figure \ref{fig:diversity-user-asst}, we observe 
that our method achieves high lexical diversity for both user and assistant utterances. This suggests that LLM's guidance can help workers compose diverse responses. 
Notably, we observe that assistant utterances in CCPE-M exhibit less diversity than those of the user. 
This could be attributed to the small number of participants acting as assistants in CCPE-M and their often short and direct responses.
This highlights that achieving diversity is non-trivial, even with trained experts.
LAPS's success in achieving diversity further demonstrates the effectiveness of our method.

\begin{table}[t!]
\centering
\caption{Lexical diversity scores of synthetic dialogue generation.
Significance against all baselines is marked by \textsuperscript{+}.
}
\setlength{\tabcolsep}{3pt}
\label{tab:eval-synthetic}
\begin{tabular}{l|l||ccc}
\hline
\multirow{2}{*}{\textbf{Domain}} & \multirow{2}{*}{\textbf{Method}} & \multirow{2}{*}{\textbf{Dist-1/2}} & \multirow{2}{*}{\textbf{Ent-4}} & \multicolumn{1}{c}{\textbf{Self-}} \\
& & & & \multicolumn{1}{c}{\textbf{BLEU\textsuperscript{$\blacktriangledown$}}} \\
\Xhline{2pt}
\multirow{3}{*}{Recipe} & Synthetic GPT-3.5 & 0.127 / 0.430 & 8.308 & 0.981 \\
 & Synthetic GPT-4 & 0.183 / 0.597 & \textbf{8.601} & 0.976 \\
 & LAPS & \textbf{0.207}\textsuperscript{+} / \textbf{0.65}\textsuperscript{+} & 8.563 & \textbf{0.955}\textsuperscript{+} \\
\hline
\hline
\multirow{3}{*}{Movie} & Synthetic GPT-3.5 & 0.138 / 0.444 & 8.331 & 0.977 \\
 & Synthetic GPT-4 & 0.178 / 0.559 & 8.481 & 0.979 \\
 & LAPS & \textbf{0.222}\textsuperscript{+} / \textbf{0.666}\textsuperscript{+} & \textbf{8.593}\textsuperscript{+} & \textbf{0.954}\textsuperscript{+} \\
\hline
\end{tabular}
\footnotesize
\begin{flushleft}
\hspace{0.7em} \textsuperscript{$\blacktriangledown$} Lower is better.
\end{flushleft}
\end{table}

Table~\ref{tab:eval-synthetic} compares LAPS- and LLM-generated conversations. The results show that synthetic dialogues are less diverse than LAPS, even with GPT-4 temperature tuning. This suggests potential diversity pitfalls in synthetic personalized dialogue generation using LLMs. One way to mitigate this issue is using a synthetic persona, as in PersonaChatGen~\citep{Lee:2022:GPD}.
However, as Table \ref{tab:eval-baselines} shows, it still falls short of human-involved LAPS, suggesting the diverse nature of human preferences.

\begin{table}[t!]
\centering
\caption{UniEval scores.
Significance against all baselines is marked by \textsuperscript{+}.
} 
\label{tab:consolidated-uni-eval-scores}
\setlength{\tabcolsep}{6pt}
\begin{tabular}{l || c c c |c}
\hline
\textbf{Dataset} & \textbf{NAT} & \textbf{UND} & \textbf{COH} & \textbf{Avg.} \\
\Xhline{2pt}
SGD & 0.794 & 0.781 & 0.758 & 0.778 \\
M2M & 0.634 & 0.616 & 0.701 & 0.650 \\
Taskmaster-1 & 0.792 & 0.779 & 0.782 & 0.784 \\
MultiWOZ & \underline{0.870} & \underline{0.860} & 0.848 & 0.859 \\
CCPE-M & 0.716 & 0.708 & 0.689 & 0.704 \\
MG-ShopDial & 0.743 & 0.730 & 0.687 & 0.720 \\
\hline
LAPS-Recipe & 0.867 & \underline{0.860} & \underline{0.891}\textsuperscript{+} & \underline{0.872}\textsuperscript{+} \\
LAPS-Movie & \textbf{0.874} & \textbf{0.868} & \textbf{0.897}\textsuperscript{+} & \textbf{0.880}\textsuperscript{+} \\
\hline
\hline
PersonaChatGen\textsuperscript{\dag} & 0.894 & 0.887 & 0.738 & 0.839 \\
\hline
\end{tabular}
\footnotesize
\begin{flushleft}
\hspace{2.0em} \textsuperscript{\dag} Scores provided as a reference, but do not represent fair comparison.
\end{flushleft}
\end{table}

\medskip \noindent \textbf{Dialogue Quality.}
Table \ref{tab:consolidated-uni-eval-scores} shows the results of the dialogue quality evaluation.
For coherence, LAPS outperforms the other datasets.
For naturalness and understandability, PersonaChatGen, which is an LLM-based fully-synthetic dataset, outperforms the other datasets.
This is consistent with the findings from~\citet{Liu:2023:GNE}, which shows that LLM-based synthetic dialogues tend to have higher scores for language-model-based automatic evaluation metrics.
Excluding fully-synthetic datasets, LAPS's performance is among the best, demonstrating that our method collects high-quality dialogues as effectively as human-human dialogue collection methods.
On average, LAPS outperforms the other datasets, highlighting the effectiveness of our method.

\begin{figure}[t]
\centering
\includegraphics[width=0.9\linewidth]{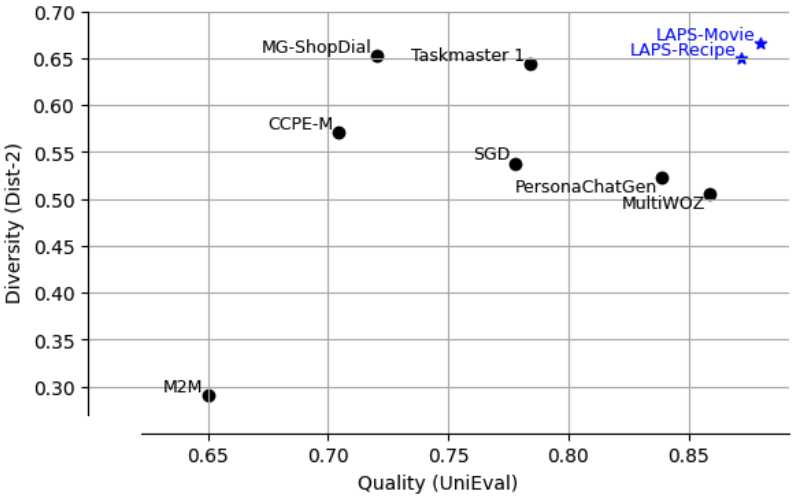}
\caption{LAPS collects diverse and high-quality dialogues compared to other dialogue collection methods.}
\label{fig:tmp-fig1-candidate}
\end{figure}

\medskip \noindent \textbf{Discussion.}
Based on these results we can answer our first and second research questions: \emph{\textbf{RQ3a:} Using LAPS, we collect large-scale multi-session human-written conversations that contain actual user preferences.} and \emph{\textbf{RQ3b:} LAPS-collected dialogues show high diversity and quality, on par with expert-involved human-human dialogues, as highlighted in Figure~\ref{fig:tmp-fig1-candidate}.}

\subsection{Personal Recommendation Evaluation}
\label{subsec:Personal Recommendation Evaluation}

\begin{table}[t!]
\centering
\caption{Preference extraction results.
Domain adapt. represents the model is first fine-tuned on the recipe domain and then further fine-tuned on the movie domain.
}
\label{tab:preference-extraction-recipe}
\setlength{\tabcolsep}{2.5pt}
\begin{tabular}{l | c | c || ccc | ccc} 
\hline
\multirow{2}{*}{\textbf{Domain}} & \textbf{Model} & \textbf{Domain} 
& \multicolumn{3}{c |}{\textbf{Exact Match}} 
& \multicolumn{3}{c}{\textbf{BERTScore}} \\
& \textbf{Size} & \textbf{Adpt.} & P & R & F & P & R & F \\
\Xhline{2pt}
\multirow{3}{*}{Recipe} & Small & \multirow{3}{*}{ -- } & 0.454 & 0.408 & 0.412 & 0.590 & 0.589 & 0.589 \\
& Base & & 0.464 & 0.476 & 0.454 & 0.622 & 0.623 & 0.622 \\
& Large & & \textbf{0.532} & \textbf{0.493} & \textbf{0.494} & \textbf{0.651} & \textbf{0.650} & \textbf{0.650} \\
\hline
\hline
\multirow{2}{*}{Movie} & \multirow{2}{*}{Large} & \texttimes & 0.453 & 0.415 & 0.425 & 0.598 & 0.592 & 0.595 \\
& & \ding{51} & \textbf{0.470} & \textbf{0.424} & \textbf{0.432} & \textbf{0.618} & \textbf{0.614} & \textbf{0.616} \\
\hline
\end{tabular}
\end{table}

\medskip \noindent \textbf{Preference Extraction.}
Table \ref{tab:preference-extraction-recipe} shows the preference extraction performance of different FlanT5 pre-trained model sizes for the recipe topic.
The results show that the performance for the Recipe domain improves as the model size increases.
Using the recipe domain as the source domain, we further fine-tune the FlanT5-Large fine-tuned model on the target movie domain.
The results show that domain adaptation consistently outperforms direct fine-tuning. This demonstrates the adaptability of our dataset to other domains through domain adaptation.

\begin{table}[t!]
\centering
\caption{Human evaluation results of recommendations.}
\label{tab:human-eval}
\setlength{\tabcolsep}{4pt}
\begin{tabular}{l | c || ccc | ccc} 
\hline
\multirow{2}{*}{\textbf{Domain}} & \textbf{Prompting} & \multicolumn{3}{c|}{\textbf{\small Rationale}} & \multicolumn{3}{c}{\textbf{\small Relevance}} \\
& \textbf{Method} & Win & Lose & Tie & Win & Lose & Tie \\
\Xhline{2pt}
\multirow{2}{*}{Recipe} 
& Standard & 17 & 29 & 4 & 18 & 22 & 10 \\
& Memory & \textbf{29} & \textbf{17} & 4 & \textbf{22} & \textbf{18} & 10 \\
\hline
\hline
\multirow{2}{*}{Movie} 
& Standard & 3 & 6 & 1 & \textbf{4} & \textbf{3} & 3 \\
& Memory & \textbf{6} & \textbf{3} & 1 & 3 & 4 & 3 \\
\hline
\end{tabular}
\end{table}

\medskip \noindent \textbf{Recommendation Human Evaluation.}
Table \ref{tab:human-eval} presents the human evaluation results for recommendation quality.
In the recipe domain, using preference memory outperforms the baseline, for both recommendation quality and rationale.
For the movie domain, the Memory method excels in rationale but not in recommendation quality. 
Given that improvements are consistently found in the rationale aspect, using preference memory is effective in improving the rationale for recommendations, a critical factor for transparent and explainable recommendations.

\begin{table}[t!]
\centering
\caption{Recommendation results.
Significance against Standard is marked by \textsuperscript{+}.
}
\label{tab:rec-autmatic}
\begin{tabular}{l | c || c | ccc} 
\hline
\multirow{2}{*}{\textbf{Domain}} & \textbf{Prompting} & \textbf{\#Prompt} & \multicolumn{3}{c}{\textbf{\small Preference Utilization}} \\
& \textbf{Method} & \textbf{Tokens} & $\text{P}_{\text{PU}}$ & $\text{R}_{\text{PU}}$ & $\text{F}_{\text{PU}}$ \\
\Xhline{2pt}
\multirow{2}{*}{Recipe} 
& Standard & 880 & \textbf{0.554} & 0.311 & 0.398 \\
& Memory & \textbf{308} & 0.470 & \textbf{0.411}\textsuperscript{+} & \textbf{0.438}\textsuperscript{+} \\
\hline
\hline
\multirow{2}{*}{Movie} 
& Standard & 957 & \textbf{0.508} & 0.364 & \textbf{0.424} \\
& Memory & \textbf{311} & 0.443 & \textbf{0.397}\textsuperscript{+} & 0.419 \\
\hline
\end{tabular}
\end{table}

\medskip \noindent \textbf{Recommendation Automatic Evaluation.}
To evaluate the effectiveness of our preference-based prompting, we compare it with the baseline standard prompting method. 
The downstream recommendation task results are depicted in Table~\ref{tab:rec-autmatic}.
The Preference Utilization results show a similar trend to the human evaluation results, demonstrating the overall win of Memory over the Standard method.
In the recipe domain, Memory outperforms Standard in $\text{F}_{\text{PU}}$ score (0.438 vs. 0.398), whereas in the movie domain, their scores show no statistical significance (0.419 vs. 0.424).
Error analysis indicates that the preference memory in the movie domain is sometimes insufficient for making recommendations, suggesting a need for improving the preference extraction method.
Nevertheless, preference-based prompting for movies reduces the prompt length threefold while maintaining a comparable $\text{F}_{\text{PU}}$ score. 
This threefold reduction has significant implications for real-world applications where inference cost is a critical factor.

\begin{figure}[t]
\centering
\includegraphics[width=0.65\linewidth]{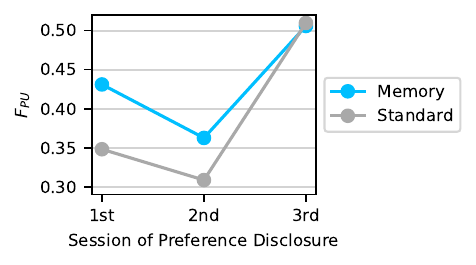}
\caption{
    Breakdown of Preference Utilization ($\text{F}_{\text{PU}}$, recipe domain).
}
\label{fig:preference-utilization-breakdown}
\end{figure}

\medskip \noindent \textbf{Preference Utilization by Session.}
Figure \ref{fig:preference-utilization-breakdown} shows the Preference Utilization scores by session in recipe recommendations. 
The x-axis represents the session where preferences are disclosed, as detailed in Section \ref{ch:laps:sec:downstream:eval-metric}.
The analysis focuses on the recommendations in the third session to examine Preference Utilization across different sessions.
The graph shows methods' struggle with utilizing preferences from earlier sessions (1st and 2nd) than those from the ongoing (3rd) session.
A similar phenomenon is reported by~\citet{Liu:2023:LIM} in a retrieval augmentation setting, referred to as \emph{LLMs' recall issue in long prompt inputs}.
The graph also demonstrates that preference-based prompting more effectively utilizes earlier session preferences than standard prompting, suggesting that preference memory can mitigate long prompt recall issues. 
We observe that this pattern (recall issues in earlier sessions and the effectiveness of preference-based prompting in addressing them) is generally consistent across the movie domain and different sessions, though not always statistically significant. Overall, our analysis shows the effectiveness of preference memory in long prompts.

\medskip \noindent \textbf{Discussion.}
Based on these results we can positively answer our last research questions: \emph{\textbf{RQ3c:}
Preference memory enhances effective utilization of user preferences in recommendations, improves the rationale of recommendations, and mitigates long prompt recall issues.}

\section{Conclusion}

In this research, we proposed a method to collect large-scale multi-session personalized conversations reflecting actual user preferences.
Our method, LAPS, employs LLMs to generate personalized guidance for human workers, reducing the cognitive load for a highly complex task.
Extensive experiments demonstrate, while being a scalable and high-quality data collection method, LAPS collects utterances as diverse as the expert-involved methods.
We further showed that utilizing extracted user preferences results in more effective personal recommendations compared to using raw user utterances of previous sessions.
In our experiment, fully-synthetic LLM-based methods do not yield diverse conversations. Using actual user preferences from LAPS as personas  is a promising avenue to explore for future.

\chapter{Conversation Evaluation}
\label{ch:face}

The previous chapters established methods for personal information extraction (Chapters~\ref{ch:conel} and~\ref{ch:crel}) and personalized response generation through large-scale personalized dataset construction (Chapter~\ref{ch:laps}), enabling personalized conversational information access (CIA) system development. However, systematic and reliable evaluation of these systems remains an open challenge. Existing automatic evaluation methods are often insufficient for assessing the dynamic nature of conversations. This necessitates an automatic, reference-free method that provides fine-grained performance insights. Therefore, this chapter addresses the research question:

\vspace{1em}
\noindent
\hspace{2.8em}%
\begin{minipage}[t]{0.83\linewidth}
\textbf{RQ4:} \emph{How to automatically evaluate personalized CIA systems in a way that accounts for natural conversation flow and provides fine-grained insights into system performance?}
\end{minipage}
\vspace{1em}

\noindent
To answer this question, this chapter introduces \textbf{FACE}: a \textbf{F}ine-grained, \textbf{A}spect-based \textbf{C}onversation \textbf{E}valuation method. FACE is reference-free and provides evaluation scores for diverse turn and dialogue-level qualities. To enable meta-evaluation, this chapter also introduces the \textbf{CRSArena-Eval}, a dataset with high-quality multi-aspect human judgments on human-system conversations. The findings show that FACE achieves a strong correlation with these human judgments, outperforming state-of-the-art methods. Unlike existing large language model (LLM)-based methods that provide single uninterpretable scores, FACE provides insights into system performance, enabling the identification and location of problems within conversations.

\section{Introduction}
\label{ch:face:sec:introduction}

Recent advances in generative LLMs have reshaped information access systems, enabling rich human-system \textit{conversational} interactions. In parallel, offline evaluation of information access systems is also evolving with the increasing adoption of LLM–based approaches~\citep{Trippas:2025:RFS}. 
In industrial settings, it is common practice to design evaluation prompts for specific aspects of a system, to reduce the high cost of human evaluation, and to help early identification of system issues during the development phase~\citep{Dey:2023:DMF,Dubois:2024:LCA}.
The research community, while maintaining skepticism, has valued the potential of LLM-based evaluation, and various LLM-based evaluation methods have been proposed for information access systems~\citep{dietz:2025:llmtropes, Balog:2025:RJA, Dietz:2024:WAR, Upadhyay:2024:LSR, Farzi:2024:ELA, Thakur:2025:AST, Aliannejadi:2025:iKAT, Upadhyay:2024:LCP}.

\begin{sidewaysfigure}
\centering
\includegraphics[width=\textwidth]{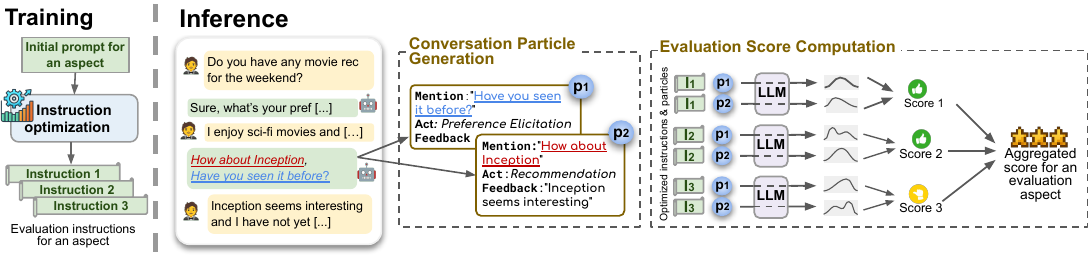}
\caption{Illustration of FACE for a turn-level aspect. Instruction optimization generates a set of diverse evaluation instructions for the given aspect (e.g., relevance) based on an initial prompt.
For evaluation, each conversation is decomposed into particles containing a dialogue act, a mention (text span from a system turn), and user feedback from the following turn. A response distribution is created for each instruction-particle pair and the weighted summation of the scores is computed. The final score is obtained by aggregating the scores across all instructions and particles.
For turn-level aspects, aggregation is performed per turn, while for dialogue-level aspects, scores of particles across the entire dialogue are aggregated.
}
\label{fig:method-illustration}
\end{sidewaysfigure}

\noindent\textbf{Research gap.}
Most research on LLM-based evaluation for information access systems revolves around assessing the relevance of retrieved passages~\citep{Upadhyay:2024:LCP, Upadhyay:2024:LSR, Dietz:2024:WAR} or comparing generated texts against gold nuggets~\citep{Pradeep:2025:GNR, Thakur:2025:AST}. Similar efforts have been made for the offline evaluation of CIA, where gold nuggets are generated for predefined trajectories of conversations~\citep{Abbasiantaeb:2025:CGE, Aliannejadi:2025:iKAT, Aliannejadi:2024:IKAT}. 
However, evaluation of conversational systems requires going beyond measuring factual accuracy over constrained conversational trajectories. 
\citet{Sakai:2023:SGF} identified twenty evaluation criteria (or aspects) beyond factual correctness for conversational systems (e.g., modesty and conciseness), along with a set of requirements that a conversational evaluation method should satisfy. 
To date, this broader perspective has not been fully realized. As highlighted in the latest IR strategic
meeting (SWIRL), developing ``reliable and informative methods for evaluating dynamic trajectories of conversations at scale'' is still an open question~\citep{Trippas:2025:RFS}.

\medskip\noindent\textbf{Research goal.} The central research question that arises here is:

\vspace{1em}
\noindent
\hspace{2.8em}%
\begin{minipage}[t]{0.83\linewidth}
\textbf{RQ4:} \emph{How to automatically evaluate personalized CIA systems in a way that accounts for natural conversation flow and provides fine-grained insights into system performance?}
\end{minipage}
\vspace{1em}

\noindent Specifically, our focus is on ``\textit{beyond factual accuracy}'' as it is an essential yet underexplored dimension of CIA evaluation. Developing such an evaluation method poses several challenges.
Reference-based evaluation methods restrict the evaluation process to predefined conversational trajectories. Even recent simulation-based approaches, such as those used in TREC iKAT 2025, remain constrained by predefined rubric questions~\citep{Aliannejadi:2025:iKAT}.
Reference-free conversational evaluation methods, on the other hand, mostly rely on LLM-based prompting and produce a single, uninterpretable score~\citep{Mehri:2020:USR,Zhong:2022:TUM,Liu:2023:GNE}, making it difficult to trace the score back to the underlying contributing factors. This violates the \textit{Specificity} requirement of conversational evaluation methods, which requires evaluation methods to ``be specific when locating problem(s) within conversations''~\citep{Sakai:2023:SGF}.

\medskip\noindent\textbf{Method.}
We propose FACE, a \underline{F}ine-grained \underline{A}spect-based \underline{C}on\-ver\-sa\-tion \underline{E}valuation method, which is a reference-free approach for evaluating conversations across multiple evaluation aspects. 
We make the following departures from existing LLM-based conversation evaluation methods. First, we mimic the human evaluation process by modeling the diverse thought processes of multiple human assessors through a pool of optimized evaluation instructions. These optimized instructions, obtained using beam search and bandit optimization, reduce the arbitrary nature of prompt engineering and make the evaluation process robust. 
Second, we introduce the concept of ``\textit{conversation particle},'' which encompasses a composite nugget design consisting of an atomic statement, a dialogue act, and user feedback. These conversation particles enable evaluating an atomic statement, while providing a broader view of the overall conversational state. By evaluating these conversation particles using a set of optimized instructions, our method achieves specificity. Additionally, by aggregating these scores at the turn or dialogue level, FACE supports both turn-level and dialogue-level evaluation aspects. An overall overview of FACE is illustrated in Figure~\ref{fig:method-illustration}.

The introduction of FACE leads us to the following sub-research questions:

\begin{itemize}
    \item \textbf{RQ4a:} \textit{How does FACE correlate with human judgments?}
    \item \textbf{RQ4b:} \textit{How generalizable are FACE-optimized instructions across different LLMs and domains?}
    \item \textbf{RQ4c:} \textit{Can fine-grained evaluation scores of FACE provide insights about system's issues?}
\end{itemize}

\medskip\noindent\textbf{Meta-Evaluation.}
To evaluate FACE, we collect 20,962 human annotations for 467 human-system conversations of the CRSArena-Dial dataset~\citep{Bernard:2025:CRS}. These conversations are obtained from interaction of real users with nine diverse conversational recommender systems. Human annotation of these conversations cover evaluation scores for seven evaluation aspects that are shown to be predictive of response quality and user satisfaction~\citep{Siro:2023:UPU, Siro:2022:UUS}. These include two turn-level aspects: \textit{relevance} and \textit{interestingness}, and five dialogue-level aspects: \textit{understanding}, \textit{task completion}, \textit{conversation efficiency}, \textit{interest arousal}, and \textit{overall impression}.
FACE is then applied to these evaluation aspects and the obtained evaluation scores are compared to human evaluation scores.

\medskip\noindent\textbf{Experiments.}
Our experiments demonstrate that FACE outperforms state-of-the-art evaluation methods by a large margin, achieving system and turn/dialogue-level Spearman of 0.9 and 0.5, respectively, without observing any system-human conversations for instruction optimization.
We demonstrate that FACE provides
a reusable instruction pool that can be robustly used across different LLMs and conversation types. Remarkably, FACE outperforms
strong baselines that utilize GPT-4 as their backbone on two chit-
chat datasets, despite using an 8B-parameter Llama 3.1 as its backbone. 
Lastly, and importantly, we present a case study of two competitive conversational systems and demonstrate how FACE scores can diagnose issues and pinpoint problems within a conversation.

\medskip

\medskip\noindent\textbf{Key contributions} of this chapter are summarized as follows:

\begin{itemize}[leftmargin=2em]
    \item We propose FACE, a conversation evaluation method that evaluates dynamic trajectories of conversations on diverse evaluation aspects and significantly outperforms state-of-the-art evaluation methods (Sec.~\ref{ch:face:sec:results:annotation_correlation}).
    \item We introduce the notion of conversation particles and demonstrate their effectiveness in reducing bias (Sec.~\ref{ch:face:sec:results:analysis}) and improving evaluation sample efficiency (Sec.~\ref{ch:face:sec:results:analysis}).
    \item We show that utilizing a pool of optimized instructions mitigates the arbitrary nature of prompt engineering for automatic evaluation and is robust across different LLMs and conversation types (Sec.~\ref{ch:face:sec:results:generalizability}).
    \item We demonstrate that FACE scores are beyond a single uninterpretable score and can be used by humans to identify and locate potential system issues (Sec.~\ref{ch:face:sec:results:interpretability}).
    \item We develop the CRSArena-Eval meta-evaluation dataset, containing 20,962 human annotations over 467 conversations, for evaluating conversational evaluators. The dataset and a meta-evaluation interface for evaluating an evaluator against CRSArena-Eval are available in our repository.
\end{itemize}

\section{Related Work}
\label{ch:face:sec:related}

\medskip \noindent \textbf{Automatic Conversation Evaluation.}
Although human annotations are the gold standard for evaluating CIA systems, they are expensive and time-consuming. Therefore, automatic evaluation methods have been proposed to scale up the evaluation process.
There are two main types of automatic evaluation methods: \textit{reference-based} and \textit{reference-free}~\citep{soudani2024survey}.

Reference-based methods use gold references to evaluate system responses, which include Recall@K, BLEU~\citep{Papineni:2002:BMA}, ROUGE~\citep{Lin:2004:RPA}, and BERTScore~\citep{Zhang:2020:BET}.
While these methods are effective for machine translation and question-answering, they have limitations in conversation evaluation, as they overlook diverse response possibilities and various evaluation aspects.
These limitations are supported by various studies showing a weak correlation between reference-based methods and human evaluations~\citep{Mehri:2020:USR,Bernard:2025:CRS,Liu:2016:HNE,Bernard:2025:LCE}.

Reference-free methods have been proposed to address these limitations~\citep{Mehri:2020:USR,Zhong:2022:TUM,Liu:2023:GNE}.
These methods evaluate system responses without relying on gold references, consider multiple aspects of system quality, and allow for the assessment of various response possibilities.
However, these methods primarily focus on turn-level evaluation, limiting their ability to assess the whole conversation, which are crucial for evaluating system performance in real-world scenarios~\citep{Siro:2022:UUS,Siro:2023:UPU}.
FACE addresses these limitations by evaluating the system based on the whole conversation and capturing multiple conversation trajectories from diverse user-system interactions.

\medskip \noindent \textbf{LLMs for Evaluation.}
Recent advancements in LLMs have led to various automatic evaluation methods, ranging from simple prompting to user simulation, that show strong performance~\citep{Upadhyay:2024:LSR,Dubois:2024:LCA,Liu:2023:GNE,Lin:2023:UMA,Zheng:2024:JLM,Dey:2023:DMF}.
For instance, G-Eval~\citep{Liu:2023:GNE} uses an LLM to generate an evaluation prompt, which is then used to assess the system's response, demonstrating a strong correlation with human evaluations in chit-chat conversations.
However, improved performance of LLM-based methods comes with challenges.
\emph{Evaluation bias} is one of them; LLMs tend to favor longer responses (length bias)~\citep{Wang:2024:HCE,Dubois:2024:LCA} and are biased toward texts from similar models (self-bias)~\citep{Xu:2024:PPL,Liu:2023:GNE,Balog:2025:RJA}.
\emph{Generalizability} is another challenge: LLMs are sensitive to handcrafted, arbitrary prompts, which are not reusable for different models~\citep{Sclar:2024:QLM,Razavi:2025:BPS}.
FACE mitigates these biases with conversation particles and instruction optimization (see Sec.~\ref{ch:face:sec:results:analysis}).

\medskip \noindent \textbf{Instruction optimization.}
To address the arbitrary nature of handcrafted prompts, various instruction optimization (or prompt optimization) methods have been proposed~\citep{Chen:2024:IEI,Kong:2024:PPR,Pryzant:2023:APO,Ye:2024:PEP,Chen:2024:RPA,Fernando:2025:PSS,Zhou:2023:LLM,Yang:2024:LLM,Yuksekgonul:2024:TAD}.
\citet{Zhou:2023:LLM} introduced APE, an automatic prompt engineering method that uses LLMs to generate prompt candidates, and perform a Monte Carlo search to find the optimal prompt.
\citet{Kong:2024:PPR} proposed PRewrite, where prompts are optimized using proximal policy optimization (PPO)~\citep{Schulman:2017:PPO}.
\citet{Pryzant:2023:APO} proposed an instruction optimization method using ``textual gradient,'' which provides natural language feedback for an LLM to optimize prompts.
These studies focus on common tasks like question answering (QA) and classification, leaving conversation evaluation unexplored.
More importantly, optimized prompts are not generalizable between LLMs~\citep{Zhou:2023:LLM}, which is crucial for evaluation tasks, where reusability is essential.
In this chapter, we present a method for applying a textual gradient approach to enhance conversation evaluation alongside effective strategies for transferring optimized prompts across different settings.

\medskip \noindent \textbf{Nugget-based Evaluation.}
To assign granular evaluation scores, multiple nugget-based evaluation approaches are proposed~\citep{Voorhees:2003:OTQ,Mayfield:2024:EMR,Pradeep:2025:GNR,Lin:2005:AEA,Ekstrand:2013:ESN,Takehi:2023:ODQ,Rajput:2011:NTC,Dietz:2024:WAR,Sakai:2023:FEC}.
Nugget-based evaluation was proposed~\citep{Voorhees:2003:OTQ} to assign granular scores to system responses for non-binary queries, wherein a nugget, which is an atomic piece of information, serves as the unit of evaluation, enabling a more traceable assessment.
Although the original nugget-based evaluation was intended as a manual method, many efforts have aimed to automate or semi-automate it~\citep{Mayfield:2024:EMR,Pradeep:2025:GNR,Lin:2005:AEA,Ekstrand:2013:ESN,Takehi:2023:ODQ,Rajput:2011:NTC,Dietz:2024:WAR,Abbasiantaeb:2025:CGE}.
One of the earlier methods is POURPRE~\citep{Lin:2005:AEA}, an automatic nugget-based evaluation method that uses n-gram co-occurrences to assess nugget presence in system responses.
More recent research employs LLMs for nugget matching; \citet{Pradeep:2025:GNR} introduced the AutoNuggetizer framework in TREC RAG 2024, where LLMs automatically create nuggets and assign them to system responses.
However, these works are reference-based and/or focus on individual responses, and more importantly, they do not target information-access conversations.

SWAN~\citep{Sakai:2023:SGF} proposes a conceptual framework for evaluating CIA systems by decomposing user-system interactions into units and evaluating them in various aspects. While SWAN is promising, its execution remains an open question, specifically how to automatically create and assess nuggets while ensuring the method's generalizability. This work addresses these challenges.

\section{Method}
\label{ch:face:sec:method}

Our Fine-Grained Aspect-based Conversation Evaluation (FACE) approach handles the one-to-many nature of conversations and provides detailed scores for each turn and dialogue-level evaluation aspect.
Figure~\ref{fig:method-illustration} illustrates the FACE method. 
Using beam search and bandit algorithms, FACE first optimizes a set of instructions for a given evaluation aspect.
During evaluation, a dialogue is decomposed into conversation particles, each containing a dialogue act (e.g., ``Recommendation''), mention (e.g., ``How about Inception''), and corresponding user feedback from the user response (e.g., ``Inception seems interesting''). Each particle is independently evaluated with optimized instructions via an LLM, generating score distributions and resulting in turn/dialogue-level scores for a given aspect.

We note, without detailed elaboration, that FACE is applicable to a broad range of evaluation aspects~\citep{Sakai:2023:SGF} and conversation types (e.g., task-oriented and chit-chat dialogues). The primary focus of this chapter, however, is on conversational recommender systems (CRSs) and seven widely recognized turn- and dialogue-level evaluation aspects, following~\citet{Siro:2022:UUS}; see Section~\ref{ch:face:sec:annotations:process} for description of these aspects.

This section describes the evaluation steps of FACE: particle generation (Sec.~\ref{ch:face:sec:method:particle_generation})  and evaluation score computation (Sec.~\ref{ch:face:sec:method:evaluator}), followed by the instruction optimization process (Sec.~\ref{ch:face:sec:method:optimizer}).

\subsection{Conversation Particle Generation}
\label{ch:face:sec:method:particle_generation}
FACE sets two goals: (1) enable reference-free evaluations to address state explosion in natural conversation evaluation, and (2) provide fine-grained scores at both turn- and dialogue-level to locate undesired system behavior within the conversation~\citep{Sakai:2023:SGF}.
To achieve these, we introduce \textit{conversation particle}, a self-contained information unit decomposed from conversations.
Each particle is composed of three parts: 
(i) Dialogue \texttt{act} is the system's action associated with the particle, such as ``recommendation'' or ``preference elicitation;'' (ii) \texttt{Mention} denotes the particle text within the system's response, like ``How about the movie A?''; and (iii) \texttt{Feedback} is the user's evaluative reply, for instance, ``The movie A seems interesting.''
Following Chapter~\ref{ch:laps}, we use 5 dialogue acts: \textit{greetings}, \textit{preference elicitation}, \textit{recommendation}, \textit{goodbye}, and \textit{others}.

We instruct an LLM to decompose system responses into particles, denoted as \textit{decomposer} in the rest of the chapter. 
Let $u_t$ be the target system response at turn $t$, $h$ the dialogue history preceding $u_t$, and $u_{t+1}$ the user's turn following $u_t$.
The decomposer $\mathcal{D}$ maps $(h, u_t, u_{t+1})$ to conversation particles $\mathbf{P}_u$:

\begin{equation*}
    \mathbf{P}_u = \mathcal{D}(h, u_t, u_{t+1}),
\end{equation*}

where each particle $p \in \mathbf{P}_u$ is a triplet (\texttt{act}, \texttt{mention}, \texttt{feedback}).
The full particle list for a dialogue, $\mathbf{P}_d$, is the union of particles from all responses with the dialogue, i.e., $\mathbf{P}_d = \bigcup_{u \in d} \mathbf{P}_u$.

Appendix~\ref{app:prompts} details prompts used for particle generation.
It is shown that LLMs are more effective than traditional methods like dependency parsing and information extraction for decomposing texts into atomic units~\citep{Pradeep:2024:INE, Alaofi:2024:GIR}.

\subsection{Evaluation Score Computation}
\label{ch:face:sec:method:evaluator}

FACE utilizes optimized instructions to generate scores for each conversation particle. 
Formally, given a particle $p$ and the evaluation instruction $I^\alpha$ for the aspect $\alpha$, an LLM generates a response $r_p^\alpha$.
To address known issues with LLM-generated scores, such as low variance and their noise~\citep{Liu:2023:GNE}, we obtain a response distribution $\{ r_{p,i}^\alpha \}_{i=1}^n$ and compute a weighted sum over the response set:
\begin{equation}
\label{eq:particle_score}
    \mathcal{S}_{particle}(I^\alpha, p) = \sum_{i=1}^n r_{p,i}^\alpha P(r_{p,i}^\alpha|I^\alpha, p; \theta),
\end{equation}
where $P(.)$ is a probability of $r_{p,i}^\alpha$ from an LLM parameterized by $\theta$, and $n$ is the number of sampled evaluation responses.

These particles scores are then aggregated per turn or conversation by taking their mean, 
\begin{equation*}
    \mathcal{S}(I^\alpha, \mathbf{P}_x) = \frac{1}{|\mathbf{P}_x|} \sum_{p \in \mathbf{P}_x} \mathcal{S}_{particle}(I^\alpha, p),
\end{equation*}
where $\mathbf{P}_x$ is the set of particles for a given turn or dialogue, depending on evaluation aspect $\alpha$; e.g., for \textit{relevance}, aggregation is performed over particles of a turn, and for \textit{task completion} aggregation is done for all particles of the dialogue.

\textbf{A unique feature of FACE} is  utilizing diverse reasoning paths for each evaluation aspect, which is obtained by selecting optimized chain-of-thought (CoT) instructions. The intuition is that evaluation requires complex reasoning, and an optimal answer can be obtained by marginalizing various thought paths~\citep{Wang:2023:SIC}.
Here, a set of top-performing optimized instructions with various CoT instructions, $\mathbf{I}^\alpha$, are applied to particles, and the resulting scores are aggregated to obtain the final score $s^\alpha$:
\begin{equation*}
    s^\alpha = FACE(\mathbf{I}^\alpha, \mathbf{P}_x) =  \frac{1}{|\mathbf{I}^\alpha|}\,\sum_{I^\alpha \in \mathbf{I}^\alpha} \mathcal{S}(I^\alpha, \mathbf{P}_x).
\end{equation*} 

\subsection{Instruction Optimization}
\label{ch:face:sec:method:optimizer}

Instruction optimizer generates diverse optimized CoT instructions for a given evaluation aspect.
The goal is to obtain a representative set of thought processes (via CoT) for an evaluation aspect, and leverage them to evaluate unseen human-system conversations. The optimization process is performed on annotated human-human conversations to capture human thinking process and reasoning when assessing dialogue quality. 
We note, at the outset, that all the optimization and selection algorithms are performed independently for each aspect.
For notational simplicity, we shall drop superscript $\alpha$ from aspect-related instruction and evaluation scores in this section.

To optimize instructions, we assume access to human evaluated dialogues, $\mathbf{H} = \{(x_i, l_i)\}_{i=1}^{m}$, where $x_i$ is either a turn or an entire dialogue depending on the aspect, and $l_i$ is its label.
Similarly, we assume access to an LLM $\mathcal{L}_1$ that generates evaluation scores, $\mathbf{S_I}=\{(x_i,\operatorname{FACE}(\mathbf{I},\mathbf{P}_{x_i}))\}_{i=1}^{m}$, where $\mathbf{P_{x_i}}$ corresponds to particles of a specific turn or conversation and $\mathbf{P} = \bigcup_{x_i} \mathbf{P_{x_i}}$ is all particles in the dialogue collection.

The optimization objective is to identify a set of optimal instructions $\mathbf{I^*}$ that maximizes the correlation between human labels and the scores generated by the automatic evaluator:
\begin{equation*}
\label{eq:objective}
  \arg\max_{\mathbf{I^*}} \mathcal{C}(\mathbf{H}, \mathbf{S}_{\mathbf{I^*}}),
\end{equation*}
where $\mathcal{C}(.)$ represents the correlation function.

\begin{algorithm}[t]
\caption{Instruction optimization of FACE}
\label{alg:optimization}
\small
\begin{algorithmic}[1]
\Require{Human evaluations $\mathbf{H}$, initial prompt $I$, iterations $K$, beam width $b$, candidate size $b'$, gradient samples $\alpha$}
\State Initialize $\mathbf{I}^{\text{pool}} \gets \emptyset$ , $\mathbf{I}_1 \gets \{I\}$
\For{$k = 1,...,K$}
    \For{each instruction $I_{k_j} \in \mathbf{I}_k$}
        \For{each particle $p \in \mathbf{P}$}
            \State // Get score
            \State $s_{p,k_j} \gets \mathcal{S}_{particle}(I_{k_j}, p)$ \Comment{Eq.~\ref{eq:particle_score}}
            \State // Textual Gradient Generation
            \State $\mathbf{G}_{p,k_j} \gets \mathcal{G}_\nabla(I_{k_j}, s_{p,k_j}, l_p; \alpha)$ \Comment{Eq.~\ref{eq:textgrad}}
            \State // Instruction Rewriting
            \State $\mathbf{I}'_{p,k_j} \gets \mathcal{R}_\delta(I_{k_j}, \mathbf{G}_{p,k_j})$ \Comment{Eq.~\ref{eq:rewrite}}
        \EndFor
    \EndFor
    \State $\mathbf{I}'_k = \bigcup_{p \in \mathbf{P}} \bigcup_j \mathbf{I}'_{p,k_j}$
    \Comment{Collect all rewrites}
    
    \State // Instruction Selection
    \State $\mathbf{I}^{\text{cand}}_k = Select^{UCB}_{b'}(\mathbf{I}'_k)$ \Comment{Eq.~\ref{eq:ucb_select} and Algorithm~\ref{alg:ucb}}
    \State $\mathbf{I}^{\text{pool}} \leftarrow \mathbf{I}^{\text{cand}}_k \cup \mathbf{I}^{\text{pool}}$ \Comment{Update pool (Eq.~\ref{eq:pool_update})}
    \State // Select top-b instructions
    \State $\mathbf{I}_{k+1} \gets \argmax_{\mathbf{I}_k \subseteq \mathbf{I}^{\text{pool}}, |\mathbf{I}_k|=b} \mathcal{C}(\mathbf{H}, \mathbf{S}_{\mathbf{I}_k})$
\EndFor
\State $\mathbf{I}^* \gets \arg\max_{\mathbf{I}^* \subseteq \mathbf{I}^{\text{pool}}} \mathcal{C}(\mathbf{H'}, \mathbf{S}_{\mathbf{I}^*})$ 
\State \Return $\mathbf{I}^*$
\end{algorithmic}
\end{algorithm}

\medskip \noindent \textbf{The optimization process} employs an LLM $\mathcal{L}_2$ to refine instructions based on the scores generated by the evaluator LLM $\mathcal{L}_1$.
FACE employs a non-parametric optimization algorithm using textual gradients~\citep{Pryzant:2023:APO}.
Here, natural language ``gradients'' (as opposed to numerical gradients) are generated to describe the shortcomings of instructions.
The gradients are used to rewrite original instructions in the opposite semantic direction.
The best instructions are iteratively selected using beam search and Upper Confidence Bound (UCB) bandits, based on correlations with human judgments. The process consists of three stages.

\medskip \noindent \textbf{(1) Textual Gradient Generation.}
This is an iterative process, where a static prompt $\nabla$ is used for generating textual gradients (lines 7-8 of Algorithm~\ref{alg:optimization}).
At each iteration $k$, the prompt $\nabla$ takes an evaluation instruction $I_{k_j}$
from the current set of instructions $\mathbf{I}_k$, its prediction score $s_{p,k_j} = \mathcal{S}_{particle}(I_{k_j}, p)$, and the corresponding human label $l_p$ for a given particle $p$.
A set of textual gradients $\mathbf{G}_{p,k_j}$ is then generated by:
\begin{equation}
    \label{eq:textgrad}
    \mathbf{G}_{p,k_j} = \mathcal{G}_\nabla(I_{k_j}, s_{p,k_j}, l_p; \alpha), 
\end{equation}
where $\mathcal{G}_\nabla(.)$ is the gradient generation function, with parameter $\alpha$ denoting the number of gradients generated per instruction-score pair.
Since human annotations are provided at the turn- or dialogue-level, $l_p$ represents human annotation for the turn or dialogue that contains the particle $p$. 

For the prompt $\nabla$, we employ reasoning templates \citep{Ye:2024:PEP}, which provide a set of items to be considered by  $\mathcal{G}_\nabla$. 
Our items include identifying inconsistencies between the predicted and human annotations, evaluating the correctness of the current task and CoT instructions, and suggesting edits to these instructions, if necessary.

\medskip \noindent \textbf{(2) Instruction Rewriting.}
This step updates each instruction using textual gradients (lines 9-10 of Algorithm~\ref{alg:optimization}).
For each particle $p$ at iteration $k$, we use the rewriting function $\mathcal{R}_\delta$ with the prompt $\delta$, which takes the current instruction $I_{k_j}$ and gradients $\mathbf{G}_{p,k_j}$ to obtain updated instructions $\mathbf{I}'_{p,k_j}$:
\begin{equation}
    \label{eq:rewrite}
    \mathbf{I}'_{p,k_j} = \mathcal{R}_\delta(I_{k_j}, \mathbf{G}_{p,k_j}).
\end{equation}
An LLM $\mathcal{L}_3$ is used for the rewriting function $\mathcal{R}_\delta$ with the prompt $\delta$ guiding it to revise the current instruction, considering the provided feedback.

We note that, while theoretically two distinct LLMs $\mathcal{L}_{2}$ and $\mathcal{L}_{3}$ are used for gradient generation and instruction rewriting, the two steps can be merged into a single LLM call by concatenating $\nabla$ and $\delta$.
This halves the number of LLM calls, resulting in a significant speedup of the optimization process.

\medskip \noindent \textbf{(3) Instruction Selection.}
The step identifies the most promising instructions for the next iteration in two stages of selecting candidate instructions using UCB bandit, and identifying the top beam based on evaluation scores on the training data. 
This step corresponds to lines 14-18 of Algorithm~\ref{alg:optimization}.

Let 
$\mathbf{I}'_k = \bigcup_{p \in \mathbf{P}} \bigcup_j \mathbf{I}'_{p,k_j}$
be the set of all rewritten instructions at the iteration $k$. The first stage identifies $b' \geq b$ promising candidate instructions $\mathbf{I}^{\text{cand}}_k$ from the rewritten instructions $\mathbf{I}'_k$ and stores them in an \textit{instruction pool}:
\begin{align}
    \mathbf{I}^{\text{cand}}_k &= Select^{UCB}_{b'}(\mathbf{I}'_k), \label{eq:ucb_select} \\
    \mathbf{I}^{\text{pool}} &\leftarrow \mathbf{I}^{\text{cand}}_k \cup \mathbf{I}^{\text{pool}}. \label{eq:pool_update}
\end{align}

The second stage then evaluates these candidate instructions by computing the correlation of the generated scores with human labels using the training set. Therefore, the instructions for the next iteration $k+1$ are generated by:
\begin{align}
    \mathbf{I}_{k+1} = \argmax_{\mathbf{I}_k \subseteq \mathbf{I}^{\text{pool}}, |\mathbf{I}_k|=b} \mathcal{C}(\mathbf{H}, \mathbf{S}_{\mathbf{I}_k}), \label{eq:instruction_selection}
\end{align}
where $\mathbf{S}_{\mathbf{I}_k}$ denotes  all predicted scores using instructions $\mathbf{I}_k$.
This process follows a beam search approach, where at each iteration we create a pool of candidates, assess their performance, and select the top ones to form the new beam for continued exploration.
Once all iterations are complete, the final optimal instructions $\mathbf{I}^*$ are selected by their correlation scores on a validation set $\mathbf{H'}$ (line 19 of Algorithm~\ref{alg:optimization}).

\medskip \noindent \textbf{The UCB selection algorithm}, denoted as $Select^{UCB}_{b'}(\cdot)$, is presented in Algorithm~\ref{alg:ucb}.
For each iteration $t$, $N_t(I)$ denotes the number of evaluations of instruction $I$ on sampled particles and $Q_t(I)$ denotes its estimated correlation.
Following~\citet{Pryzant:2023:APO}, it samples a subset of particles and their corresponding human annotations, then selects the instruction that maximizes the UCB criterion $Q_t(I) + \beta \sqrt{\log t / N_t(I)}$, where $\beta$ is an exploration constant.
The selected instruction is evaluated on the sampled particles, and its estimated effectiveness is updated based on the correlation with human annotations.
After $T$ iterations, the algorithm returns the candidate instructions $\mathbf{I}^{\text{cand}}_{k}$ containing the top $b'$ instructions according to their final estimated effectiveness $Q_T$ with the $SelectTopInstructions_{b'}(Q_T)$ function.
This forms a set of promising candidates for the next stage of the selection process.
We note that, for efficient execution of the UCB process, we approximate it by dividing $T$ into multiple small batches and processing each batch in parallel.

\begin{algorithm}[t]
\caption{$Select^{UCB}_{b'}(\cdot)$ - Candidate Selection with UCB Bandits}
\label{alg:ucb}
\small
\begin{algorithmic}[1]
\Require{All rewritten instructions $\mathbf{I}'_k$, particles $\mathbf{P}$, human annotations $\mathbf{H}$, number of UCB iterations $T$, and the number of selected instructions $b'$.}
\State Initialize $N_t(I) \gets 0, Q_t(I) \gets 0, \forall I \in \mathbf{I}'_k$
\For{$t = 1,...,T$}
    \State // Sample particles and corresponding labels uniformly
    \State $\mathbf{P}_{smp} \subset \mathbf{P}$, $\mathbf{H}_{smp} \subset \mathbf{H}$
    \State // Select the instruction with the highest UCB criterion
    \State $I \gets \arg\max_{I \in \mathbf{I}'_k} \{Q_t(I) + \beta\sqrt{\frac{\log t}{N_t(I)}}\}$
    \State // Evaluate the instruction $I$ on the sampled particles
    \State $\mathbf{S}_{smp} \gets \{\mathcal{S}_{particle}(I, p)\}_{p \in \mathbf{P}_{smp}}$
    \State // Compute correlation and update UCB criterion
   \State Observe reward $r \gets \mathcal{C}(\mathbf{S}_{smp}, \mathbf{H}_{smp})$
    \State $N_t(I) \gets N_t(I) + |\mathbf{P}_{smp}|$
    \State $Q_t(I) \gets Q_t(I) + \frac{r - Q_t(I)}{N_t(I)}$
\EndFor
\State \Return $\mathbf{I}^{\text{cand}}_k \gets SelectTopInstructions_{b'}(Q_T)$
\end{algorithmic}
\end{algorithm}

\section{Human Annotation Collection}
\label{ch:face:sec:annotations}

To assess the correlation of automatic conversation evaluation methods with human judgments, we need a dataset and crowdsource human annotations on a set of human-system conversations; we thus create the CRSArena-Eval dataset to enable this meta-evaluation, focusing on recommendation conversations as a case study, which we describe in this section.
To achieve this, we first collect human-system conversations (Section~\ref{ch:face:sec:annotations:dialogue}), then crowdsource human annotations on these conversations (Section~\ref{ch:face:sec:annotations:process}).

\subsection{Dialogue Collection}
\label{ch:face:sec:annotations:dialogue}

To create our meta-evaluation dataset, we first collect multi-turn conversations between users and various systems, on which we later collect human annotations as described in Section~\ref{ch:face:sec:annotations:process}.
Similarly to~\citet{Mehri:2020:USR}, our goal in creating such a dataset is not to train and benchmark the best-performing systems, but rather to collect responses with varying quality of systems and obtain reliable human judgments of these responses.
This section describes the first human-system dialogue collection process.

\medskip \noindent \textbf{Dialogue Collection Platform.}
To collect human-system conversations, inspired by Chatbot Arena~\citep{Chiang:2024:CAO}, we develop a platform named \textit{CRS Arena}.
Our CRS Arena displays two CRSs side-by-side in a battle-style interface where users interact with them one after the other, resulting in a collection of conversations and user feedback.
While users can declare a winner, these declarations cannot directly serve as evaluation data due to the lack of evaluation aspects and granularity, which we will further collect annotations on in Section~\ref{ch:face:sec:annotations:process}.

\medskip \noindent \textbf{CRSs.}
The resulting dataset, called \textit{CRSArena-Dial}, consists of dialogues with multi-turn interactions between users and nine state-of-the-art CRSs trained on OpenDialKG~\citep{Moon:2019:OEC} and ReDial~\citep{Li:2018:TDC} datasets.
Specifically, the CRSArena-Dial dataset consists of human conversation with nine state-of-the-art CRSs, including KBRD~\citep{Chen:2019:KBR}, BARCOR~\citep{Wang:2022:BTU}, UniCRS~\citep{Wang:2022:UCR}, ChatGPT~\citep{Wang:2023:REC}, and CRB-CRS~\citep{Manzoor:2022:RBC}, each developed based on OpenDialKG~\citep{Moon:2019:OEC} and ReDial~\citep{Li:2018:TDC} datasets, except CRB-CRS, which is solely on the ReDial dataset.

\medskip \noindent \textbf{Collected Dialogues.}
The CRSArena-Dial dataset is collected through two distinct crowdsourcing environments: an open setup with public access and a closed setup with access restricted to a selected group of crowd workers.
A key feature of this dataset is that it contains authentic conversations between real users and fully automated CRSs, as opposed to the Wizard-of-Oz setting commonly used in dialogue corpora creation.
While the minimally regulated environment may introduce some noise, the dataset is highly representative of how regular users naturally express preferences and seek recommendations.

\subsection{Dialogue Annotation}
\label{ch:face:sec:annotations:process}

In this section, we describe the process of collecting human annotations on the CRSArena-Dial dataset described in Section~\ref{ch:face:sec:annotations:dialogue}.

\medskip \noindent \textbf{Dataset.}
We collect human annotations on the CRSArena-Dial dataset as described in Section~\ref{ch:face:sec:annotations:dialogue}.
To ensure the quality of dialogues, we excluded seven dialogues that were unsuitable for our annotation, such as those with only a single user utterance, resulting in 467 dialogues with a total of 2,235 system responses for our annotations.

\medskip \noindent \textbf{Evaluation Aspects.}
Over 50 aspects have been proposed for evaluating conversational systems in the literature.
We follow~\citet{Siro:2022:UUS} and collect annotations for seven evaluation aspects, whose effectiveness for CRS evaluation has been rigorously validated~\citep{Siro:2022:UUS,Siro:2023:UPU}.
These aspects cover both system and user-centric features of a conversation.
The aspects and their descriptions (used as instructions to annotators) are as follows:

\medskip

{
\textbf{Turn-level Aspects:}
\begin{itemize}
    \item \textit{Relevance (0--3):} Does the assistant's response make sense and meet the user's interests?
    \item \textit{Interestingness (0--2):} Does the response make the user want to continue the conversation?
\end{itemize}

\textbf{Dialogue-level Aspects:}
\begin{itemize}
    \item \textit{Understanding (0--2):} Does the assistant understand the user's request and try to fulfill it?
    \item \textit{Task Completion (0--2):} Does the assistant make suggestions that the user finally accepts?
    \item \textit{Efficiency (0--1):} Does the assistant suggest items matching the user's interests within the first three interactions?
    \item \textit{Interest Arousal (0--2):} Does the assistant try to spark the user's interest in something new?
    \item \textit{Overall Impression (0--4):} What is the overall impression of the assistant's performance?
\end{itemize}

\medskip
}
\noindent
These instructions and scales are based on~\citet{Siro:2022:UUS}, with minor adjustments from~\citet{Sakai:2023:SGF} for clarity. Turn-level aspects are evaluated for each system turn, while dialogue-level aspects are evaluated for the entire dialogue.

Although we select seven thoroughly assessed aspects supported by the existing literature~\citep{Siro:2023:UPU}, 
we note that FACE can be applied to a broad range of aspects and the choice of aspects is often context-dependent.
Therefore, FACE algorithm allows researchers and practitioners to explore diverse aspects relevant to their own systems.

\subsection{Interface}
\label{ch:face:sec:interface}

To crowdsource high-quality annotations, we built an interface, through multiple pilot experiments, to overcome the widely reported challenges in the literature~\citep{Bernard:2023:MGS,Eickhoff:2011:HCY}. Those challenges include workers
(1) skipping reading context, (2) misunderstanding aspect definitions, (3) getting distracted by overwhelming information on the annotation page, and (4) annotating randomly without focus.
Our interface requires users to pass a quiz on aspect definitions and enforces the annotation of each turn before evaluating the entire dialogue. It further includes hidden tests with expert-verified answers, and workers who fail to meet the required agreement are excluded.

\subsection{Participants and Quality Control}
We recruited Prolific\footnote{\url{https://www.prolific.co/}} workers from English-speaking countries with a $ 100\% $ approval rate and $ \geq 1000 $ previous submissions.
Considering some workers exhibit behavior aimed at just maximizing financial gain~\citep{Eickhoff:2011:HCY}, we filtered out those with a history of subpar submissions to ensure quality.
Furthermore, workers with <30\% agreement with experts on hidden tests were excluded.
Each batch of work contained annotations for 20 dialogues, taking around 40 minutes to complete, at the cost of £6. Three annotations per annotation task were collected. In case of disagreement, additional annotations were collected until ties were resolved.

\subsection{Analysis}
\label{ch:face:sec:annotations:analysis}
We now analyze our collected annotations, referred to as \emph{CRSArena-Eval}.

\medskip \noindent \textbf{Statistics.}
A total of 20,962 annotations were collected from 109 workers, spanning 467 dialogues and 2,235 system turns.
On average, each dialogue has 14.6 annotation tasks, each annotated by three workers.
Noteworthy, our annotation interface made the process highly efficient, requiring only 8 seconds per annotation.
For 92\% of the tasks, an agreement was achieved by the first three annotators and for the remaining 8\% of tasks additional annotations were collected to resolve ties.
In total, these processes produced 6,805 final labels.

\begin{table}[t!]
\caption{Inter-annotator agreement of CRSArena-Eval and AB-ReDial~\citep{Siro:2023:UPU} based on Pearson's $r$, Spearman's $\rho$, and Krippendorff's $\alpha$ correlations.}
\label{tab:agreement}
\setlength{\tabcolsep}{3.5pt}
\begin{tabular}{l || c c c | c c c}
\hline
\multirow{2}{*}{\textbf{Aspect}} & \multicolumn{3}{c|}{\textbf{CRSArena-Eval}} & \multicolumn{3}{c}{\textbf{AB-ReDial}} \\
 & $r$ & $\rho$ & $\alpha$ & $r$ & $\rho$ & $\alpha$ \\
\Xhline{2pt}
\multicolumn{7}{l}{\textbf{Turn-level}} \\
\hline
Relevance         & \textbf{0.613} & \textbf{0.611} & \textbf{0.612} & 0.527 & 0.502 & 0.526 \\
Interestingness   & \textbf{0.386} & \textbf{0.386} & \textbf{0.375} & 0.209 & 0.217 & 0.209 \\
\hline
\multicolumn{7}{l}{\textbf{Dialogue-level}} \\
\hline
Understanding     & \textbf{0.505} & \textbf{0.477} & \textbf{0.496} & 0.321 & 0.313 & 0.318 \\
Task Completion   & \textbf{0.481} & \textbf{0.440} & \textbf{0.482} & 0.345 & 0.321 & 0.346 \\
Efficiency        & \textbf{0.297} & \textbf{0.297} & \textbf{0.289} & 0.225 & 0.225 & 0.226 \\
Interest Arousal  & 0.242 & 0.241 & 0.226 & \textbf{0.254} & \textbf{0.291} & \textbf{0.247} \\
Overall Impression & \textbf{0.573} & \textbf{0.526} & \textbf{0.572} & 0.321 & 0.300 & 0.321 \\
\hline
\end{tabular}
\end{table}

\begin{figure}[t]
\centering
\includegraphics[width=1.0\linewidth]{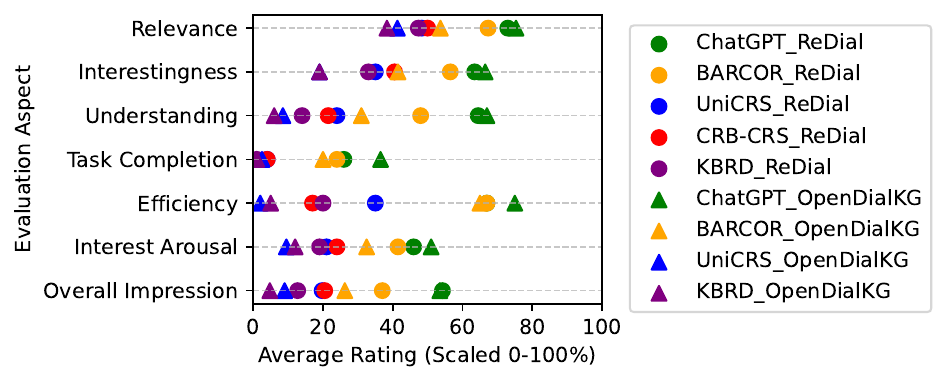}
\caption{Distribution of human annotation scores for seven aspects across nine systems in CRSArena-Eval.}
\label{fig:crs_distribution}
\end{figure}

\medskip \noindent \textbf{Inter-annotator Agreement.}
Given the ordinal nature of judgments, we report inter-annotator agreement using Pearson's $r$ and Spearman's $\rho$, following~\citet{Mehri:2020:USR}, along with Krippendorff's $\alpha$. 
Average scores across all aspects are $r = 0.443$, $\rho = 0.425$, and $\alpha = 0.436$, indicating moderate agreement.
For comparison, the same aspects on the AB-ReDial dataset~\citep{Siro:2022:UUS,Siro:2023:UPU} yield average agreements of $r = 0.328$, $\rho = 0.325$, and $\alpha = 0.327$, showing higher agreement of our CRSArena-Eval annotations.

Table~\ref{tab:agreement} presents the breakdown of inter-annotator agreement for each aspect, showing that CRSArena-Eval consistently achieves higher agreement than AB-ReDial in all aspects except interest arousal.
These results demonstrate that CRSArena-Eval provides higher quality annotations even compared to existing high-quality annotations of the AB-ReDial dataset, showing both the difficulty of the task and high quality annotations of our dataset.

\medskip \noindent \textbf{System Score Distribution.}
Figure~\ref{fig:crs_distribution} shows the distribution of collected scores for the nine CRSs in CRSArena-Eval.
It shows no system reaches the high end of the scale, indicating that existing CRSs do not fully satisfy users.
This aside, the scores cover a broad range, reflecting differing system quality, which is crucial to assess the ability of automatic evaluators to distinguish system performance~\citep{Mehri:2020:USR}.

\section{Experimental Setup}
\label{ch:face:sec:setup}

\medskip \noindent \textbf{Datasets.}
For instruction optimization, we use the  \textit{AB-ReDial}~\citep{Siro:2022:UUS,Siro:2023:UPU} dataset, which contains annotations of human-human conversations from the  ReDial dataset (cf. Sec.~\ref{ch:face:sec:annotations:process}). 
The annotations are obtained for the seven evaluation aspects and are per turn/dialogue.
This ensures no human-system conversations are involved in the optimization process.
We use 60\% of AB-ReDial for training and the rest for validation.
For evaluation, we use the \textit{CRSArena-Eval} dataset (cf. Sec.~\ref{ch:face:sec:annotations}).
CRSArena-Eval (RD)/(KG) denote the subset of the dataset for systems developed using ReDial~\citep{Li:2018:TDC} and OpenDialKG~\citep{Moon:2019:OEC}, respectively.

\medskip \noindent \textbf{Settings.}
Unless indicated otherwise Llama-3.1-8B-Instruct~\citep{Dubey:2024:LHM} is used for FACE.
All experiments were performed using the SGLang~\citep{Zheng:2024:SGE} library for its prediction efficiency. The temperature of 0.6 is set across all experiments unless otherwise stated.
Following~\citet{Pryzant:2023:APO}, we run the instruction optimization for $K = 6$ iterations and stored $b' = 16$ instructions in the instruction pool, resulting in $96$ instructions.
We set parameters $\alpha=2$, $c=1$, $b = 4$,  and use the batch size of $B = |\mathbf{I}'_k|/2$ and $T = 5B$ iterations.
A sampling size of $n = 5$ is used to create a score distribution (Sec.~\ref{ch:face:sec:method:evaluator}).
All hyperparameters are obtained using the validation set or following~\citet{Pryzant:2023:APO}.
The final selection of the optimal instruction set $\mathbf{I^*}$ (line 18 of Algorithm~\ref{alg:optimization}) is done using the validation set and results in a set of 16 instructions.

\medskip \noindent \textbf{Prompts.}
Inspired by TREC RAG 2024~\citep{Pradeep:2025:GNR}, the decomposer prompt starts with: \textit{``Your task is to extract conversation nuggets [...]''} followed by CoT prompts and their format.
The textual gradient prompt $\nabla$ begins with \textit{``Examine the original instructions, predicted nugget score, and gold score.''} and then identifies inconsistency between predicted and gold scores, followed by suggestions to the instruction if necessary.
The instruction rewriting prompt $\delta$ starts with \textit{``Propose new instructions of \~50 words based on [...]''} followed by the guidance on how to rewrite based on the textual gradient, inspired by~\citet{Ye:2024:PEP}. 
The seed evaluation instruction $I$ is as follows \textit{``Given the dialogue, evaluate the quality of the target nugget based on the given aspect. Step 1: [...]''}.
Examples of prompts are provided in Appendix~\ref{app:prompts}.

\medskip \noindent \textbf{Baselines.}
Multiple automatic evaluators are used as our baselines:
\textit{LLM\textsuperscript{Direct}} directly prompts the LLM to annotate the given turn/dialogue using the same instructions as human annotations for CRSArena-Eval;
\textit{LLM\textsuperscript{CoT+ICL}} adds CoT~\citep{Kojima:2024:LLM,Wei:2022:CTP} and in-context learning (ICL)~\citep{Brown:2020:LMF} with two examples to the prompt of LLM\textsuperscript{Direct}; 
\textit{UniEval}~\citep{Zhong:2022:TUM} and \textit{G-Eval}~\citep{Liu:2023:GNE} are state-of-the-art reference-free conversation evaluation methods.
Since UniEval covers limited aspects, we only report those overlapping with ours.
For fair comparison, we use Llama-3.1-8B-Instruct as the backbone for LLM-based methods LLM\textsuperscript{Direct}, LLM\textsuperscript{CoT+ICL}, and G-Eval in Section~\ref{ch:face:sec:results:annotation_correlation}.
To show generalizability to different LLMs in Section~\ref{ch:face:sec:results:generalizability}, other LLMs are used for our experiments.

\needspace{4\baselineskip}
\medskip \noindent \textbf{Metrics.}
For correlation metrics, we use Pearson's and Spearman's to appropriately handle the annotation scales. Following~\citet{Mehri:2020:USR}, correlation significance is computed by p-value derived from t-distribution using Python's SciPy library~\citep{Virtanen:2020:FAS}.

\section{Results}
\label{ch:face:sec:results}
In this section, we address the research questions outlined in Section~\ref{ch:face:sec:introduction} through a series of experiments:
\textbf{RQ4a:} \textit{How does FACE correlate with human judgments?}
\textbf{RQ4b:} \textit{How generalizable are FACE-optimized instructions across different LLMs and domains?}
\textbf{RQ4c:} \textit{Can fine-grained evaluation scores of FACE provide insights about system's issues?}

\subsection{FACE Annotation Correlation}
\label{ch:face:sec:results:annotation_correlation}

\begin{sidewaystable}
\centering
\caption{Annotation correlations based on Pearson's $r$ and Spearman's $\rho$. 
All columns show the correlations averaged over all aspects.
All FACE correlations are statistically significant with $p < 0.01$.}
\label{tab:correlation_main}
\setlength{\tabcolsep}{3.5pt}
\resizebox{\textwidth}{!}{
\begin{tabular}{l||cccc|cccccccccc|cc}
\hline
\multirow{3}{*}{\textbf{Methods}} & 
\multicolumn{4}{c|}{\textbf{Turn-level}} & 
\multicolumn{10}{c|}{\textbf{Dialogue-level}} & \multicolumn{2}{c}{\multirow{2}{*}{\textbf{All}}}\\
& \multicolumn{2}{c}{\textbf{Rel.}} & \multicolumn{2}{c|}{\textbf{Int.}} & 
\multicolumn{2}{c}{\textbf{Und.}} & \multicolumn{2}{c}{\textbf{Task}} & 
\multicolumn{2}{c}{\textbf{Eff.}} & \multicolumn{2}{c}{\textbf{Int.}} & 
\multicolumn{2}{c|}{\textbf{Overall}} & & \\
 & $r$ & $\rho$ & $r$ & $\rho$ & $r$ & $\rho$ & $r$ & $\rho$ & $r$ & $\rho$ & $r$ & $\rho$ & $r$ & $\rho$ & $r$ & $\rho$\\
\Xhline{2pt}
\multicolumn{17}{c}{\textbf{CRSArena-Eval (RD)}} \\
\hline
LLM\textsuperscript{Direct} & 0.464 & 0.455 & 0.248 & 0.260 & 0.522 & 0.482 & 0.405 & 0.363 & 0.101 & 0.101 & 0.217 & 0.203 & 0.564 & 0.522 & 0.360 & 0.341\\
LLM\textsuperscript{CoT+ICL} & 0.453 & 0.446 & 0.175 & 0.177 & 0.481 & 0.457 & 0.425 & 0.400 & 0.174 & 0.174 & 0.188 & 0.174 & 0.498 & 0.472 & 0.342 & 0.329\\
UniEval & 0.311 & 0.288 & 0.182 & 0.242 & 0.246 & 0.225 & -- & -- & -- & -- & -- & -- & 0.395 & 0.387 & -- & --\\
G-Eval & 0.490 & 0.471 & 0.302 & 0.289 & 0.490 & 0.444 & 0.351 & 0.364 & \textbf{0.488} & 0.482 & 0.332 & 0.325 & 0.577 & 0.577 & 0.433 & 0.422 \\
\hline
FACE w/o train & 0.468 & 0.462 & 0.279 & 0.290 & 0.605 & 0.574 & 0.482 & 0.392 & 0.339 & 0.423 & 0.235 & 0.255 & 0.617 & 0.555 & 0.432 & 0.422 \\
FACE & \textbf{0.549} & \textbf{0.550} & \textbf{0.443} & \textbf{0.437} & \textbf{0.650} & \textbf{0.635} & \textbf{0.570} & \textbf{0.453} & 0.484 & \textbf{0.534} & \textbf{0.447} & \textbf{0.430} & \textbf{0.712} & \textbf{0.668} & \textbf{0.551} & \textbf{0.530} \\
\hline
\multicolumn{17}{c}{\textbf{CRSArena-Eval (KG)}} \\
\hline
LLM\textsuperscript{Direct} & 0.452 & 0.452 & 0.238 & 0.231 & 0.599 & 0.546 & 0.538 & 0.481 & 0.137 & 0.137 & 0.425 & 0.378 & 0.655 & 0.557 & 0.435 & 0.397 \\
LLM\textsuperscript{CoT+ICL} & 0.419 & 0.408 & 0.203 & 0.190 & 0.562 & 0.520 & 0.475 & 0.434 & 0.114 & 0.114 & 0.309 & 0.279 & 0.599 & 0.521 & 0.383 & 0.352 \\
UniEval & 0.416 & 0.428 & 0.262 & 0.401 & 0.563 & 0.541 & -- & -- & -- & -- & -- & -- & 0.618 & 0.659 & -- & -- \\
G-Eval & 0.533 & 0.505 & 0.334 & 0.316 & 0.535 & 0.475 & 0.430 & 0.422 & 0.485 & 0.463 & 0.424 & 0.403 & 0.656 & 0.642 & 0.485 & 0.461 \\
\hline
FACE w/o train & 0.492 & 0.486 & 0.308 & 0.324 & 0.664 & 0.611 & 0.426 & 0.411 & 0.240 & 0.419 & 0.297 & 0.322 & 0.672 & 0.557 & 0.443 & 0.447 \\
FACE & \textbf{0.543} & \textbf{0.527} & \textbf{0.471} & \textbf{0.453} & \textbf{0.719} & \textbf{0.677} & \textbf{0.593} & \textbf{0.484} & \textbf{0.518} & \textbf{0.543} & \textbf{0.449} & \textbf{0.404} & \textbf{0.766} & \textbf{0.679} & \textbf{0.580} & \textbf{0.538} \\
\hline
\end{tabular}
}
\end{sidewaystable}

\medskip \noindent \textbf{Annotation Correlation.}
Table~\ref{tab:correlation_main} shows the annotation correlation results, demonstrating that FACE, on average,  outperforms all baselines by a large margin. 
We note that FACE is not optimized on any human-system conversations (cf. Sec.~\ref{ch:face:sec:setup}), highlighting its strong generalization to unseen systems.
Although one can argue that FACE can capture some information from the human-human ReDial dataset during the instruction optimization process, results on  CRSArena-Eval (KG) shows generalization and robustness of FACE to unseen recommendation datasets. 

The results also demonstrate that even without instruction optimization (FACE w/o train), FACE remains competitive with the state-of-the-art method, G-Eval, suggesting the effectiveness of our particle-based approach. 
The benefit of training is more pronounced for challenging aspects such as \textit{Interestingness}, \textit{Efficiency}, and \textit{Interest Arousal}, which indicates that FACE effectively optimizes instructions for aspects that LLMs struggle to capture using a single thought process.

\begin{table}[t!]
\centering
\setlength{\tabcolsep}{3.7pt}
\begin{tabular}{l|cc|cc|cc}
\hline
\textbf{Methods} & \multicolumn{2}{c|}{\textbf{Turn-level}} & \multicolumn{2}{c|}{\textbf{Dial-level}} & \multicolumn{2}{c}{\textbf{All}} \\
 & $r$ & $\rho$ & $r$ & $\rho$ & $r$ & $\rho$ \\
\Xhline{2pt}
R@1 & -0.197 & 0.060 & -0.120 & 0.081 & -0.142 & 0.075 \\
R@10 & -0.192 & 0.048 & -0.111 & 0.071 & -0.134 & 0.064 \\
iEvaLM\textsuperscript{free} & 0.191 & 0.191 & 0.232 & 0.232 & 0.257 & 0.184 \\
iEvaLM\textsuperscript{attr} & 0.527 & 0.527 & 0.553 & 0.553 & 0.575 & 0.517 \\
Distinct-3 & 0.716 & 0.841 & 0.665 & 0.780 & 0.680 & 0.798 \\
Distinct-4 & 0.654 & 0.800 & 0.609 & 0.760 & 0.622 & 0.771 \\
LLM\textsuperscript{Direct} & 0.860 & 0.822 & 0.872 & 0.799 & 0.868 & 0.806 \\
G-Eval & 0.740 & 0.840 & 0.893 & 0.830 & 0.850 & 0.833 \\
\hline
FACE & \textbf{0.930} & \textbf{0.842} & \textbf{0.913} & \textbf{0.837} & \textbf{0.918} & \textbf{0.838} \\
\hline
\end{tabular}
\caption{System ranking correlations on CRSArena-Eval, averaged over corresponding aspects.  All FACE correlations are statistically significant with $p < 0.05$.} 
\label{tab:ranking_correlation}
\end{table}

\medskip \noindent \textbf{Ranking Correlation.}
Table~\ref{tab:ranking_correlation} shows the correlation of system rankings created by different automatic evaluation methods.
System ranking correlations are calculated by averaging each system's score, ranking systems, and measuring correlation with system rankings based on human judgments.
We obtain correlation for reference-based metrics  by computing system rankings from reported scores~\citep{Wang:2023:REC}.
As baselines, we report Recall,  and Distinct-n~\citep{Li:2016:DPO} as reference-based metrics, and LLM\textsuperscript{Direct} and G-Eval as top-performing reference-free metrics from Table~\ref{tab:correlation_main}.

From the results, we notice that recall metrics are insufficient, which is in line with the literature~\citep{Bernard:2025:CRS}.
Surprisingly, iEvaLM baselines struggle to achieve high ranking correlations; this could be due to their reliance on a recall-based approach to evaluate the simulation results.
While the results show that FACE outperforms all baselines, we note that system ranking correlations should be used with caution, as they are less informative than annotation correlation, especially for competitive systems~\citep{Faggioli:2023:PLL, dietz:2025:llmtropes}.

\medskip \noindent
Overall, we can answer \textbf{(RQ4a)}: FACE achieves high annotation and system ranking correlations with human judgments, outperforming state-of-the-art methods by a large margin.

\begin{table}[t!]
\centering
\setlength{\tabcolsep}{2.0pt}
\begin{tabular}{l|l|c||cc|cc|cc}
\hline
\multirow{2}{*}{\textbf{Methods}} & \multirow{2}{*}{\textbf{LLM}} & \multirow{2}{*}{\textbf{Size}} &
\multicolumn{2}{c|}{\scriptsize\textbf{CRS-RD}} &
\multicolumn{2}{c|}{\scriptsize\textbf{CRS-KG}} &
\multicolumn{2}{c}{\textbf{Avg.}} \\
 & & &  $r$ & $\rho$ &  $r$ & $\rho$ & $r$ & $\rho$  \\
\Xhline{2pt}
LLM\textsuperscript{Direct} & Llama & 8B & 0.564 & 0.522 & 0.655 & 0.557 & 0.610 & 0.540 \\
G-Eval & Llama & 8B & 0.577 & 0.577 & 0.656 & 0.642 & 0.617 & 0.610 \\
FACE & Llama & 8B & \textbf{0.712} & 0.668 & \textbf{0.766} & 0.679 & \textbf{0.739} & 0.674 \\
\hline
FACE* & Gemma & 9B & 0.689 & \textbf{0.687} & 0.718 & \textbf{0.703} & 0.704 & \textbf{0.695} \\
FACE* & Gemma & 2B & 0.647 & 0.603 & 0.728 & 0.646 & 0.688 & 0.625 \\
FACE* & Qwen & 7B & 0.698 & 0.664 & 0.764 & 0.693 & 0.731 & 0.679 \\
FACE* & Qwen & 3B & 0.643 & 0.632 & 0.725 & 0.674 & 0.684 & 0.653 \\
FACE* & Qwen & 1.5B & 0.557 & 0.606 & 0.605 & 0.635 & 0.581 & 0.621 \\
\hline
\end{tabular}
\caption{Results on generalizability of FACE to other LLMs. 
CRS-RD and -KG represent CRSArena-Eval (RD) and (KG), respectively.
All FACE annotation correlations are statistically significant with $p < 0.01$.}
\label{tab:llm_generalizability}
\end{table}

\subsection{Generalizability of FACE}
\label{ch:face:sec:results:generalizability}
We hypothesize that the pool of optimized instructions by FACE can be reused for different LLMs and domains. To assess this hypothesis, we take the instruction pool and re-select the top instructions (line 18 of Algorithm~\ref{alg:optimization}) for different LLMs and datasets. We denote this adapted FACE as \emph{FACE*}.

\medskip \noindent \textbf{Generalization to other LLMs.}
To assess generalizability of FACE to other LLMs, we use the AB-ReDial validation set and re-select instructions for five LLMs: Gemma 2 (9B and 2B) and Qwen 2.5 (7B, 3B, 1.5B). 
Table~\ref{tab:llm_generalizability} shows annotation correlation results for top-performing baselines of Table~\ref{tab:correlation_main}. The results indicate that adapting to Gemma 9B and Qwen 7B achieves performance comparable to FACE.
Interestingly, our method is highly effective for small models: FACE adapted to Gemma 2B outperforms G-Eval, which uses an LLM with 4x more parameters. A similar observation can be made for Qwen 3B and 1.5B.

\medskip \noindent \textbf{Generalization to chitchat conversations.}
To examine generalizability of FACE to another type of conversations, we evaluate FACE on the existing chitchat datasets: (1) \emph{USR-Persona}~\citep{Mehri:2020:USR} (based on PersonaChat~\citep{Zhang:2018:PDA}), containing personalized chit-chats, and (2) \emph{USR-Topical}~\citep{Mehri:2020:USR} (based on Topical-Chat~\citep{Gopalakrishnan:2019:TTK}), containing knowledge-grounded conversations. These datasets provide annotations for six evaluation aspects, of which ``maintains context'' is the only aspect that is similar to ours. We use the instruction pool for the relevance aspect and re-select optimal instructions using the validation set of USR-Persona for USR-Topical evaluation and vice versa, to ensure that the test set is completely unseen.

Table~\ref{tab:domain_generalizability} presents the results of FACE generalizability to chitchat conversations, with G-Eval results obtained using GPT-3.5 and 4.
On average, FACE outperforms all baselines except G-Eval with GPT-4, while FACE${\text{*}_{\hspace{-0.3em}\text{72B}}}$ outperforms all baselines.
This is especially striking, considering that FACE is completely blind to category of conversations and uses an LLM with a lower number of parameters than GPT. Additionally, using an open model for evaluation has the added value of reproducibility.

Based on the results of Tables~\ref{tab:llm_generalizability} and~\ref{tab:domain_generalizability}, we answer our second research question \textbf{(RQ4b)}: FACE-optimized instructions are highly generalizable to different LLMs and domains, by performing a simple adaptation of FACE to a new LLM/domain. The adaptation to larger models can even surpass the state-of-the-art method with GPT-4 as a backbone on chit-chat conversations.

\subsection{FACE Interpretability}
\label{ch:face:sec:results:interpretability}
\begin{table}[t!]
\centering
\setlength{\tabcolsep}{3.2pt}
\begin{tabular}{l||cc|cc|cc}
\hline
\multirow{2}{*}{\textbf{Methods}} &
\multicolumn{2}{c|}{\small\textbf{USR-Persona}} &
\multicolumn{2}{c|}{\small\textbf{USR-Topical}} &
\multicolumn{2}{c}{\textbf{Avg.}} \\
 &  $r$ & $\rho$ &  $r$ & $\rho$ & $r$ & $\rho$  \\
\Xhline{2pt}
ROUGE-L & 0.114 & 0.091 & 0.193 & 0.203 & 0.154 & 0.147 \\
BLEU-4 & 0.147 & 0.151 & 0.131 & 0.235 & 0.139 & 0.193 \\
METEOR & 0.250 & 0.256 & 0.250 & 0.302 & 0.250 & 0.279 \\
BERTScore & 0.188 & 0.157 & 0.214 & 0.233 & 0.201 & 0.195 \\
Dial-M & 0.400 & 0.390 & 0.370 & 0.400 & 0.385 & 0.395 \\
USR & 0.607 & 0.528 & 0.416 & 0.377 & 0.512 & 0.453 \\
UniEval & 0.616 & 0.580 & \textbf{0.595} & \textbf{0.613} & 0.605 & 0.597 \\
    G-Eval\textsuperscript{GPT-3.5} & 0.441 & 0.458 & 0.519 & 0.544 & 0.480 & 0.501 \\
G-Eval\textsuperscript{GPT-4} & 0.607 & 0.670 & 0.594 & 0.605 & 0.601 & 0.638 \\
\hline
FACE & 0.473 & 0.544 & 0.498 & 0.506 & 0.486 & 0.525 \\
FACE${\text{*}_{\hspace{-0.3em}\text{72B}}}$ & \textbf{0.681} & \textbf{0.697} & 0.570 & 0.582 & \textbf{0.625} & \textbf{0.639} \\
\hline
\end{tabular}
\caption{Results on generalizability of FACE to chitchat conversations. All FACE correlations are statistically significant with $p < 0.01$.}
\label{tab:domain_generalizability}
\end{table}

\begin{figure}[t]
\centering
\includegraphics[width=1.0\linewidth]{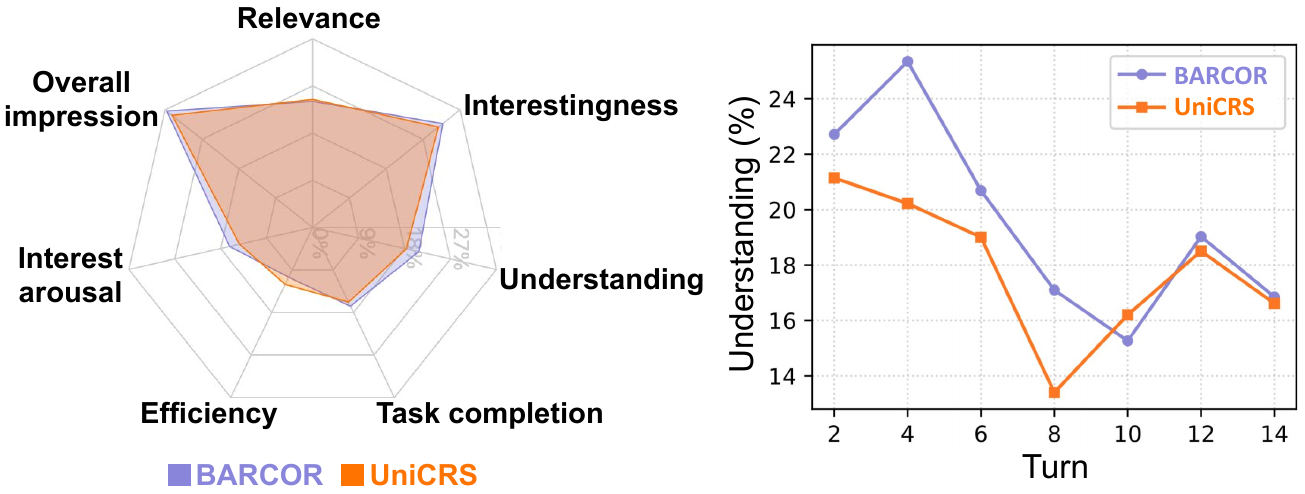}
\caption{Breakdown analysis comparing BARCOR and UNICRS.
Left: FACE evaluation results for each aspect. 
Right: Scores of Understanding aspect per system turn.
}
\label{fig:interpretability}
\end{figure}

We present a preliminary small-scale case study to demonstrate how FACE fine-grained scores can assist humans in identifying issues in CIA systems. Specifically, we compare FACE scores for two systems that are difficult to diagnose and receive contradictory evaluations from human and existing metrics:
BARCOR~\citep{Wang:2022:BTU} and UniCRS~\citep{Wang:2022:UCR}.
Humans and FACE prefer BARCOR (cf. Fig.~\ref{fig:crs_distribution} and \citep{Bernard:2025:CRS}), while recall-based metrics favor UniCRS~\citep{Wang:2023:REC}.
The left radar chart in Figure~\ref{fig:interpretability} shows FACE analysis for BARCOR and UniCRS for seven evaluation aspects.
Although the overall impression indicates similar performance, FACE scores show that UniCRS excels in relevance and efficiency, while BARCOR is better in user understanding and keeping users interested.
The right graph shows the detailed analysis of the user understanding aspect, where particle scores are aggregated for each turn. It is evident from the plot that BARCOR obtains higher scores in earlier turns. This indicates that BARCOR understands user preferences early on, which may explain its higher human preference.
These insights suggest that, while UniCRS excels in recommendations, overall performance can be improved by focusing on user understanding.

\begin{sidewaystable}
\caption{Example conversations from UniCRS and BARCOR, with particles scored in the top quartile shown in green and those in the bottom quartile shown in red.}
\label{tab:unicrs_barcor_example}
\centering
\begin{minipage}[t]{0.47\textwidth}
    \centering
    \begin{tabularx}{\linewidth}{l X}
    \hline
    \textbf{Role} & \textbf{Utterance} \\
    \hline
    User & Give me a title for the best fantasy based film \\
    Asst & I would recommend the following items 1 \hlgreen{The Lord of the Rings} 2 \hlgreen{Pan's Labyrinth (2006)} 3 \hlgreen{Avatar (2009)} \\
    User & which of the three do you think has the overall most comprehensive plotline \\
    Asst & \hlred{yes it does} \\
    User & Can you choose one that is most comprehensive \\
    Asst & \hlred{Yes, it does.} \\
    User & Do you not understand what I am asking? \\
    ... & ... \\
    \end{tabularx}
    \vspace{1ex}
    \textbf{UniCRS}
\end{minipage}
\hfill
\begin{minipage}[t]{0.52\textwidth}
    \centering
    \begin{tabularx}{\linewidth}{l X}
    \hline
    \textbf{Role} & \textbf{Utterance} \\
    \hline
    User & Hello! Can you give me a movie recommendation, please? \\
    Asst & \hlgreen{What kind of movies do you like?} \\
    User & Sci-fi, for instance \\
    Asst & I would recommend the following items: 1: \hlgreen{Blade Runner 2049 (2017)} 2: \hlgreen{Star Wars: Episode VIII -- The Last Jedi (2017)} 3: \hlgreen{Arrival (2016)} \\
    User & Thanks, those are great recommendations. Could you give me any comedy movies? \\
    Asst & \hlgreen{Have you seen The Hangover (2009)?} \\
    User & I haven't, but I have heard of it. How many Hangover movies are there? \\
    ... & ... \\
    \end{tabularx}
    \vspace{1ex}
    \textbf{BARCOR}
\end{minipage}
\end{sidewaystable}

To further understand each particle's contribution, we analyze particle scores per dialogue act and assess how they influence the overall impression of the systems. We found that particles associated with ``preference elicitation'' scored higher for BARCOR than UniCRS (45.5 vs. 39.0; scaled 0-100\%), which shows BARCOR's superior user understanding stems from more effective preference elicitation.
Table~\ref{tab:unicrs_barcor_example} shows example conversations from both CRSs, illustrating the general phenomenon captured by FACE scores. The examples demonstrate that while  UniCRS provides good initial recommendations, it fails to understand the user and elicit further preferences, whereas BARCOR successfully elicits preferences and recommends items matching user interests.
This difference can be overlooked when using item-level recall metrics, and it demonstrates the strength of FACE's interpretability.

Overall, we answer \textbf{(RQ4c)} positively: FACE can provide valuable insights into systems' behavior, which are useful for system improvement.

\section{Analysis}
\label{ch:face:sec:results:analysis}

\begin{figure}[t]
\centering
\includegraphics[width=0.90\linewidth]{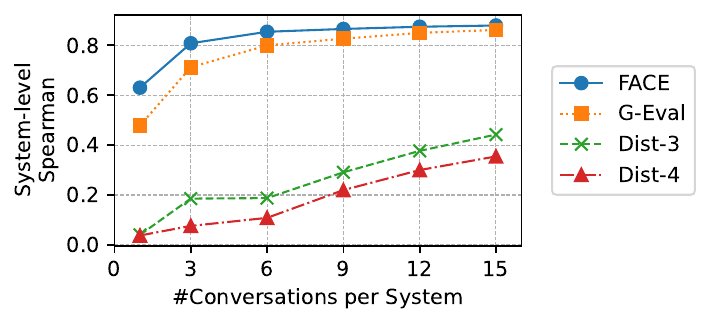}
\caption{Sample efficiency of different evaluation methods for the overall aspect on CRSArena-Eval.}
\label{fig:sample_efficiency}
\end{figure}

\textbf{Sample Efficiency.}
To determine the systems' score and create a system ranking, the evaluation method needs to be fed with a sample of user-system conversations.  To measure how many samples are needed to find a system ranking with a high correlation with human judgments, we plot system ranking correlations for various conversation counts per system in Figure~\ref{fig:sample_efficiency}.
The results indicate that FACE has strong sample efficiency; it achieves a Spearman correlation of 0.8 with gold rankings using only 3 dialogues per system, making it twice as efficient as the best-performing existing method, G-Eval.
Given that collecting human-system conversations require cost and effort, FACE's sample efficiency significantly enhances actual usability.

\medskip \noindent \textbf{Bias Analysis.}
We conduct analyses to see whether FACE exhibits length bias and self-bias (cf. Sec.~\ref{ch:face:sec:related}).
For length bias~\citep{Dubois:2024:LCA,Wang:2024:HCE}, using the CRSArena-Eval dataset, we examine the correlation between a system's average word count in conversations and the overall score.
We find that Pearson's correlations are 0.824 and 0.868 for FACE and humans, respectively. This indicates no sign of length bias compared to humans, which is in line with existing work~\citep{Chiang:2024:CAO,Dubois:2024:LCA} that report humans also favour longer responses, highlighting the nuanced nature of the LLM length bias.

For self-bias, where LLMs prefer system responses over human ones, we use FACE to evaluate pairs of system- and human-generated responses and see if they show any preferences compared to gold human annotators.
We examine two conversation types: USR-Persona for chit-chat, and a combination of CRSArena-Eval and AB-ReDial for a recommendation.
We could not find evidence for self-bias in FACE; e.g., for USR-Persona, FACE aligns with human preferences 77.8\% of the time when humans prefer human-generated responses and 71.4\% of the time when they prefer system-generated responses.

\section{Conclusion}

We present FACE, a fine-grained, aspect-based evaluation method for CIA systems.
It addresses the shortcomings of existing metrics, such as focusing on fixed dialogue history with reference-based metrics, limited generalizability of LLM-based metrics, and relying on non-granular scores with limited insights.
FACE is shown to strongly correlate with human judgments, generalize across LLMs and domains, and provide insights for system improvement.
Future work needs to address current limitations (see Section~\ref{ch:conclusion:sec:limitations}) by further examining evaluation biases, assessing effectiveness across broad domains, and exploring how FACE can help expert evaluators.

\chapter{Conclusion}
\label{ch:conclusion}

This chapter begins by describing our answers to the research questions laid out in Chapter~\ref{ch:intro}, followed by future work in the area of personalized conversational information access (CIA) systems.

\section{Answers to Research Questions}

The main research question that we set out to address, as discussed in Chapter~\ref{ch:intro}, was:

\vspace{1em}
\noindent
\hspace{2.8em}%
\begin{minipage}[t]{0.83\linewidth}
    \textbf{RQ Main:} \emph{How can we develop personalized CIA systems and evaluate them effectively?}
\end{minipage}
\vspace{1em}

\noindent To answer this question, we focused on the four sub-questions from three key areas: personal information extraction, personalized response generation, and conversation evaluation.

\subsection{Personal Information Extraction}

To personalize CIA systems, understanding the user's context is crucial.
One of the promising techniques for extracting personal information from conversations is Entity Linking (EL), which is proven to be effective in documents and short queries.
However, while EL for those use cases has been extensively studied, research on its application in conversational settings has been limited.
Therefore, we began investigating the characteristics of EL in conversations by addressing the first research question:

\vspace{1em}
\noindent
\hspace{2.8em}%
\begin{minipage}[t]{0.83\linewidth}
    \textbf{RQ1:} \textit{What does Entity Linking in conversations entail?}
\end{minipage}
\vspace{1em}

\noindent To address this question, we introduced ConEL in Chapter~\ref{ch:conel}, a dataset of conversations annotated with named entities, general concepts, and personal entities.
Through our analysis of this dataset, we discovered that human annotators identified not only named entities but also general concepts as important entities in conversations, with personal entities showing a particularly high presence in social chat conversations.
Based on these findings, we concluded that \textit{entity linking in conversations entails linking entities that include general concepts, named entities, and personal entities to enhance machine understanding of users' utterances, in contrast to traditional approaches that focus mainly on named entities.}

Building upon this dataset, we subsequently analyzed the performance of traditional EL systems.
Our evaluation revealed that traditional EL tools were suboptimal for conversational contexts, as they particularly struggled to handle concepts and personal entities, which proved crucial for enabling personalized conversations.

This finding led us to the second research question:

\vspace{1em}
\noindent
\hspace{2.8em}%
\begin{minipage}[t]{0.83\linewidth}
\textbf{RQ2:} \textit{How can we develop an Entity Linking method to accurately identify important entities for personalized conversational systems?}
\end{minipage}
\vspace{1em}

\noindent To address this question, we proposed CREL in Chapter~\ref{ch:crel}, a conversational EL method designed for personalized conversational systems.
CREL employs coreference resolution to recognize and link personal entity mentions to their corresponding entries in a knowledge graph, thereby addressing the challenges of personal contextual understanding.
To enable the development and evaluation of CREL, we created ConEL-2, an extension of the ConEL dataset that contains 11 times more dialogues with personal entity annotations.
Our evaluation results demonstrate that CREL significantly outperforms traditional EL systems, particularly in handling personal entities, which is crucial for personalized CIA systems.

Overall, we defined the task of EL in conversations and demonstrated that existing EL systems were insufficient for conversational contexts; we then proposed CREL, an EL method for conversations that effectively addressed personal contextual challenges and significantly outperformed traditional EL systems.

\subsection{Personalized Response Generation}

In the previous section, we addressed the challenges of EL in conversational settings, which is a crucial step for personalized response generation.
However, even with an effective EL method, generating responses that effectively reflect a user's personal preferences remains a key challenge.

\vspace{1em}
\noindent
\hspace{2.8em}%
\begin{minipage}[t]{0.83\linewidth}
\textbf{RQ3:} \emph{How can users' personal preferences be utilized to generate personalized responses?}
\end{minipage}
\vspace{1em}

\noindent
To investigate this research question, we must first address a significant bottleneck: the limited availability of realistic large-scale dialogue datasets for training and evaluating personalized response generation models. Without datasets that accurately reflect real-world user interactions and preferences, it becomes impossible to effectively investigate how personal preferences can be leveraged in response generation.

In the first half of Chapter~\ref{ch:laps}, we proposed LAPS, a framework that leverages large language models (LLMs) to efficiently guide human workers in composing multi-session and multi-domain conversations that capture diverse user preferences.
Our experiments confirmed that this approach significantly accelerates dataset construction while maintaining the diversity and quality of human-written dialogues, effectively addressing the scalability constraints of existing approaches.
The results demonstrated that LAPS-produced conversations exhibit the same level of naturalness and diversity as expert-created ones, highlighting the framework's effectiveness in collecting real-world personalized conversational data.

In the second half of Chapter~\ref{ch:laps}, we addressed RQ3 by training a preference extraction model and performing personalized response generation using semi-structured user preferences (termed \textit{preference memory}), which are constructed from the preferences extracted by the model.
The experiments demonstrated that utilizing preference memory enhances effective preference utilization, provides explanations for recommendations, and mitigates LLMs' recall limitations during response generation.
This suggests the effectiveness of using extracted personal preferences for generating personalized responses.

\subsection{Conversation Evaluation}

The evaluation of personalized CIA systems is crucial, yet existing methods often fail to assess the whole conversation across diverse interaction trajectories. This highlights the necessity for more effective, automatic evaluation, which motivated our fourth research question:

\vspace{1em}
\noindent
\hspace{2.8em}%
\begin{minipage}[t]{0.83\linewidth}
    \textbf{RQ4:} \emph{How to automatically evaluate personalized CIA systems in a way that accounts for natural conversation flow and provides fine-grained insights into system performance?}
\end{minipage}
\vspace{1em}

\noindent To answer this question, we introduced FACE in Chapter~\ref{ch:face}.
FACE is a fine-grained, aspect-based evaluation method for CIA systems that captures the natural conversation flow and provides detailed insights into system performance.
FACE employs an algorithm that first decomposes system responses into fine-grained ``conversation particles.''
It then evaluates these particles using an LLM guided by a diverse set of instructions. These instructions are iteratively refined through a prompt optimization process that leverages ``textual gradients,'' which is a form of natural language feedback, and uses beam search to select the instructions that best correlate with human judgments.

To enable the evaluation of automatic evaluators (meta-evaluation), we created CRSArena-Eval, a new meta-evaluation dataset containing 20,962 human annotations on 467 conversations with nine diverse conversational recommender systems (CRSs). The annotations cover seven key turn- and dialogue-level aspects, allowing for a robust meta-evaluation of how well automatic evaluators correlate with human judgments.

Evaluation of FACE on CRSArena-Eval demonstrates that FACE closely aligns with human judgments, significantly outperforming traditional automatic evaluation methods.
In addition, we demonstrated FACE's generalizability, showing its optimized instructions can be effectively adapted to other LLMs and to a different conversational domain, namely chit-chat.
Moreover, we showed that FACE's fine-grained scores can be used to identify the issue within the conversation, and provide insights for system improvement.

\section{Positioning of the Thesis}

This section positions the contributions of this thesis in relation to existing work in the CIA field.

\medskip \noindent \textbf{Personal Information Extraction.}
Chapters~\ref{ch:conel} and \ref{ch:crel} advance personal information extraction by addressing EL challenges in conversations, overcoming limitations of existing EL methods for well-written documents~\citep{vanHulst:2020:REL,Cao:2021:GENRE} that perform suboptimally in conversational contexts. Recent advances by \citet{Li:2023:EHP,Aliannejadi:2024:IKAT} demonstrate the growing importance of personal preference extractions, positioning this work as foundational for extracting preferences to populate such personal knowledge bases/graphs from conversational text. The methods developed here can serve as a basis for further exploration into more context-aware personalized CIA systems.

\medskip \noindent \textbf{Personalized Response Generation.}
Chapter~\ref{ch:laps} contributes a scalable, LLM-augmented method for collecting personalized dialogue data, addressing the scalability challenge of prior human-written collections from studies~\citep{Radlinski:2019:CCP,Bernard:2023:MGS}. This work is positioned to complement the recent LLM post-training paradigm~\citep{Ouyang:2022:TLM,Zhou:2023:LMA} by providing a method for collecting the authentic preference-grounded data that can be used for model alignment to further enhance personalized CIA systems.

\medskip \noindent \textbf{Conversation Evaluation.}
Chapter~\ref{ch:face} introduces a fine-grained automatic evaluation method that addresses the limitations of existing approaches~\citep{Zhang:2020:BET,Mehri:2020:USR,Liu:2023:GNE,Zhong:2022:TUM}, which produce uninterpretable scores and fail to assess the whole conversation or diverse interaction trajectories. This work is positioned to strengthen the LLM-as-a-judge paradigm through automated instruction optimization, addressing LLM biases identified by multiple studies~\citep{Wang:2024:HCE,Dubois:2024:LCA,Xu:2024:PPL,Liu:2023:GNE,dietz:2025:llmtropes,Faggioli:2023:PLL,Clarke:2024:LRA}, while providing interpretable scores for developing advanced user simulators and meta-evaluation benchmarks.

\section{Limitations}
\label{ch:conclusion:sec:limitations}

This section discusses the limitations of this thesis.

\medskip \noindent \textbf{General Limitations.}
As we have discussed in Chapter~\ref{ch:intro}, one of the limitations of this thesis is that our focus has been restricted to text-based interactions. However, the growing popularity of voice assistants and visual chatbots signals a shift toward multi-modal interactions~\citep{Zamani:2023:CIS}.
To solve this limitation, exploration of multi-modal extensions incorporating voice or visual inputs is needed.

\medskip \noindent \textbf{Ethical Risks of Social Chat.}
Although this thesis focuses on information access, several chapters involve social chat as a conversational setting for analysis and evaluation.
We acknowledge that such interaction is not entirely innocuous; social chat systems can foster over-reliance and emotional dependency, posing risks to user autonomy and privacy~\citep{Akbulut:2024:AMR}.
While the mitigation of these risks fall outside the scope of this thesis, we acknowledge that the importance of recognizing these potential harms is fundamental to the ethical design of conversational systems.

\medskip \noindent \textbf{Personal Information Extraction.}
In Chapter~\ref{ch:crel}, we proposed CREL, a conversational EL method for conversations.
However, the current method only focuses on personal entity mentions starting with possessive pronouns such as ``my'' or ``our''. Therefore, it cannot capture personal entities expressed through complex references, such as ``the car I owned.''
The mitigation strategy might involve training a personal entity mention detection model using the same approach as mention detection in traditional EL systems~\citep{vanHulst:2020:REL}, but with a dataset specifically targeting personal mentions that can be built upon the same annotation strategy we developed in ConEL (Chapter~\ref{ch:conel}) and ConEL-2 (Chapter~\ref{ch:crel}). Another limitation is that our method may not generalize to domain-specific knowledge bases with long-tail entities~\citep{hoveyda-2024-real}, and requires domain adaptation.

\medskip \noindent \textbf{Personalized Response Generation.}
In Chapter~\ref{ch:laps}, we introduced LAPS, a scalable approach for creating personalized conversational datasets by utilizing LLMs to guide human annotators. A limitation, however, is that creating task descriptions and specific LLM prompts still requires manual effort for each new task setting or domain.
Future research should aim to moderate this process to allow researchers to more conveniently adapt frameworks to various settings, thus improving scalability and usability further.

\medskip \noindent \textbf{Conversation Evaluation.}
In Chapter~\ref{ch:face}, we introduced FACE, a fine-grained, aspect-based conversation evaluation method. A few limitations are as follows:
(1) While FACE is a general method applicable to various conversations, this chapter evaluated it only on CRSs and one aspect for chit-chat; further exploration in other domains/aspects is needed.
(2) Similarly, the proposed meta-evaluation dataset, CRSArena-Eval, specifically targets CRSs, which is a subset of the broader field of information-access conversations which often involve more complex user needs and mixed initiatives~\citep{Zamani:2023:CIS}; thus, incorporating broader types of information-access conversations is needed.
(3) Although FACE did not exhibit bias in our analysis (Sec.~\ref{ch:face:sec:results:analysis}), evaluating unknown or emerging biases~\citep{Fang:2025:LLM} remains underexplored.
(4) Knowing the limitations of LLM-based evaluation~\citep{dietz:2025:llmtropes,Clarke:2024:LRA,Faggioli:2023:PLL,Takehi:2024:LRA}, we emphasize that FACE may not replace expert human evaluations.
Instead, it facilitates research and development on conversational systems by offering a scalable and efficient method for evaluating some potential known issues of CRSs~\citep{Dubois:2024:ASF}.

\section{Future Directions}

This section discusses several promising future research directions in the area of personalized CIA systems.

\subsection{Personal Information Extraction}

\medskip \noindent \textbf{Representation of Personal Information.}
Chapters~\ref{ch:conel} and \ref{ch:crel} demonstrated the extraction of personal information from conversations using EL methods.
However, determining \textit{how} to store personal information for CIA systems remains an open question.
While knowledge graphs (KGs), as employed in Chapters~\ref{ch:conel} and \ref{ch:crel}, represent one possible approach, alternative storage methods are being explored, including personalized textual knowledge bases (PTKBs)~\citep{Aliannejadi:2024:IKAT} and storing personal information within the parametric knowledge of neural networks~\citep{Sukhbaatar:2015:EEM,Zhang:2018:PDA}.
Textual representations, including KGs and PTKBs, offer higher transparency, control, and scrutability~\citep{Radlinski:2022:NLU}, while neural representations could provide easier optimization.
Each approach has its advantages and disadvantages, and therefore further research is needed to determine the optimal approach for storing personal information for effective personalization in CIA systems.

\medskip \noindent \textbf{Privacy-Aware Personal Information Extraction.}
Another important future direction involves addressing privacy concerns related to personal information.
With the growing popularity of agentic information-access systems~\citep{Trippas:2025:RFS,Hoveyda:2025:AOM}, which autonomously perform tasks on behalf of users, these systems are increasingly exposed to sensitive privacy information.
For instance, when a system accesses a user's medical records, it may inadvertently expose and store sensitive information about that user.
Therefore, future research should explore strategies for determining when personal information should \textit{not} be stored while still maximizing the benefits of personalization.

\subsection{Personalized Response Generation}

\medskip \noindent \textbf{Optimizing Personal Information Utilization.}
While Chapters~\ref{ch:conel}, \ref{ch:crel}, and \ref{ch:laps} have explored approaches to extract and utilize personal information, such as personal knowledge graphs (PKGs), a significant challenge remains in determining \textit{how} and \textit{when} to effectively retrieve and apply this PKG information within conversational systems. In the current LLM paradigm, conversational systems typically retrieve relevant PKG details and naively insert them into the model's input context~\citep{Aliannejadi:2024:IKAT}.
Future research could investigate more dynamic methods, potentially employing techniques similar to reinforcement learning from human feedback (RLHF)~\citep{Ouyang:2022:TLM} to train models that selectively retrieve and utilize PKG information based on conversational context and user preferences, thereby enhancing responses to better align with user needs.

\medskip \noindent \textbf{Factuality and Transparency.}
While CIA systems may generate responses that users find appealing, the prevalence of factual fabrications~\citep{Zamani:2023:CIS,Joko:2026:WIA} and sycophantic behavior~\citep{Amirshahi:2025:ERR,Sharma:2024:USL} in LLMs poses significant challenges to ensuring factual accuracy and transparency.
For instance, a system that generates an incorrect cure for serious diseases may receive positive user feedback despite being potentially harmful.
Similarly, a system that produces biased responses based on users' personal information might elicit favorable reactions while undermining fairness principles.
One potential solution involves implementing uncertainty estimation techniques~\citep{Kadavath:2022:LMK}, which assign confidence scores to generated responses, thereby enabling users to assess the reliability of the information they receive.
However, current uncertainty estimation methods have proven unreliable when external knowledge is incorporated into LLM contexts~\citep{Soudani:2025:WUE}, a common scenario in CIA systems.
Consequently, future research should prioritize developing more robust uncertainty estimation methods that enable CIA systems to provide confidence scores alongside their responses, thus enhancing factuality and transparency while empowering users to make informed decisions.

\subsection{Conversation Evaluation}

\medskip \noindent \textbf{Automatic Evaluation of Personalized CIA Systems.}
In Chapter~\ref{ch:face}, we proposed FACE, a fine-grained, aspect-based evaluation method for conversational systems.
However, as mentioned in the limitation section (Section~\ref{ch:conclusion:sec:limitations}), the evaluation was limited to CRSs and only one aspect for chit-chat, therefore the expansion to cover more general CIA systems is needed.
When expanding to general CIA systems, however, several challenges arise.
(1)~\textit{Differing knowledge levels among users} present a significant challenge, as users possess varying degrees of background knowledge that result in different information gains from identical responses~\citep{Trippas:2025:RFS}. For example, a novice user may require comprehensive explanations, whereas an expert may prefer concise yet highly technical responses.
(2)~\textit{Varying user preferences and priorities}, as different users may prioritize different aspects of the response. One might prioritize comprehensiveness, while another might value conciseness or creativity.
These challenges necessitate \textit{personalized evaluation criteria}, requiring the evaluation process itself to be also personalized.

\needspace{4\baselineskip}
\medskip \noindent \textbf{Realistic User Simulations for Evaluation.}
In Chapter~\ref{ch:face}, we utilized human-system dialogues to evaluate conversational systems, yet a key challenge is that collecting such dialogues, while easier than collecting expert annotations, could still require efforts. An alternative approach could involve the use of user simulations to generate user-system interactions~\citep{Bernard:2025:USD,Trippas:2025:RFS,Verberne:2015:USI}. Existing user simulation methods, however, struggle to produce conversations that are naturally diverse, frequently suffer from biases, and have difficulty accurately reflecting real user behaviors~\citep{Balog:2023:USE}.
For example, as discussed in Chapter~\ref{ch:laps}, LLM-simulated users tend to have lower lexical diversity compared to real humans.
Therefore, future research should focus on developing realistic user simulators, with key areas to explore including:
(1)~ensuring a broad range of simulated conversations to capture a realistic distribution of user behaviors and preferences, while
(2)~mitigating biases in simulations, such as over/under-representation of certain cultures, demographics, genders, or medical conditions, and
(3)~maximizing the effectiveness of simulations in terms of providing better feedback for system development and evaluation.

\appendix
\chapter{Resources}
\label{app:resources}

This appendix provides the links to the code repositories and datasets used in this thesis.
Note that some repositories include both code and datasets from their respective chapters.

\section{Code Repositories}
The code repositories of the methods proposed in this thesis are publicly available at:
\begin{itemize}
    \item \textbf{CREL} (Chapter~\ref{ch:crel}): The implementation of our conversational entity linking method. The repository also includes the ConEL-2 dataset.
    \begin{itemize}
        \item \href{https://github.com/informagi/conversational-entity-linking-2022}{\texttt{informagi/conversational-entity-linking-2022}}
    \end{itemize}
    \item \textbf{LAPS} (Chapter~\ref{ch:laps}): The implementation of our method for constructing large-scale personalized conversational datasets. The repository also includes the LAPS dataset.
    \begin{itemize}
        \item \href{https://github.com/informagi/laps}{\texttt{informagi/laps}}
    \end{itemize}
    \item \textbf{FACE} (Chapter~\ref{ch:face}): The implementation of our fine-grained, aspect-based conversation evaluation method. The repository also includes the CRSArena-Eval dataset.
    \begin{itemize}
        \item \href{https://github.com/informagi/face}{\texttt{informagi/face}}
    \end{itemize}
\end{itemize}

\section{Datasets}
The datasets released as part of this thesis are publicly available at:
\begin{itemize}
    \item \textbf{ConEL dataset} (Chapter~\ref{ch:conel}): The initial conversational entity linking dataset, annotated with named entities, personal entities, and concepts, comprises a total of 125 dialogues and 821 user utterances. The repository also contains \textbf{online resources listing around 130 conversational datasets} with a detailed comparison of their characteristics.
    \begin{itemize}
        \item \href{https://github.com/informagi/conversational-entity-linking}{\texttt{informagi/conversational-entity-linking}}
    \end{itemize}
    \item \textbf{ConEL-2 dataset} (Chapter~\ref{ch:crel}): A larger-scale resource for training and evaluating conversational EL systems. This dataset extends original ConEL with 290 multi-turn conversations, comprising 1,327 total utterances, and is annotated with a total of 2,833 entities, including personal entities, concepts, and named entities.
    \begin{itemize}
        \item \href{https://github.com/informagi/conversational-entity-linking-2022}{\texttt{informagi/conversational-entity-linking-2022}}
    \end{itemize}
    \item \textbf{LAPS dataset} (Chapter~\ref{ch:laps}): A large-scale collection of multi-session, human-written dialogues focused on personalization. The dataset spans the recipe and movie domains, containing 1,406 conversations and 11,215 extracted user preferences.
    \begin{itemize}
        \item \href{https://github.com/informagi/laps}{\texttt{informagi/laps}}
    \end{itemize}
    \item \textbf{CRSArena-Eval} (Chapter~\ref{ch:face}): A meta-evaluation dataset for automatic evaluation methods. It consists of 467 human-system dialogues containing 2,235 system turns across nine conversational recommender systems, annotated with 20,962 high-quality human judgments on seven different evaluation aspects.
    \begin{itemize}
        \item \href{https://github.com/informagi/face}{\texttt{informagi/face}}
    \end{itemize}
\end{itemize}

\chapter{Prompt Examples of FACE}
\label{app:prompts}

This appendix provides the prompts used in our FACE method in Chapter~\ref{ch:face}.\footnote{Note that in all prompts, the term ``nugget'' refers to ``conversation particle'' as defined in the paper. All used and optimized prompts can be found in our GitHub repository.}

\section{Prompts Used in FACE}

\bigskip \noindent \textbf{\large Conversation Particle Generation Prompt}

\vspace{0.5em}
\begin{adjustwidth}{1.5em}{1.5em}
\small\itshape
    \noindent Dialogue History: \texttt{\{dialogue\_history\}}\\
    Target Assistant Turn: \texttt{\{target\_turn\}}\\
    User's Response: \texttt{\{user\_response\}}

    \vspace{2mm}

    \noindent \textbf{Your task is to extract conversation nuggets, which are minimal, atomic units of information or facts from the target assistant turn.}
    
    \vspace{2mm}
    
    \noindent Each nugget consists of:
    \begin{itemize}
        \item \texttt{"dialogue\_act"}: one of the following labels: \texttt{"greeting," "preference elicitation," "recommendation," "goodbye," or "others."}
        \item \texttt{"nugget\_mention"}: the atomic unit of information from the target assistant turn. [...]
        \item \texttt{"user\_feedback"}: the excerpt of user feedback against the given nugget. [...]
    \end{itemize}
    
    \vspace{2mm}
    
    \noindent The output must be a JSON list of nuggets. [...]

    \vspace{2mm}

    \noindent Must think step by step:
    \begin{enumerate}
        \item Explain the dialogue history, the target assistant turn, and the user feedback.
        \item How many conversation nuggets are found in the target assistant turn?
        \item For each nugget, discuss the meaning of the user feedback.
        \item Output in JSON format.
    \end{enumerate}
\end{adjustwidth}
\vspace{0.5em}

\bigskip \noindent \textbf{\large Textual Gradient Prompt $\nabla$}

\vspace{0.5em}
\begin{adjustwidth}{1.5em}{1.5em}
\small\itshape
    \noindent \textbf{Examine the original instructions, predicted nugget score, and gold dialogue (or turn) score.}
    
    \begin{itemize}
        \item Based on the gold dialogue (or turn) score, is the predicted nugget score reasonable?
        \item Does original instructions describe how to use the nugget's information correctly?
        \item Necessary to edit the original instructions?
    \end{itemize}
\end{adjustwidth}
\vspace{0.5em}

\bigskip \noindent \textbf{\large Instruction Rewriting Prompt $\delta$}

\vspace{0.5em}
\begin{adjustwidth}{1.5em}{1.5em}
\small\itshape
    \noindent \textbf{Propose new instructions of \textasciitilde50 words based on the feedback.}
    \begin{itemize}
        \item Note that the full dialogue can be changed, thus your new instructions must be general enough to handle different contexts.
        \item Note that the task is "nugget" evaluation, not "turn" or "dialogue" evaluation; thus, the new instructions should focus on how to use the nugget.
        \item Must provide "task description" and explicit "step-by-step instructions" for the nugget evaluation; in step-by-step instructions labeling each step as "Step 1," "Step 2," and so on.
        \item Break down the evaluation into smaller steps and provide a checklist (``Does the nugget...?'' or ``Is this nugget...?'') for each step.
        \item [...]
    \end{itemize}
\end{adjustwidth}
\vspace{0.5em}

\bigskip \noindent \textbf{\large Initial Prompt (Before Optimization)}

\vspace{0.5em}
\begin{adjustwidth}{1.5em}{1.5em}
\small\itshape
    \noindent \textbf{Task description:} Given the dialogue, evaluate the quality of the target nugget based on the \texttt{\{evaluation\_aspect\}}.

    \vspace{0.5em}
    \noindent \textbf{Step-by-step instructions:}
    \begin{itemize}
        \item \textbf{Step 1:} Read the dialogue history, target nugget, and user's response.
        \begin{itemize}
            \item What does the target nugget convey?
        \end{itemize}
        
        \item \textbf{Step 2:} Carefully read the grading criteria.
        \begin{itemize}
            \item What are the grading criteria?
        \end{itemize}
        
        \item \textbf{Step 3:} Evaluate the target nugget.
        \begin{itemize}
            \item Which grade should be assigned to the target nugget?
        \end{itemize}
    \end{itemize}
\end{adjustwidth}
\vspace{0.5em}

\section{Instructions Optimized by FACE}

Here, we provide the optimized instruction examples for the dialogue-level overall impression aspect and the turn-level relevance aspect.
Please note that, in the actual process, FACE optimizes \textbf{multiple} instructions for each aspect, as shown in Figure~\ref{fig:method-illustration}.

\medskip \noindent \textbf{Optimized Instructions for Overall Impression Aspect (Dialogue-level)}
\vspace{0.5em}
\begin{adjustwidth}{1.5em}{1.5em}
\small\itshape
    \noindent \textbf{Task description:} Evaluate the nugget based on its relevance, accuracy, and usefulness.

    \vspace{0.5em}
    \noindent \textbf{Step-by-step instructions:}
    \begin{itemize}
        \item \textbf{Step 1:} Check if the nugget is relevant to the conversation.
        \begin{itemize}
            \item Does the nugget relate to the dialogue context?
            \item Is the nugget a direct response to the user's question or concern?
            \item Is the nugget related to the user's preferences or interests?
        \end{itemize}
        
        \item \textbf{Step 2:} Evaluate the nugget's accuracy.
        \begin{itemize}
            \item Is the information in the nugget accurate based on the dialogue?
            \item Does the nugget correctly represent the conversation?
        \end{itemize}
        
        \item \textbf{Step 3:} Assess the nugget's usefulness.
        \begin{itemize}
            \item Does the nugget provide a helpful or relevant suggestion?
            \item Does the nugget address the user's needs or concerns?
            \item Does the nugget facilitate a meaningful continuation of the conversation?
        \end{itemize}
    \end{itemize}
\end{adjustwidth}
\vspace{0.5em}

\noindent \textbf{Optimized Instructions for Relevance Aspect (Turn-level Aspect)}

\vspace{0.5em}
\vspace{0.5em}
\begin{adjustwidth}{1.5em}{1.5em}
\small\itshape

    \noindent \textbf{Task description:} Evaluate the quality of the target nugget based on its relevance to the user's request.

    \vspace{0.5em}
    \noindent \textbf{Step-by-step instructions:}
    \begin{itemize}
        \item \textbf{Step 1:} Identify the user's request and the nugget's suggestion.
        \begin{itemize}
            \item Step 1.1: Does the nugget's suggestion directly address the user's request?
            \item Step 1.2: Is the nugget's genre or category aligned with the user's interest?
        \end{itemize}
        
        \item \textbf{Step 2:} Assess the nugget's relevance.
        \begin{itemize}
            \item Step 2.1: Does the nugget's information accurately address the user's need?
            \item Step 2.2: Is the nugget's suggestion consistent with the user's preferences or interests?
        \end{itemize}
    \end{itemize}
\end{adjustwidth}
\vspace{0.5em}

\cleardoublepage
\phantomsection
\addcontentsline{toc}{chapter}{Bibliography}
\bibliographystyle{abbrvnat}
{
	\scriptsize
	\bibliography{./refs.bib}
}

\clearpage
\begingroup
\let\cleardoublepage\clearpage
\chapter*{Summary}
\addcontentsline{toc}{chapter}{Summary}
\markboth{Summary}{Summary}

Conversational interactions have reshaped information retrieval systems, as users increasingly favor direct answers over traditional hyperlinks.
This paradigm shift has reinforced the importance of conversational information access (CIA) systems that ground responses in retrieved knowledge to provide reliable information.
To build more personalized and reliable CIA systems, this thesis addresses key challenges by focusing on three areas: (1) personal context extraction, (2) personalized response generation, and (3) effective and interpretable system evaluation.

First, we tackle personal context extraction.
We introduce the ConEL dataset and define the problem of Entity Linking (EL) in conversations.
We find that traditional tools struggle with the unique challenge of identifying personal entities and concepts.
To address this, we propose CREL, a novel EL method tailored for this setting.
We find that CREL significantly outperforms these tools, accurately identifying all entity types vital for personalization.

Second, we focus on personalized response generation.
Effectively utilizing extracted personal context to generate personalized responses remains challenging without large-scale dialogue datasets capturing realistic user preferences across sessions.
We introduce LAPS, a scalable method using large language models (LLMs) to guide human workers in composing high-quality dialogues that capture diverse, real user preferences.
Using the dataset from LAPS, we investigate how personal preferences can be utilized for personalizing responses.
We show that using semi-structured preference memory, compared to appending raw conversation history, improves the accuracy of preference utilization and explanations in responses while mitigating LLMs' recall issues.

Finally, we address the need for effective and interpretable system evaluation.
We introduce FACE, an automatic, reference-free evaluation method that provides a fine-grained, aspect-based assessment of entire conversations.
It decomposes system responses into conversation particles and uses an LLM with optimized instructions to generate interpretable scores.
We find that FACE's evaluations align closely with human judgments, offering actionable insights while outperforming existing evaluation methods.

In this thesis, alongside methodological contributions, we place a significant focus on resource contributions.
Specifically, we release the ConEL, LAPS, and CRSArena-Eval datasets for conversational entity linking, training personalized dialogue systems, and the meta-evaluation of conversation evaluators, respectively.
These resource contributions, coupled with the method and theoretical advancements the thesis introduces, further the field of personalized CIA research.

\endgroup

\end{document}